\documentclass[a4paper,11pt]{article}
\usepackage{a4wide,amsmath,amssymb,bm,bbm}
\usepackage[makeroom]{cancel}
\usepackage[english]{babel}
\usepackage[utf8]{inputenc}

\usepackage{color}

\usepackage{hyperref,enumerate}
\hypersetup{
    colorlinks,
    linkcolor={black},
    citecolor={black},
    urlcolor={black}
}

\DeclareMathOperator{\tr}{tr}
\renewcommand{\o}{\overline}
\renewcommand{\u}{\underline}

\makeatletter

\@addtoreset{equation}{section} \makeatother

\begin{document}
\setcounter{page}{0}

\hfill
\vspace{30pt}

\begin{center}
{\huge{\bf {The $\alpha'^2$ correction from double field theory}}}

\vspace{80pt}

Stanislav Hronek,\quad Linus Wulff\quad and\quad Salomon Zacar\'ias

\vspace{15pt}

\small {\it Department of Theoretical Physics and Astrophysics, Faculty of Science, Masaryk University\\ 611 37 Brno, Czech Republic}
\\
\vspace{12pt}
\texttt{436691@mail.muni.cz,\,wulff@physics.muni.cz,\,zacarias.salomon84@gmail.com}\\

\vspace{80pt}

{\bf Abstract}
\end{center}
\noindent
It is known that the order $\alpha'$ correction to the tree-level effective action for the bosonic and heterotic string can be described in the framework of Double Field Theory (DFT). Here we determine the DFT action and transformations at order $\alpha'^2$ by a direct calculation. The result is vastly simpler than previous proposals. We show that this correction reproduces the known $\alpha'^2$ correction to the heterotic string effective action. The relation of our action to an (implicit) all order proposal coming from the so-called generalized Bergshoeff-de Roo identification is also discussed.

\clearpage
\tableofcontents

\section{Introduction}
String theory compactified on a $d$-torus has an $O(d,d;\mathbbm Z)$ T-duality symmetry. When we take the string coupling to zero this is `enhanced' to a continuous $O(d,d;\mathbbm R)$ symmetry of the dimensionally reduced low-energy effective action (to all orders in $\alpha'$) \cite{Meissner:1991zj,Sen:1991zi,Meissner:1991ge}. Double Field Theory (DFT) \cite{Siegel:1993th,Hull:2009mi,Hohm:2010pp} is an attempt to re-formulate the (tree-level) string effective action so that this symmetry becomes manifest already \emph{before} dimensional reduction.\footnote{One can also consider a DFT description of the dimensionally reduced theory only, but this is not what we have in mind here.} To do this one doubles the spacetime dimension $D\rightarrow 2D$, which allows for a formulation with an $O(D,D)$ global symmetry acting on the coordinates. An $O(D,D)$ invariant `section condition' is then imposed which, when solved, reduces the physics to be $D$-dimensional.

A priori it is not clear that such a re-formulation of the string low-energy effective action should be possible. But surprisingly it was found that supergravity can indeed be reformulated with a manifest $O(D,D)$ symmetry. The next surprise was that it is also possible to cast the first $\alpha'$ correction to the heterotic and bosonic string in manifestly $O(D,D)$ invariant DFT form \cite{Marques:2015vua} (see also \cite{Hohm:2013jaa,Bedoya:2014pma,Hohm:2014xsa,Coimbra:2014qaa,Deser:2014wva,Deser:2017fko}). However, in hindsight the existence of an $O(D,D)$ invariant description of the first $\alpha'$ correction is not so surprising. The reason is that this correction can be generated from an \emph{uncorrected} DFT action by a version of a trick originally introduced by Bergshoeff and de Roo \cite{Bergshoeff:1988nn}. They considered supergravity coupled to vectors and observed that the supersymmetry transformations of the vectors and a suitably shifted spin connection take the same form. It is therefore consistent to identify them and in doing so generate a higher-derivative Riemann squared correction from the standard $F^2$ kinetic term for the gauge field. They applied this trick to the heterotic string effective action to generate the first two $\alpha'$ corrections. The same trick can be applied in DFT. One first has to couple DFT to vectors, which is easily done by starting in a higher dimension and dimensionally reducing. These vectors can then be identified with components of the generalized spin connection. This was carried out in \cite{Baron:2018lve,Baron:2020xel} and it was shown that one recovers the first $\alpha'$ correction of \cite{Marques:2015vua} (a somewhat different approach was discussed earlier in \cite{Lee:2015kba}). Unlike in the supergravity setting this `generalized Bergshoeff-de Roo trick can be implemented exactly, i.e. there is no need to add additional $\alpha'$ corrections. Instead the recursive form of the identification leads to an infinite series of $\alpha'$ corrections. It is natural to think of these as the T-duality completion of the first order correction required by the Green-Schwarz anomaly cancellation mechanism for the heterotic string \cite{Green:1984sg} (though this works also for the bosonic string). Since this formulation implicitly contains all orders in $\alpha'$ it is in principle straightforward to extract the action and transformations to any order one is interested in {(of course the string effective action contains also other corrections, the first being the $\zeta(3)\alpha'^3$ Riemann$^4$ correction, which likely cannot be captured by DFT \cite{Hronek:2020xxi})}. This was carried out to order $\alpha'^2$ in \cite{Baron:2020xel} by solving the implicit identification recursively. Unfortunately, doing this in the most straightforward way leads to enormously long expressions already at this order. For example the DFT action they found for the heterotic case consisted of around 200 terms at order $\alpha'^2$! It was later shown in \cite{Hronek:2021nqk} that these expressions were highly redundant and many terms either canceled or could be removed by field redefinitions and that going to supergravity these expressions produced the correct cubic terms (Riemann cubed in the bosonic case and no cubic term in the heterotic case).

It is therefore clear that the story of the $\alpha'^2$ corrections in DFT could be simplified a lot, but it is not clear exactly how much. There also remains the question of matching the full expression to supergravity. Here we will address these two questions focusing mainly on the heterotic case. Instead of starting from the generalized Bergshoeff-de Roo approach we construct the $\alpha'^2$ correction directly by starting with the known action at order $\alpha'$ and the corresponding correction to the double Lorentz transformations. The second order action should then be such that its lowest order variation cancels against the corrected variation of the order $\alpha'$ action, after some suitable corrections to the transformations at order $\alpha'^2$. We find that this problem is relatively easy to solve in the heterotic case leading to quite a simple action and transformations for DFT up to order $\alpha'^2$. In the notation of \cite{Baron:2020xel} the action up to this order takes the form
\begin{equation}
S=\int dX\,e^{-2d}L=\int dX\,e^{-2d}\left(\mathcal R+a\mathcal R^{(0,1)}+b\mathcal R^{(1,0)}+a^2\mathcal R^{(0,2)}+ab\mathcal R^{(1,1)}+b^2\mathcal R^{(2,0)}\right)\,,
\label{eq:action}
\end{equation}
where the parameters $a,b$ are proportional to $\alpha'$ with $(a,b)=(-\alpha',0)$ for the heterotic string and $(a,b)=(-\alpha',-\alpha')$ for the bosonic string. Here we find that in the heterotic case the full Lagrangian up to this order can be written as
\begin{equation}
\begin{aligned}
L=&\,{}
-2(\partial^a-F^a)(\partial^b-F^b)\mathcal M^{ab}
+2\partial^aF^b\mathcal M^{ab}
-F_a{}^{bc}F_a{}^{de}\mathcal M^{bd}\mathcal M^{ce}
\\
&{}
+\tfrac13(F^{abc}+aM^{abc})(F^{def}+aM^{def})\mathcal M^{ad}\mathcal M^{be}\mathcal M^{cf}
-\frac{a}{2}\mathcal R^{ab}{}_{cd}\mathcal R^{ef}{}_{cd}\mathcal M^{ae}\mathcal M^{bf}
\\
&{}
-aF^{abC}\partial_CF^f{}_{de}F^g{}_{de}\mathcal M^{af}\mathcal M^{bg}
-\frac{a^2}{4}F^a{}_{de}\mathcal R^{bc}{}_{de}F^a{}_{fg}\mathcal R^{bc}{}_{fg}
+\frac{a^2}{2}F^a{}_{fg}\partial^bF^c{}_{fg}F^b{}_{de}\partial^aF^c{}_{de}\,,
\end{aligned}
\label{eq:La2}
\end{equation}
where we have defined
\begin{equation}
M^{ab}=F^a{}_{cd}F^b{}_{dc}\,,\qquad M^{abc}=F^a{}_{de}F^b{}_{ef}F^c{}_{fd}
\label{eq:Ms}
\end{equation}
and introduced the new ``metric''
\begin{equation}
\mathcal M^{ab}=[(\eta+\frac{a}{2}M)^{-1}]^{ab}=\eta^{ab}-\frac{a}{2}M^{ab}+\frac{a^2}{4}(M^2)^{ab}+\ldots\,.
\label{eq:Mcal}
\end{equation}
The DFT notation is explained in the next section. The form of this action, {which mostly consists of dressing the lower order action by $\mathcal M$}, suggests that it probably captures several of the terms at higher orders in $\alpha'$ as well, but it cannot be the full answer at order $\alpha'^3$ since we know that there should be a quartic Riemann term present in that case \cite{Bergshoeff:1989de}. The correction to the double Lorentz transformations takes the form
\begin{equation}
\delta E^{aM}E_{bM}=\frac{a}{2}\partial_b\u\lambda_{cd}F^a{}_{cd}+a^2(D_d-F_d)\left(\partial_{[b}\u\lambda_{|fg|}\o D_{d]}F^a{}_{fg}\right)\,.
\label{eq:DLT}
\end{equation}
{There is a price to be paid for this} simplicity of the action and transformations, namely the Lagrangian is not invariant but transforms by a total derivative
\begin{equation}
\delta L=4a^2(\partial_c-F_c)\left(\mathcal R^b{}_d\partial_{[c}\u\lambda_{|fg|}\o D_{d]}F^b{}_{fg}\right)
\label{eq:delta-L}
\end{equation}
and the double Lorentz transformations close only up to equations of motion at this order. {On the other hand, the Lagrangian and transformations obtained by the generalized Bergshoeff-de Roo identification \cite{Baron:2018lve} must be invariant and close off-shell. In section \ref{sec:off-shell} we show how to obtain a Lagrangian and transformations with these properties, which take a more complicated form, directly from our results and verify that the transformations agree with those found in \cite{Baron:2018lve}.

The above DFT action and transformations are shown to reproduce the action and transformations of the heterotic string (setting the gauge fields and fermions to zero) to order $\alpha'^2$. We also discuss a version of the generalized Bergshoeff-de Roo identification, which leads to some simplifications compared to \cite{Baron:2020xel}, which leads directly (upon certain field redefinitions) to the above action and transformations. Finally we derive also the DFT action for the case of the bosonic string (up to total derivative terms), but that case turns out to be considerably more involved.
}

The rest of this paper is organized as follows. Section \ref{sec:DFT} gives a summary of the double field theory notation and identities that we will use in the rest of the paper. In section \ref{sec:R01} we describe the first $\alpha'$ correction to DFT, while section \ref{sec:R02} contains the extension of this result to the next order for the heterotic case. We show, in section \ref{sec:match}, that the DFT action and transformations reproduce the tree-level effective action for the heterotic string up to order $\alpha'^2$. Section \ref{sec:gBdR} gives a brief description of the generalized Bergshoeff-de Roo identification and we find the field redefinitions relating {a version of this identification} to the results of section \ref{sec:R02}. Finally we describe, in section \ref{sec:R11}, the additional terms needed in the DFT description of the bosonic string at order $\alpha'^2$. We end with some conclusions. Certain details of the calculations in the last two sections can be found in the appendix.

\section{DFT notation and identities}\label{sec:DFT}
We will use the so-called flux formulation of DFT \cite{Geissbuhler:2011mx,Geissbuhler:2013uka} (see also \cite{Hohm:2010xe}). The basic fields are the generalized vielbein, which we parametrize as
\begin{equation}
E_A{}^M=
\frac{1}{\sqrt2}
\left(
\begin{array}{cc}
e^{(+)a}{}_m-e^{(+)an}B_{nm} & e^{(+)am}\\
-e^{(-)}_{am}-e^{(-)}_a{}^nB_{nm} & e^{(-)}_a{}^m
\end{array}
\right)
\label{eq:E}
\end{equation}
and the generalized dilaton
\begin{equation}
e^{-2d}=e^{-2\Phi}\sqrt{-G}\,.
\end{equation}
In this formulation there is a global $O(D,D)$ symmetry which rotates the doubled coordinate indices $M,N,\ldots$ and a local double Lorentz $O(D-1,1)\times O(D-1,1)$ symmetry rotating the doubled Lorentz indices $A,B,\ldots$. The two vielbeins $e^{(\pm)}$ for the metric $G_{mn}$, transform only under the first, respectively second, Lorentz group factor. The standard supergravity description is recovered by fixing the gauge $e^{(+)}=e^{(-)}=e$, leaving only the diagonal copy of the Lorentz group (and solving the section condition to remove the dual coordinates). In this formulation a global $O(D,D)$ symmetry will be manifest. Instead, the local double Lorentz symmetry $O(D-1,1)\times O(D-1,1)$, required for consistency, will not be manifest and needs to be verified explicitly. {Let us note that, as we will see later, when we include higher derivative corrections the fields $e,B,\Phi$ above will be related to the standard supergravity fields by certain non-covariant field redefinitions.}

There are two constant metrics, the $O(D,D)$ metric $\eta^{AB}$ and the generalized metric $\mathcal H^{AB}$, which take the form\footnote{The $O(D,D)$ metric with lower indices takes the same form, while changing to coordinate indices it takes the form
$$
\eta^{MN}=
\left(
\begin{array}{cc}
	0 & \delta_m^n\\
	\delta_n^m & 0
\end{array}
\right)
$$
and similarly with lower indices.
}
\begin{equation}
\eta^{AB}=
\left(
\begin{array}{cc}
	\eta_{ab} & 0\\
	0 & -\eta^{ab}
\end{array}
\right)\,,\qquad
\mathcal H^{AB}=
\left(
\begin{array}{cc}
	\eta_{ab} & 0\\
	0 & \eta^{ab}
\end{array}
\right)\,,
\label{eq:eta}
\end{equation}
where $\eta=(-1,1,\ldots,1)$ is the $D$-dimensional Minkowski metric. We use the $O(D,D)$ metric to raise and lower doubled indices. The projection operators
\begin{equation}
P_\pm^{AB}=\frac12\left(\eta^{AB}\pm\mathcal H^{AB}\right)\,,
\label{eq:Ppm}
\end{equation}
are easily seen to project on upper and lower indices respectively. The analog of the spin connection and derivative of the dilaton are the ``generalized fluxes''
\begin{equation}
F_{ABC}=3\partial_{[A}E_B{}^ME_{C]M}\,,\qquad
F_A=2\partial_Ad-\partial^ME_{AM}
\,,\qquad
\partial_A\equiv E_A{}^M\partial_M\,.
\label{eq:fluxes}
\end{equation}
They are the only generalized diffeomorphism\footnote{The generalized diffeomorphisms encode the standard diffeomorphisms plus $B$-field gauge transformations. The generalized vielbein transforms as
$$
\delta E_A{}^ME_{BM}=2\partial_{[A}V_{B]}-F_{ABC}V^C\,,
$$
with the parameters encoded in the generalized vector $V_A$.
} scalars that can be constructed from one derivative of $E$ and $d$. They can be seen to satisfy the following Bianchi identities
\begin{equation}
4\partial_{[A}F_{BCD]}=3F_{[AB}{}^EF_{CD]E}\,,\qquad
2\partial_{[A}F_{B]}=-(\partial^C-F^C)F_{ABC}\,,
\end{equation}
while the commutator of two derivatives with a Lorentz index is
\begin{equation}
[\partial_A,\partial_B]=F_{ABC}\partial^C\,.
\end{equation}
In terms of these derivatives the section condition takes the form\footnote{The $\otimes$ notation means that the two derivatives act on different factors in an expression.}
\begin{equation}
\partial^A\otimes\partial_A=0\,,\qquad(\partial^A-F^A)\partial_A=0\,.
\end{equation}
Under a (general) variation of the generalized vielbein the fluxes transform as
\begin{equation}
\delta_E F_{ABC}=3\partial_{[A}\delta E_{BC]}+3\delta E_{[A}{}^DF_{BC]D}\,,\quad
\delta_E F_A=\partial^B\delta E_{BA}+\delta E_{AB}F^B\,,\quad\delta E_{AB}\equiv \delta E_A{}^ME_{BM}\,.
\label{eq:deltaEF}
\end{equation}

Using the (constant) projection operators (\ref{eq:Ppm}) we can split the capital Lorentz indices into upper/lower indices, $A=({}^a,\,{}_a)$, in an invariant way. Under this splitting we see that for example $F_{ABC}$ consists of four independent fields
\begin{equation}
F^{abc}\,,\qquad F_a{}^{bc}\,,\qquad F^a{}_{bc}\,,\qquad F_{abc}\,.
\end{equation}
Note that we are not allowed to raise/lower these indices in DFT. Furthermore we see from (\ref{eq:eta}) that a contracted $A$ index leads to a sum of a term with two upper indices contracted with the usual Minkowski metric minus a term with two lower indices contracted with the Minkowski metric. It is convenient to use a convention where repeated upper(lower) indices are understood to be contracted with the $D$-dimensional Minkowski metric, e.g.
\begin{equation}
F^AF_A=F^aF^a-F_aF_a\,.
\label{eq:F2}
\end{equation}

The nontrivial symmetry in this formalism is the local double Lorentz symmetry, since this is not manifest and must be checked by hand. This consists of two factors of the usual Lorentz group and we denote the corresponding infinitesimal parameters $\o\lambda^{ab}$ and $\u\lambda_{ab}$, respectively. The generalized fluxes transform as follows
\begin{equation}
\begin{aligned}[t]
\delta F^{abc}=&3\partial^{[a}\o\lambda^{bc]}+3\o\lambda^{[a|d|}F^{bc]d}\,,
\\
\delta F_a{}^{bc}=&\partial_a\o\lambda^{bc}-\u\lambda_{ad}F_d{}^{bc}+2\o\lambda^{[b|d}F_a{}^{d|c]}\,,
\\
\delta F^a=&\partial^b\o\lambda^{ba}+\o\lambda^{ab}F^b\,,
\end{aligned}
\qquad
\begin{aligned}[t]
\delta F_{abc}=&3\partial_{[a}\u\lambda_{bc]}-3\u\lambda_{[a|d|}F_{bc]d}\,,
\\
\delta F^a{}_{bc}=&\partial^a\u\lambda_{bc}+\o\lambda^{ad}F^d{}_{bc}-2\u\lambda_{[b|d}F^a{}_{d|c]}\,,
\\
\delta F_a=&-\partial_b\u\lambda_{ba}-\u\lambda_{ab}F_b\,.
\end{aligned}
\label{eq:deltaFs}
\end{equation}
Notice that contracted lower indices are always accompanied by extra minus signs as in (\ref{eq:F2}).

At the two-derivative level there is only one double Lorentz invariant, dubbed the generalized Ricci scalar\footnote{This is equal to the same expression with the positions of the indices reversed, as follows by computing $\partial^AF_A$ using the definition of $F_A$. This can therefore be thought of as an extra Bianchi identity for $F_A$.}
\begin{equation}
\mathcal R=4\partial^aF^a-2F^aF^a-F_a{}^{bc}F_a{}^{bc}+\tfrac13F^{abc}F^{abc}\,.
\end{equation}
Indeed, the DFT action
\begin{equation}
S=\int dX\,e^{-2d}\mathcal R
\end{equation}
reproduces the usual NSNS sector of the supergravity action (see below). Using (\ref{eq:deltaEF}) the variation with respect to the generalized vielbein and dilaton gives
\begin{equation}
\delta_E\mathcal R
=
4(\partial^a-F^a)(D_b-F_b)\delta E^a{}_b
-4\delta E^a{}_b\mathcal R^a{}_b
\,,
\qquad
\delta_d(e^{-2d}\mathcal R)
=e^{-2d}
\left(
8(\partial^a-F^a)\partial^a\delta d
-2\delta d\mathcal R
\right)\,,
\label{eq:deltaR}
\end{equation}
where the first term in both expressions gives a total derivative {when we take the $e^{-2d}$ factor into account}, so the DFT equations of motion are
\begin{equation}
\mathcal R^a{}_b=0\,,\qquad\mathcal R=0\,.
\end{equation}
Here we have introduced the generalized Ricci tensor
\begin{equation}
\mathcal R^a{}_b=\partial^aF_b-(\partial_c-F_c)F^a{}_{bc}+F_c{}^{da}F^d{}_{cb}\,.
\end{equation}
Defining $\mathcal R_a{}^b$ in the same way but with upper and lower indices exchanged (and remembering the extra minus signs for each pair of contracted lower indices) one finds from the Bianchi identities that $\mathcal R_b{}^a=\mathcal R^a{}_b$.

It is useful to define also a DFT analog of the Riemann tensor. Following \cite{Hronek:2020skb} we define\footnote{Defining $\mathcal R_{ab}{}^{cd}$ in the same way with upper and lower indices exchanged (and an extra sign for each pair of contracted lower indices) we have from the Bianchi identities that $\mathcal R_{cd}{}^{ab}=-\mathcal R^{ab}{}_{cd}$.}
\begin{equation}
\mathcal R^{ab}{}_{cd}=2\partial^{[a}F^{b]}{}_{cd}-F^{abe}F^e{}_{cd}+2F^{[a}{}_{ce}F^{b]}{}_{ed}\,.
\label{eq:Riem}
\end{equation}
It is important to note that, unlike the generalized Ricci tensor, this object does not transform in a covariant way. Instead one finds its double Lorentz transformation to be
\begin{equation}
\delta\mathcal R^{ab}{}_{cd}
=
2\o\lambda^{[a|e|}\mathcal R^{|e|b]}{}_{cd}
-2\u\lambda_{[c|e}\mathcal R^{ab}{}_{e|d]}
-\partial^e\o\lambda^{ab}F^e{}_{cd}
-F_e{}^{ab}\partial_e\u\lambda_{cd}\,,
\label{eq:deltaRie}
\end{equation}
where the last two terms would be absent for an object transforming covariantly. {This is consistent with general arguments that there is no standard notion of a Riemann tensor in DFT \cite{Hohm:2011si}}. Nevertheless this `generalized Riemann tensor' turns out to be very useful.

We also find it convenient to introduce ``semi-covariant'' derivatives acting on a generalized vector as follows
\begin{equation}
\begin{aligned}
&D_aV^b=\partial_aV^b-F_a{}^{bc}V^c\,,\qquad
D^aV^b=\partial^aV^b-\frac12F^{abc}V^c\,,\\
&D_aV_b=\partial_aV_b+\frac12F_{abc}V_c\,,\qquad
D^aV_b=\partial^aV_b+F^a{}_{bc}V_c\,.
\end{aligned}
\label{eq:Ds}
\end{equation}
Note that for example $D_aV^b$ transforms covariantly under the $\o\lambda$ transformation if $V^b$ transforms covariantly, but $D^aV^b$ does not. These derivatives satisfy
\begin{align}
[D^a,D^b]V_c=&\,{}-F_d{}^{ab}\partial_dV_c+\mathcal R^{ab}{}_{cd}V_d\,,\label{eq:comm-uul}\\
[D_c,D^a]V^b=&\,{}-\frac12\u D^dF_c{}^{ab}V^d+\u D^{(a}F_c{}^{b)d}V^d\,.\label{eq:comm-luu}
\end{align}
In the last expression we have defined $\u D$ to be the semi-covariant derivative with the connection acting only on the lower indices, e.g. 
\begin{equation}
\u D^dF_c{}^{ab}=\partial^dF_c{}^{ab}+F^d{}_{ce}F_e{}^{ab}\,.
\end{equation}
Similarly we define $\o D$ to act only on the upper indices. The commutators with the opposite placement of indices are easily found by exchanging upper and lower indices and keeping in mind that contracted lower indices come with an extra minus sign. Using these semi-covariant derivatives the Bianchi identities take a simpler form, in particular we have
\begin{align}
2D^{[a}F^{b]cd}=&\,{}-2\partial^{[c}F^{d]ab}-F_e{}^{ab}F_e{}^{cd}-2F_e{}^{a[c}F_e{}^{d]b}\,,\label{eq:bianchi-uuuu}\\
D_aF^{bcd}=&\,{}3\u D^{[b}F_a{}^{cd]}\label{eq:bianchi-luuu}\,,\\
\mathcal R^{ab}{}_{cd}=&\,{}-\mathcal R_{cd}{}^{ab}\,.
%
%
\end{align}
We can also use these derivatives to write
\begin{equation}
\mathcal R^{ab}{}_{cd}=(D^{[a}+\o D^{[a})F^{b]}{}_{cd}
\,,\qquad
\mathcal R^a{}_b=D^aF_b-\o D_cF^a{}_{bc}\,.
\end{equation}
Finally we have the Bianchi identity for the generalized Riemann tensor which takes the form
\begin{equation}
3D^{[a}\mathcal R^{bc]}{}_{de}=-D^fF^{abc}F^f{}_{de}-3F_f{}^{[ab}\o D_fF^{c]}{}_{de}\,.
\label{eq:dR}
\end{equation}
We also find the following expressions for the divergence of the generalized Riemann and Ricci tensors
\begin{align}
(\u D^a-F^a)\mathcal R^{ab}{}_{cd}
=&\,
2D_{[c}\mathcal R^b{}_{d]}
-F^e{}_{cd}\partial^eF^b
-F_e{}^{ab}\o D_eF^a{}_{cd}\,,
\label{eq:div-R}
\\
(D_b-F_b)\mathcal R^a{}_b=&\,\frac14\partial^a\mathcal R\,,\qquad (D^a-F^a)\mathcal R^a{}_b=\frac14\partial_b\mathcal R\,.
\label{eq:div-Ric}
\end{align}

\subsection{Reduction to (super)gravity}
To reduce the DFT expressions to (super)gravity one should do two things:\footnote{{As already mentioned, beyond the leading order in $\alpha'$ additional (non-covariant) field redefinitions are needed to make contact with standard supergravity.}}
\begin{itemize}
	\item[1.] Solve the section condition to remove the doubling of coordinates by setting $\partial_M=(0,\,\partial_m)$.
	\item[2.] Gauge fix the double Lorentz transformations down to the diagonal copy, which becomes the usual Lorentz group, by setting the two vielbeins in (\ref{eq:E}) equal, $e^{(+)}=e^{(-)}=e$.
\end{itemize}
Doing this one finds
\begin{equation}
\partial_a\rightarrow\frac{1}{\sqrt2}\partial_a\,,\qquad\partial^a\rightarrow\frac{1}{\sqrt2}\partial^a\,.
\end{equation}
The factor of $\sqrt2$ appears since on the LHS we define $\partial_A=E_A{}^M\partial_M$ while on the RHS we have $\partial_a=e_a{}^m\partial_m$. This should hopefully not lead to confusion since we will never mix doubled and standard fields in the same expression. The generalized fluxes become
\begin{equation}
\begin{aligned}
&F^a\rightarrow\frac{1}{\sqrt2}(2\partial^a\Phi+\omega_b{}^{ab})\,,\qquad
F_a\rightarrow\frac{1}{\sqrt2}(2\partial_a\Phi+\omega_{ba}{}^b)\,,
\\
&F^a{}_{bc}\rightarrow\frac{1}{\sqrt2}\omega^{(-)a}{}_{bc}\,,\qquad
F_a{}^{bc}\rightarrow-\frac{1}{\sqrt2}\omega_a^{(+)bc}\,,
\\
&F_{abc}\rightarrow\frac{1}{\sqrt2}(3\omega^{(-)}_{[abc]}+H_{abc})\,,\qquad
F^{abc}\rightarrow-\frac{1}{\sqrt2}(3\omega^{(+)[abc]}-H^{abc})\,,
\end{aligned}
\label{eq:Fsg}
\end{equation}
where $\omega^{(\pm)}=\omega\pm\frac12H$ denotes the torsionful spin connections. Finally, the generalized Riemann and Ricci tensor/scalar reduce to
\begin{equation}
\begin{aligned}
\mathcal R^{ab}{}_{cd}\rightarrow&\,\frac12\left(R^{(-)ab}{}_{cd}+\omega^{(+)eab}\omega^{(-)}_{ecd}\right)\,,\\
\mathcal R^a{}_b\rightarrow&\,\frac12\left(2\nabla^{(-)a}\partial_b\Phi+R^{(-)ac}{}_{bc}\right)\,,\\
\mathcal R\rightarrow&\,R+4\nabla^a\partial_a\Phi-4\partial^a\Phi\partial_a\Phi-\frac{1}{12}H^{abc}H_{abc}\,.
\end{aligned}
\label{eq:Rsg}
\end{equation}
Here we see explicitly the non-covariance of the generalized Riemann tensor, though to leading order in fields it reduces to the curvature of the torsionful connection. We also see that the generalized Ricci tensor contains the equations of motion for the metric and $B$-field as its symmetric and anti-symmetric part respectively, while the generalized Ricci scalar coincides with the usual Lagrangian for the NSNS sector fields (up to a total derivative).

\section{First order correction: \texorpdfstring{$\mathcal R^{(0,1)}$}{R(0,1)}}\label{sec:R01}
Here we recall the form of the first $\alpha'$ correction to the DFT action for the heterotic case. To the first order in $\alpha'$ the DFT action takes the form
\begin{equation}
S=\int dX\,e^{-2d}\left(\mathcal R+a\mathcal R^{(0,1)}\right)\,.
\end{equation}
The expression for $\mathcal R^{(0,1)}$ was first found in \cite{Marques:2015vua} (see also \cite{Baron:2017dvb}), but we will write it in the simpler form found in \cite{Hronek:2020skb}, which makes use of the generalized Riemann tensor
\begin{equation}
\begin{aligned}
\mathcal R^{(0,1)}=&\,{}
(\partial^a-F^a)(\partial^b-F^b)M^{ab}
-\tfrac12\mathcal R^{ab}{}_{cd}\mathcal R^{ab}{}_{cd}
+F^{abC}F^a{}_{de}\partial_CF^b{}_{de}
\\
&{}
-\left(\partial^aF^b-F_c{}^{da}F_c{}^{db}+\frac12F^{acd}F^{bcd}\right)M^{ab}
+\frac{2}{3}F^{abc}M^{abc}
\,,
\end{aligned}
\label{eq:R01}
\end{equation}
where we have written it even more compactly using the $M$'s defined in (\ref{eq:Ms}) in terms of traces of the ``generalized spin connection''. Let us calculate the double Lorentz variation of this object.
Splitting the transformation into $\o\lambda$ and $\u\lambda$ terms, 
\begin{equation}
\delta\mathcal R^{(0,1)}=\o\delta\mathcal R^{(0,1)}+\u\delta\mathcal R^{(0,1)}\,,
\end{equation}
we find, using (\ref{eq:deltaFs}) and (\ref{eq:deltaRie}), that
\begin{equation}
\o\delta\mathcal R^{(0,1)}=0
\end{equation}
while
\begin{align}
\u\delta\mathcal R^{(0,1)}=&\,{}
-\left(2\partial^{[a}F^{b]}+(\partial_C-F_C)F^{abC}\right)F^a{}_{cd}\partial^b\u\lambda_{cd}
\nonumber\\
&{}
-2(\partial^a-F^a)(D_b-F_b)\left[F^a{}_{cd}\partial_b\u\lambda_{cd}\right]
+2\mathcal R^a{}_bF^a{}_{cd}\partial_b\u\lambda_{cd}\,.
\end{align}
The first term vanishes by the Bianchi identity for $F^a$. Comparing the remaining terms to the variation of the generalized Ricci scalar $\mathcal R$ in (\ref{eq:deltaR}) we see that they can be canceled by modifying the double Lorentz transformations by a term proportional to $\alpha'$ as
\begin{equation}
\delta' E^{aM}E_{bM}=-\delta' E_b{}^ME^a{}_M=a\hat\lambda^a{}_b=\frac{a}{2}F^a{}_{de}\partial_b\u\lambda_{de}\,.
\label{eq:deltaEprime}
\end{equation}
We have discussed the correction to the action and transformations for the heterotic case only but it is easy to obtain the most general correction by noting that exchanging upper and lower indices everywhere (taking care of the extra minus sign for contracted lower indices) we get another solution $\mathcal R^{(1,0)}$ with corresponding correction to the transformations, which we can add to the above with a new coefficient $b$. This gives the complete 2-parameter deformation at this order.

\section{Second order correction: \texorpdfstring{$\mathcal R^{(0,2)}$}{R(0,2)}}\label{sec:R02}
We will now find the $\alpha'^2$ correction to DFT for the heterotic case, i.e. $\mathcal R^{(0,2)}$ in (\ref{eq:action}). To do this we first need to work out the terms obtained from inserting the order $\alpha'$ correction to the double Lorentz transformation (\ref{eq:deltaEprime}) in the order $\alpha'$ correction to the Lagrangian (\ref{eq:R01}). Our task is then to find an $\mathcal R^{(0,2)}$ such that its lowest order double Lorentz variation cancels against these terms, up to terms that can be canceled by modifying the double Lorentz transformations at order $\alpha'^2$.

It will be convenient to first consider a general variation of the order $\alpha'$ Lagrangian. For the generalized Riemann squared term we find, using (\ref{eq:Riem}),
\begin{equation}
\mathcal R^{ab}{}_{cd}\delta\mathcal R^{ab}{}_{cd}=
\mathcal R^{ab}{}_{cd}
\left(
2D^a\delta F^b{}_{cd}
+2\delta E^{aC}\partial_CF^b{}_{cd}
-\delta F^{abe}F^e{}_{cd}
\right)\,,
\end{equation}
where $\delta E_{AB}\equiv \delta E_A{}^ME_{BM}$ is a general variation of the generalized vielbein. Writing the first term as a total derivative plus a term involving the divergence of $\mathcal R$ we find, using the expression for the divergence in (\ref{eq:div-R}),
\begin{equation}
\begin{aligned}
\mathcal R^{ab}{}_{cd}\delta\mathcal R^{ab}{}_{cd}=&\,{}
2(\partial^a-F^a)\left[\mathcal R^{ab}{}_{cd}\delta F^b{}_{cd}\right]
-4D_c\mathcal R^b{}_d\delta F^b{}_{cd}
+2\partial^aF^bF^a{}_{cd}\delta F^b{}_{cd}
\\
&{}
+\mathcal R^{ab}{}_{cd}
\left(
2\delta E^{aC}\partial_CF^b{}_{cd}
-F^{abe}\delta F^e{}_{cd}
-\delta F^{abe}F^e{}_{cd}
\right)
+2F_e{}^{ab}\o D_eF^a{}_{cd}\delta F^b{}_{cd}\,.
\end{aligned}
\end{equation}
Using this result we find for the general variation of the order $\alpha'$ Lagrangian in (\ref{eq:R01})
\begin{equation}
\begin{aligned}
\delta\mathcal R^{(0,1)}
=&\,
(\partial_C-F_C)\left[\delta E^{aC}(\partial^b-F^b)M^{ab}-F^{abC}F^b{}_{de}\delta F^a{}_{de}+\partial^a\delta E^{bC}M^{ab}-F^{aC}{}_D\delta E^{bD}M^{ab}\right]
\\
&{}
+(\partial^a-F^a)\left[\delta[(\partial^b-F^b)M^{ab}]-2\mathcal R^{ab}{}_{cd}\delta F^b{}_{cd}\right]
+4D_c\mathcal R^b{}_d\delta F^b{}_{cd}
-2\mathcal R^{ab}{}_{de}\delta E^{aC}\partial_CF^b{}_{de}
\\
&{}
+\delta F_c{}^{ab}F^b{}_{de}\partial_cF^a{}_{de}
+\delta F^{abc}F^a{}_{de}\partial^bF^c{}_{de}
-\delta E_{CD}F^{abD}F^a{}_{de}\partial^CF^b{}_{de}
+2F_c{}^{da}\delta F_c{}^{db}M^{ab}
\\
&{}
-\frac{4}{3}\delta F^{abc}M^{abc}
-2\partial_CF^a\delta E^{bC}M^{ab}
-\partial^a\delta E^{bC}\partial_CM^{ab}
-\delta E^{bC}F^a{}_{CD}\partial^DM^{ab}\,.
\label{eq:deltaL1}
\end{aligned}
\end{equation}
It can be checked that this reproduces the variations found in the previous section (although starting from the above expressions leads to a longer calculation).

Specifying now to the $\alpha'$ modification to the double Lorentz transformation (\ref{eq:deltaEprime}), dropping equation of motion terms and total derivatives, and keeping for the moment only the leading terms in the number of $F$'s we find
\begin{equation}
\delta'\mathcal R^{(0,1)}
\sim
2\mathcal R^{ab}{}_{cd}\partial_eF^b{}_{cd}\hat\lambda^a{}_e
-2\partial^a\hat\lambda^b{}_cF^a{}_{de}\partial_cF^b{}_{de}
+\mathcal O(F^4)\,.
\end{equation}
These terms have to be canceled by adding terms of order $\alpha'^2$ to the action. We consider the following basis of quartic terms constructed only out of $F^a{}_{bc}$
\begin{align}
&aF^a{}_{fg}\partial^bF^c{}_{fg}F^c{}_{de}\partial^bF^a{}_{de}
+bF^a{}_{fg}\partial^bF^c{}_{fg}F^a{}_{de}\partial^bF^c{}_{de}
+cF^a{}_{fg}\partial^bF^c{}_{fg}F^a{}_{de}\partial^cF^b{}_{de}
\nonumber\\
&{}
+dF^a{}_{fg}\partial^bF^c{}_{fg}F^c{}_{de}\partial^aF^b{}_{de}
+eF^a{}_{fg}\partial^bF^c{}_{fg}F^b{}_{de}\partial^aF^c{}_{de}
+f\partial^aF^b{}_{cd}\partial^eF^b{}_{cd}F^a{}_{fg}F^e{}_{fg}
\nonumber\\
&{}
+g\partial^aF^b{}_{cd}\partial^bF^e{}_{cd}F^a{}_{fg}F^e{}_{fg}
+h\partial^bF^a{}_{cd}\partial^bF^e{}_{cd}F^a{}_{fg}F^e{}_{fg}\,.
\end{align}
After a bit of work one finds that to cancel the terms from the $\delta'$ variation we need to take
\begin{equation}
b=a-\tfrac12\,,\quad
c=-a+\tfrac12\,,\quad
d=-a\,,\quad
e=\tfrac12\,,\quad
f=-\tfrac12\,,\quad
g=1-a\,,\quad
h=a-\tfrac12\,.
\end{equation}
The freedom in choosing $a$ corresponds only to the freedom to integrate by parts and make field redefinitions. Taking $a=0$ and writing things in terms of the generalized Riemann tensor we find
\begin{equation}
L_4=
\frac12\mathcal R^{ab}{}_{de}\mathcal R^{bc}{}_{de}F^a{}_{fg}F^c{}_{fg}
-\frac14F^a{}_{de}\mathcal R^{bc}{}_{de}F^a{}_{fg}\mathcal R^{bc}{}_{fg}
+\frac12F^a{}_{fg}\partial^bF^c{}_{fg}F^b{}_{de}\partial^aF^c{}_{de}\,.
\end{equation}
However, at the next order in fields these terms will have a non-zero $\o\lambda$ variation, which must be canceled by terms with 5 $F$'s since the $\delta'$ terms involve only $\u\lambda$. We find
\begin{equation}
\begin{aligned}
\o\delta L_4=&\,
\partial^b\o\lambda^{ha}F^b{}_{de}\partial^h[F^c{}_{de}F^a{}_{fg}]F^c{}_{fg}
+\mathcal O(F^5)
\\
=&\,
\frac12\partial^b[\partial^a\o\lambda^{bc}M^{cd}M^{da}]
-\frac12\partial^b\partial^a\o\lambda^{ad}M^{bc}M^{cd}
-\frac32\partial^{[a}\o\lambda^{bc]}\partial^bF^f{}_{de}F^a{}_{de}M^{cf}
\\
&{}
\hspace{-20pt}
+\frac32\partial^{[a}\o\lambda^{bc]}\partial^bF^a{}_{de}F^f{}_{de}M^{cf}
+\frac12\partial_a\o\lambda^{bc}\partial_aF^b{}_{de}F^f{}_{de}M^{cf}
-\frac12\partial_a\o\lambda^{bc}\partial_aF^f{}_{de}F^b{}_{de}M^{cf}
+\mathcal O(F^5)\,.
\end{aligned}
\end{equation}
It is now easy to see that to cancel these terms we need to add
\begin{align}
L_5=&\,
\frac12\partial^aF^bM^{ac}M^{cb}
+\frac12F^{abc}\partial^bF^f{}_{de}F^a{}_{de}M^{cf}
-\frac12F^{abc}\partial^bF^a{}_{de}F^f{}_{de}M^{cf}
\nonumber\\
&{}
-\frac12F_a{}^{bc}\partial_aF^b{}_{de}F^f{}_{de}M^{cf}
+\frac12F_a{}^{bc}\partial_aF^f{}_{de}F^b{}_{de}M^{cf}\,.
\end{align}
But now $\o\delta(L_4+L_5)$ gives rise also to terms of order $F^5$ and to cancel these we will need to add $F^6$ terms to the Lagrangian. One finds
\begin{align}
\o\delta(L_4+L_5)=&\,
\frac12(\partial^c-F^c)[\partial^a\o\lambda^{cb}M^{ad}M^{db}]
-\frac32F^{abc}\partial^{[d}\o\lambda^{ab]}M^{ce}M^{ed}
-\frac32F^{cde}\partial^{[c}\o\lambda^{ab]}M^{ad}M^{be}
\nonumber\\
&{}
+3\partial^{[a}\o\lambda^{bc]}M^{bcd}M^{da}
+F_a{}^{bc}\partial_a\o\lambda^{bd}M^{ce}M^{ed}
+\frac12F_a{}^{bc}\partial_a\o\lambda^{de}M^{bd}M^{ce}\,,
\end{align}
which is easily seen to be canceled by adding
\begin{align}
L_6=&
\frac14F^{acd}F^{bcd}M^{ae}M^{eb}
+\frac14F^{abc}F^{cde}M^{ad}M^{be}
-F^{acd}M^{bcd}M^{ab}
-\frac12F_c{}^{da}F_c{}^{db}M^{ae}M^{eb}
\nonumber\\
&{}
-\frac14F_c{}^{ab}F_c{}^{de}M^{ad}M^{be}\,.
\end{align}
This takes care of the $\o\lambda$ variation.

We now look at the remaining $\u\lambda$-terms. With a bit of work one finds, using (\ref{eq:deltaL1}), that
\begin{equation}
\begin{aligned}
\delta'\mathcal R^{(0,1)}+\u\delta(L_4&+L_5+L_6)
=
(\partial^a-F^a)[\delta'[(\partial^b-F^b)M^{ab}]-2\mathcal R^{ab}{}_{cd}\partial_c\u\lambda_{fg}\o D_dF^b{}_{fg}]
\\
&{}
+(\partial_C-F_C)\left[
-F^{abC}F^b{}_{de}\delta' F^a{}_{de}
-F^{abC}\partial^{[a}\u\lambda_{de}F^{f]}{}_{de}M^{fb}
+F_a{}^{bC}\hat\lambda^d{}_aM^{db}
\right]
\\
&{}
-(\partial_c-F_c)(\partial^b-F^b)[\hat\lambda^a{}_cM^{ab}]
+2(\partial^c\u\lambda_{fg}F^b{}_{fg}F^c{}_{de}+2\delta' F^b{}_{de})D_d\mathcal R^b{}_e
\\
&{}
-2\partial^a\u\lambda_{de}F^b{}_{ef}F^c{}_{fd}M^{abc}\,.
%
\end{aligned}
\label{eq:deltaR01f}
\end{equation}
All terms are either total derivatives or involve the equations of motion except the last one, which is canceled by a term
\begin{equation}
L'_6=\frac13M^{abc}M^{abc}\,.
\end{equation}
Collecting the terms in $L_4$, $L_5$, $L_6$ and $L'_6$ together we have
\begin{equation}
\begin{aligned}
\mathcal R^{(0,2)}=&\,{}
-\frac12\mathcal R^{ac}{}_{de}\mathcal R^{cb}{}_{de}M^{ab}
-\frac14F^a{}_{de}\mathcal R^{bc}{}_{de}F^a{}_{fg}\mathcal R^{bc}{}_{fg}
+\frac12F^a{}_{fg}\partial^bF^c{}_{fg}F^b{}_{de}\partial^aF^c{}_{de}
\\
&{}
+\frac12F^{abC}\partial_CF^a{}_{de}F^f{}_{de}M^{bf}
-\frac12F^{abC}\partial_CF^f{}_{de}F^a{}_{de}M^{bf}
+\frac14F^{abC}F_C{}^{de}M^{ad}M^{be}
\\
&{}
+\frac12(\partial^aF^b-F_c{}^{da}F_c{}^{db}+\frac12F^{acd}F^{bcd})M^{ae}M^{eb}
-F^{abc}M^{abd}M^{cd}
+\frac13M^{abc}M^{abc}
\\
&{}
+\mbox{total derivatives}
\,.
%
\end{aligned}
\end{equation}
Finally we must look at the equation of motion terms in the variation (\ref{eq:deltaR01f}). They are
\begin{equation}
4\partial_{[d}\u\lambda_{|fg|}\o D_{e]}F^b{}_{fg}D_d\mathcal R^b{}_e\,.
\end{equation}
We can now read off, using (\ref{eq:deltaR}), the required $\alpha'^2$ modification to the double Lorentz transformations which cancels these terms
\begin{equation}
\delta''E^a{}_b=a^2(D_d-F_d)\left(\partial_{[b}\u\lambda_{|fg|}\o D_{d]}F^a{}_{fg}\right)\,.
\label{eq:deltaEprime2}
\end{equation}
The full action up to this order then takes the form (\ref{eq:La2}). Note that the first term in the Lagrangian is a total derivative. Its second order contribution has been added in order to simplify the transformation of the Lagrangian, which is non-zero at this order and takes the form of a total derivative (\ref{eq:delta-L}). The corrected double Lorentz transformations up to this order take the form (\ref{eq:DLT}).

It is important to note that the corrected double Lorentz transformations we have found close only on-shell at the second order in $\alpha'$. Indeed, a short calculation gives
\begin{equation}
[\delta_{\u\lambda},\delta_{\u\lambda'}]E^{aM}E_{bM}
=
%
%
D^aV_b
-D_bV^a
+\delta_{\tilde\lambda}E^{aM}E_{bM}
+a^2\mathcal R^a{}_c\tr(\partial_{[b}\u\lambda\partial_{c]}\u\lambda')
+\mathcal O(a^3)\,.
\end{equation}
The first two terms are a generalized diffeomorphism with parameters
\begin{equation}
\begin{aligned}
V_a=&\,
\frac{a}{4}\tr(\u\lambda'\partial_a\u\lambda)
-\frac{a^2}{2}(D_c-F_c)\tr(\partial_a\u\lambda'\partial_c\u\lambda)
-(\u\lambda\leftrightarrow\u\lambda')
\,,\\
V^a=&\,
\frac{a}{4}\tr(\u\lambda'\partial^a\u\lambda)
-\frac{a^2}{4}F^a{}_{cd}\tr(\partial_c\u\lambda'\partial_d\u\lambda)
-(\u\lambda\leftrightarrow\u\lambda')\,,
\end{aligned}
\end{equation}
the next term a double Lorentz transformation with parameter
\begin{equation}
\tilde\lambda_{cd}=[\u\lambda,\u\lambda']_{cd}-a\tr(\partial_{[c}\u\lambda'\partial_{d]}\u\lambda)\,,
\end{equation}
while the last term is not of this form but vanishes on-shell. Note that the generalized dilaton does not transform under the above generalized diffeomorphism, as follows from the section condition and Bianchi identities.\footnote{Indeed,
$$
\begin{aligned}
\delta d=&
\frac12(\partial_a-F_a)V_a-\frac12(\partial^a-F^a)V^a
=
-\frac{a^2}{4}\partial_a\partial_c\tr(\partial_a\u\lambda'\partial_c\u\lambda)
+\frac{a^2}{4}(\partial_c-F_c)[F_a\tr(\partial_a\u\lambda'\partial_c\u\lambda)]
\\
&{}
+\frac{a^2}{4}(\partial_a-F_a)[F_c\tr(\partial_a\u\lambda'\partial_c\u\lambda)]
+\frac{a^2}{8}(\partial^A-F^A)[F_{Acd}\tr(\partial_c\u\lambda'\partial_d\u\lambda)]
-\frac{a^2}{4}\partial_{[c}F_{a]}\tr(\partial_a\u\lambda'\partial_c\u\lambda)
-(\u\lambda\leftrightarrow\u\lambda')
=0\,.
\end{aligned}
$$
} This is of course consistent with the fact that the generalized dilaton does not transform under the corrected double Lorentz transformations up to this order.

\subsection{Alternative action and transformations}
When we relate this to supergravity in the next section it turns out to be more convenient to first modify the action and transformations of the DFT description slightly. The first step is to rewrite the second order correction to the double Lorentz transformations (\ref{eq:deltaEprime2}) as
\begin{equation}
\begin{aligned}
a^{-2}\delta''E^a{}_b
=&
-\frac12D_b(D_d-F_d)(\partial_d\u\lambda_{fg}F^a{}_{fg})
+\frac12(\o D_d-F_d)\o D_d(\partial_b\u\lambda_{fg}F^a{}_{fg})
-\frac12F_{bcd}D_c(\partial_d\u\lambda_{fg}F^a{}_{fg})
\\
&{}
+\frac12(\partial^c-F^c)(F^c{}_{bd}\partial_d\u\lambda_{fg}F^a{}_{fg})
+\frac12F^c{}_{bd}D_d(\partial^c\u\lambda_{fg}F^a{}_{fg})
+\frac12\mathcal R^{ac}{}_{bd}\partial_d\u\lambda_{fg}F^c{}_{fg}
\\
&{}
+\frac14F_{bcd}F^e{}_{cd}\partial^e\u\lambda_{fg}F^a{}_{fg}
-\frac12\mathcal R^d{}_b\partial^d\u\lambda_{fg}F^a{}_{fg}\,.
\end{aligned}
\label{eq:deltapp1}
\end{equation}
We recognize the first term as a generalized diffeomorphism with parameter 
\begin{equation}
V^a=\frac{a^2}{2}(D_d-F_d)(\partial_d\u\lambda_{fg}F^a{}_{fg})\,.
\end{equation}
We can remove this term by performing the opposite generalized diffeomorphism with parameter $-V^a$. This leads to an extra term in the variation of the Lagrangian,
\begin{equation}
(\delta L)_1=-(\partial^A-F^A)(-V_A\mathcal R)
=
\frac{a^2}{2}(\partial^a-F^a)((D_d-F_d)(\partial_d\u\lambda_{fg}F^a{}_{fg})\mathcal R)
\end{equation} 
and a transformation of the generalized dilaton
\begin{equation}
\delta d=\frac12(\partial^A-F^A)(-V_A)=
-\frac{a^2}{4}(\partial^a-F^a)(D_d-F_d)(\partial_d\u\lambda_{fg}F^a{}_{fg})\,.
\label{eq:deltad}
\end{equation}
Next we note that the last term in (\ref{eq:deltapp1}) is proportional to the equations of motion, since it involves the generalized Ricci tensor. This term can also be removed by noting that, using (\ref{eq:deltaR}), it can be replaced with the terms
\begin{equation}
(\delta L)_2=
2a^2(\partial^a-F^a)(D_b-F_b)(\mathcal R^d{}_b\partial^d\u\lambda_{fg}F^a{}_{fg})
-2a^2\mathcal R^d{}_b\mathcal R^a{}_b\partial^d\u\lambda_{fg}F^a{}_{fg}
\end{equation}
in the variation of the action. Furthermore, the second of these terms can be removed by adding to the Lagrangian the term
\begin{equation}
\Delta L_1=-a^2\mathcal R^a{}_c\mathcal R^b{}_cM^{ab}\,,
\end{equation}
while the first term becomes, using (\ref{eq:div-Ric}),
\begin{equation}
(\delta L)'_2=
\frac{a^2}{2}(\partial^a-F^a)\left(\partial^d\mathcal R\partial^d\u\lambda_{fg}F^a{}_{fg}\right)
+2a^2(\partial^a-F^a)\left[\mathcal R^d{}_bD_b(\partial^d\u\lambda_{fg}F^a{}_{fg})\right]\,.
\end{equation}
In total the Lagrangian now transforms as
\begin{equation}
\begin{aligned}
\delta(L+\Delta L_1)
=&
2a^2(\partial^a-F^a)\left[\mathcal R^b{}_d\partial^a\u\lambda_{fg}\o D_dF^b{}_{fg}\right]
+2a^2(\partial^a-F^a)\left[\mathcal R^d{}_b\o D_b\partial^d\u\lambda_{fg}F^a{}_{fg}\right]
\\
&{}
+2a^2(\partial^a-F^a)\left[\mathcal R^d{}_b\partial^d\u\lambda_{fg}\o D_bF^a{}_{fg}\right]
-2a^2(\partial_a-F_a)\left[\mathcal R^b{}_d\partial_d\u\lambda_{fg}\o D_aF^b{}_{fg}\right]
\\
&{}
+\frac{a^2}{2}(\partial^a-F^a)(D_d-F_d)\left[(\partial_d\u\lambda_{fg}F^a{}_{fg})\mathcal R\right]
\,.
\end{aligned}
\end{equation}
The first two terms are partially canceled by adding to the Lagrangian the term
\begin{equation}
\Delta L_2=-2a^2(\partial^a-F^a)\left[\mathcal R^b{}_dF^a{}_{fg}\o D_dF^b{}_{fg}\right]\,.
\end{equation}

To summarize, we can take the transformations of the fields to be
\begin{equation}
\begin{aligned}
a^{-2}\delta''_{\mathrm{alt}}E^a{}_b
=&
\frac12(\o D_d-F_d)\o D_d(\partial_b\u\lambda_{fg}F^a{}_{fg})
-\frac12F_{bcd}D_c(\partial_d\u\lambda_{fg}F^a{}_{fg})
+\frac12(\partial^c-F^c)(F^c{}_{bd}\partial_d\u\lambda_{fg}F^a{}_{fg})
\\
&{}
+\frac12F^c{}_{bd}D_d(\partial^c\u\lambda_{fg}F^a{}_{fg})
+\frac12\mathcal R^{ac}{}_{bd}\partial_d\u\lambda_{fg}F^c{}_{fg}
+\frac14F_{bcd}F^e{}_{cd}\partial^e\u\lambda_{fg}F^a{}_{fg}
\end{aligned}
\label{eq:deltaEalt}
\end{equation}
and
\begin{equation}
\delta'' d=
-\frac{a^2}{4}(\partial^a-F^a)(D_d-F_d)(\partial_d\u\lambda_{fg}F^a{}_{fg})\,,
\label{eq:delta-d}
\end{equation}
with the Lagrangian
\begin{equation}
L_{\mathrm{alt}}=L
-a^2\mathcal R^a{}_c\mathcal R^b{}_cM^{ab}
-2a^2(\partial^a-F^a)\left[\mathcal R^b{}_dF^a{}_{fg}\o D_dF^b{}_{fg}\right]
\end{equation}
now transforming as
\begin{equation}
\begin{aligned}
\delta L_{\mathrm{alt}}
=&
2a^2(\partial^a-F^a)\left[\mathcal R^d{}_b\partial^d\u\lambda_{fg}\o D_bF^a{}_{fg}\right]
-2a^2(\partial_a-F_a)\left[\mathcal R^b{}_d\partial_d\u\lambda_{fg}\o D_aF^b{}_{fg}\right]
\\
&{}
+4a^2(\partial^a-F^a)\left[\mathcal R^b{}_dF^a{}_{fg}\partial_d\u\lambda_{fh}F^b{}_{hg}\right]
+\frac{a^2}{2}(\partial^a-F^a)(D_d-F_d)\left[(\partial_d\u\lambda_{fg}F^a{}_{fg})\mathcal R\right]
\,.
\end{aligned}
\label{eq:deltaLalt}
\end{equation}
While this Lagrangian and transformations look more complicated, this form turns out to be more straightforward for reproducing the corresponding supergravity expressions, as we will {see in the next section. But before that we turn to the question of finding transformations that close off-shell.}

{
\subsection{Invariant action with transformations that close off-shell}\label{sec:off-shell}
The $\alpha'^2$-corrected DFT we have constructed has the unfamiliar property that the Lagrangian is only double Lorentz invariant up to a total derivative term and the double Lorentz transformations close only on-shell. One would expect there to exist an alternative formulation where the Lagrangian is invariant and the transformation close off-shell, as happens at order $\alpha'$. Indeed, the generalized Bergshoeff-de Roo identification of \cite{Baron:2018lve} would automatically lead to a Lagrangian and transformations with these properties. Here we will see how such a formulation can be derived from our results {by performing field redefinitions and generalized diffeomorphisms as well as modifying the transformations and Lagrangian}. It turns out that there is a price to pay in that the transformations and, in particular, the action become considerably more complicated. The calculations involved in this section are somewhat long and the results are not used elsewhere in the paper so this section can be skipped by readers not interested in the detailed relation between the two formulations.

To find the right form of the Lagrangian and transformations involves a little bit of guess work. One way to get to the answer is to start with a slightly different question. The transformations we found at order $\alpha'^2$ contain several terms linear in the fields, the same order as the first $\alpha'$ correction. In the previous section we saw that we could remove one of these terms by a generalized diffeomorphism obtaining the transformations (\ref{eq:deltaEalt}) and (\ref{eq:delta-d}) with fewer linear terms. It is natural to ask whether we could remove more, or perhaps all, of the terms linear in the fields at order $\alpha'^2$. We will now show that the answer is yes (later we will see that there is a small caveat). For simplicity we will first do the calculation dropping all terms of higher order in the number of fields. Then we have from (\ref{eq:deltaEalt}) and (\ref{eq:delta-d})
\begin{equation}
\delta''_{\mathrm{alt}}E^a{}_b=\frac{a^2}{2}\partial^2(\partial_b\u\lambda_{ef}F^a{}_{ef})+\mathcal O(F^2)\,,\qquad
\delta'' d=-\frac{a^2}{4}\partial^a\partial_d(\partial_d\u\lambda_{ef}F^a{}_{ef})+\mathcal O(F^2)\,.
\end{equation}
It is easy to see, using the section condition to raise the $d$ index, that
\begin{equation}
\delta'' d=\frac{a^2}{8}\delta(\partial^a\partial^bM^{ab})+\mathcal O(F^2)\,,
\end{equation}
so the leading part of the transformation of the generalized dilaton can be removed by a field redefinition. For the leading term in the transformation of the generalized vielbein we have
\begin{equation}
\begin{aligned}
\partial^2(\partial_b\u\lambda_{ef}F^a{}_{ef})
=&\,
\partial^2\partial_b\u\lambda_{ef}F^a{}_{ef}
+2\partial^c\partial_b\u\lambda_{ef}\partial^cF^a{}_{ef}
+\partial_b\u\lambda_{ef}\partial^2F^a{}_{ef}
\\
=&\,
\partial^2\partial_b\u\lambda_{ef}F^a{}_{ef}
+2\partial^c\partial_b\u\lambda_{ef}\partial^cF^a{}_{ef}
+\partial_b\u\lambda_{ef}\partial^c\mathcal R^{ca}{}_{ef}
+\partial_b\u\lambda_{ef}\partial^a\partial^cF^c{}_{ef}
+\mathcal O(F^2)
\\
=&\,
\partial^a(\partial_b\u\lambda_{ef}\partial^cF^c{}_{ef})
+\partial_b(\partial^2\u\lambda_{ef}F^a{}_{ef})
-\partial^2\u\lambda_{ef}\partial_bF^a{}_{ef}
-\partial^a\partial_b\u\lambda_{ef}\partial^cF^c{}_{ef}
\\
&{}
+2\partial^c\partial_b\u\lambda_{ef}\partial^cF^a{}_{ef}
+\partial_b\u\lambda_{ef}\partial^c\mathcal R^{ca}{}_{ef}
+\mathcal O(F^2)
\\
=&\,
\partial^a(\partial_b\u\lambda_{ef}\partial^cF^c{}_{ef})
+\partial_b(\partial^2\u\lambda_{ef}F^a{}_{ef})
-\delta(\partial_bF^a{}_{ef}\partial^cF^c{}_{ef})
+2\partial^c\partial_b\u\lambda_{ef}\partial^cF^a{}_{ef}
\\
&{}
+2\partial_b\u\lambda_{ef}\partial_e\mathcal R^a{}_f
+\mathcal O(F^2)\,,
\end{aligned}
\end{equation}
where we used (\ref{eq:div-R}) in the last step. We see that the first three terms can be removed by a generalized diffeomorphism and a field redefinition. The fourth term can be further rewritten as
\begin{equation}
\begin{aligned}
2\partial^c\partial_b\u\lambda_{ef}\partial^cF^a{}_{ef}
=&\,
\partial^c\partial_b\u\lambda_{ef}\partial^cF^a{}_{ef}
+\partial^c\partial_b\u\lambda_{ef}\mathcal R^{ca}{}_{ef}
+\partial^c\partial_b\u\lambda_{ef}\partial^aF^c{}_{ef}
+\mathcal O(F^2)
\\
=&\,
\partial^a(\partial^c\partial_b\u\lambda_{ef}F^c{}_{ef})
+\partial_b(\partial^c\u\lambda_{ef}\partial^cF^a{}_{ef})
-\delta(F^c{}_{ef}\partial_b\partial^cF^a{}_{ef})
\\
&{}
+\delta(\partial_bF^c{}_{ef}R^{ca}{}_{ef})
+\mathcal O(F^2)\,.
\end{aligned}
\end{equation}
After suitable field redefinitions we therefore have
\begin{equation}
\delta''E^a{}_b=
\frac{a^2}{2}\partial^a[\partial^c(\partial_b\u\lambda_{ef}F^c{}_{ef})]
+\frac{a^2}{2}\partial_b[\partial^c(\partial^c\u\lambda_{ef}F^a{}_{ef})]
+a^2\partial_b\u\lambda_{ef}\partial_e\mathcal R^a{}_f
+\mathcal O(F^2)\,.
\end{equation}
It is not hard to see that performing the opposite generalized diffeomorphism produces terms in $\delta''d$ which can again be canceled by a field redefinition and terms in the variation of the Lagrangian which are of higher order. This leaves the last term in $\delta''E$, which if we remove it and put it into the variation of the Lagrangian via (\ref{eq:deltaR}) gives, using (\ref{eq:deltaLalt}),
\begin{equation}
\delta L_{\mathrm{alt}}
=
-4\partial^a\partial_b[a^2\partial_b\u\lambda_{ef}\partial_e\mathcal R^a{}_f]
+\mathcal O(F^2)
=
-4a^2\delta(\partial^a\partial^b[F^b{}_{ef}\partial_e\mathcal R^a{}_f])
+\mathcal O(F^2)\,,
\end{equation}
which can be removed by modifying the Lagrangian. This shows that we can indeed get rid of all the linear terms in the transformations at order $\alpha'^2$. Now we will carry out the corresponding calculation without dropping the higher order terms.

For the leading term in the transformation (\ref{eq:deltaEalt}) we have
\begin{equation}
\begin{aligned}
(\o D_c-F_c)\o D_c(\partial_b\u\lambda_{ef}F^a{}_{ef})
&=\,{}
D^aU_b
-D_bU^a
+\delta(\Delta E_1)^a{}_b
+2\partial_c\partial_b\u\lambda_{ef}\o D_cF^a{}_{ef}%
+2\partial_b\u\lambda_{ef}D_e\mathcal R^a{}_f%
\\
&{}
+\partial^c\u\lambda_{ef}F^a{}_{ef}\mathcal R^c{}_b%
-(\partial^c-F^c)(\partial_d\u\lambda_{ef}F^a{}_{ef}F^c{}_{bd})%
+\partial_d\u\lambda_{ef}\partial^cF^a{}_{ef}F^c{}_{bd}%
\\
&{}
-\partial_d\partial^c\u\lambda_{ef}F^a{}_{ef}F^c{}_{bd}%
+2\partial_b\u\lambda_{ef}R^{ca}{}_{eg}F^c{}_{gf}%
+2\partial_b\u\lambda_{ef}\partial^cF^a{}_{eg}F^c{}_{gf}%
\\
&{}
+2\partial_b\partial^c\u\lambda_{ef}F^c{}_{eg}F^a{}_{gf}%
+2\partial^c\u\lambda_{ef}\partial_bF^c{}_{eg}F^a{}_{gf}%
+\partial^d\u\lambda_{ef}F^a{}_{ef}F^c{}_{bg}F_g{}^{cd}
\end{aligned}
\end{equation}
where we have defined
\begin{equation}
(\Delta E_1)^a{}_b=F^a{}_{ef}\partial_b(\partial^c-F^c)F^c{}_{ef}
\end{equation}
and
\begin{equation}
U^A=\partial^A\u\lambda_{ef}(\partial^c-F^c)F^c{}_{ef}\,,
\end{equation}
while the fourth term can be rewritten as
\begin{equation}
\begin{aligned}
2\partial_c\partial_b\u\lambda_{ef}\o D_cF^a{}_{ef}
=&\,
D^aW_b-D_bW^a
+\delta(\Delta E_2)^a{}_b
-2\partial_c\u\lambda_{ef}F^d{}_{ef}\mathcal R^{ad}{}_{bc}%
-2\partial_c\u\lambda_{ef}\partial^dF^a{}_{ef}F^d{}_{bc}%
\\
&{}
-2\partial_b\u\lambda_{ef}\mathcal R^{ca}{}_{eg}F^c{}_{gf}%
-2\partial_b\u\lambda_{ef}\partial^cF^a{}_{eg}F^c{}_{gf}%
-2\partial_b(\partial^c\u\lambda_{ef}F^c{}_{eg})F^a{}_{gf}%
\,,
\end{aligned}
\label{eq:dLdF}
\end{equation}
where
\begin{equation}
(\Delta E_2)^a{}_b=\o D_bF^c{}_{ef}\mathcal R^{ca}{}_{ef}+\o D_bF^c{}_{ef}\partial^cF^a{}_{ef}-\u D^cF_b{}^{da}F^c{}_{ef}F^d{}_{ef}
\end{equation}
and
\begin{equation}
\begin{aligned}
W^a=&\,\partial^c\partial^a\u\lambda_{ef}F^c{}_{ef}+\partial_c\u\lambda_{ef}F^d{}_{ef}F_c{}^{ad}+\partial^c\o\lambda^{ad}F^c{}_{ef}F^d{}_{ef}\,,\\
W_b=&\,\partial^c\partial_b\u\lambda_{ef}F^c{}_{ef}+\partial_c\u\lambda_{ef}F^d{}_{ef}F^d{}_{bc}\,.
\end{aligned}
\end{equation}
Using these results the transformation of the generalized vielbein (\ref{eq:deltaEalt}) can be written
\begin{equation}
\begin{aligned}
a^{-2}\delta''E^a{}_b
=&
\frac12D^a(U_b+W_b)
-\frac12D_b(U^a+W^a)
+\frac12\delta(\Delta E_1+\Delta E_2)^a{}_b
\\
&{}
+\partial_b\u\lambda_{ef}D_e\mathcal R^a{}_f
+\frac12\partial^c\u\lambda_{ef}F^a{}_{ef}\mathcal R^c{}_b
\\
&{}
-\frac12\partial_d\u\lambda_{ef}F^c{}_{ef}\mathcal R^{ac}{}_{bd}
-\frac12\partial_c\u\lambda_{ef}\partial^dF^a{}_{ef}F^d{}_{bc}
+\frac12\partial^D\u\lambda_{ef}\o D_cF^a{}_{ef}F_{bcD}\,,
\end{aligned}
\end{equation}
while the transformation of the generalized dilaton (\ref{eq:delta-d}) can be written
\begin{equation}
a^{-2}\delta'' d=\frac18(\partial_A-F_A)W^A+\delta\Delta d
\end{equation}
with
\begin{equation}
\Delta d=\frac18(\partial^a-F^a)(\partial^b-F^b)M^{ab}\,.
\label{eq:Deltad}
\end{equation}
We can set the transformation of the generalized dilaton to zero by performing the field redefinition
\begin{equation}
d\rightarrow d+\Delta d
\end{equation}
and a compensating generalized diffeomorphism with parameter
\begin{equation}
V^A=-\frac12U^A-\frac14W^A
\label{eq:gen-diff2}
\end{equation}
(note that the $U^A$ piece does not affect the generalized dilaton since $(\partial_A-F_A)U^A=0$ by the section condition). After this the transformation of the generalized vielbein becomes
\begin{equation}
\begin{aligned}
a^{-2}\delta''E^a{}_b
=&
\frac14D^aW_b
-\frac14D_bW^a
+\frac12\delta(\Delta E_1+\Delta E_2)^a{}_b
+\partial_b\u\lambda_{ef}D_e\mathcal R^a{}_f
+\frac12\partial^c\u\lambda_{ef}F^a{}_{ef}\mathcal R^c{}_b
\\
&{}
-\frac12\partial_d\u\lambda_{ef}F^c{}_{ef}\mathcal R^{ac}{}_{bd}
-\frac12\partial_c\u\lambda_{ef}\partial^dF^a{}_{ef}F^d{}_{bc}
+\frac12\partial^D\u\lambda_{ef}\o D_cF^a{}_{ef}F_{bcD}\,.
\end{aligned}
\end{equation}
The $W$-terms in the transformation are awkward since $W^a$ involves $\o\lambda$, the Lorentz factor that should not be modified. We can get rid of these terms by using (\ref{eq:dLdF}) at the price of introducing another term linear in the fields into the transformations, which now take the form
\begin{equation}
\begin{aligned}
a^{-2}\delta''E^a{}_b
=&
\Delta E^a{}_b
+\partial_b\u\lambda_{ef}D_e\mathcal R^a{}_f
+\frac12\partial^c\u\lambda_{ef}F^a{}_{ef}\mathcal R^c{}_b
+\frac12\partial_b\partial_c\u\lambda_{ef}\o D_cF^a{}_{ef}
\\
&{}
+\frac12\partial_b\u\lambda_{ef}\mathcal R^{ca}{}_{eg}F^c{}_{gf}
+\frac12\partial_b\u\lambda_{ef}\partial^cF^a{}_{eg}F^c{}_{gf}
+\frac12\partial_b(\partial^c\u\lambda_{ef}F^c{}_{eg})F^a{}_{gf}\,,
\end{aligned}
\end{equation}
where
\begin{equation}
\begin{aligned}
\Delta E^a{}_b
=
\frac12F^a{}_{ef}\partial_b(\partial^c-F^c)F^c{}_{ef}
+\frac14\o D_bF^c{}_{ef}\mathcal R^{ca}{}_{ef}
+\frac14\o D_bF^c{}_{ef}\partial^cF^a{}_{ef}
-\frac14\u D^cF_b{}^{da}F^c{}_{ef}F^d{}_{ef}\,.
\end{aligned}
\label{eq:DeltaE}
\end{equation}
These terms are removed by the field redefinition\footnote{Note the sign of the second term, which is due to the contraction of two lower indices.}
\begin{equation}
E^{aM}\rightarrow E^{aM}-\Delta E^a{}_bE_b{}^M
\end{equation}
The last four terms in the transformation agree precisely with \cite{Baron:2018lve} (exchanging over and underlined indices in their eq. (3.33)). The second and third term, which are proportional to the generalized Ricci tensor {and lead to the transformations closing only on-shell}, can be removed at the expense of introducing additional terms in the variation of the Lagrangian via (\ref{eq:deltaR}). Doing this and taking also the generalized diffeomorphism (\ref{eq:gen-diff2}) into account one finds from (\ref{eq:deltaLalt})
\begin{equation}
\begin{aligned}
a^{-2}\delta L_{\mathrm{alt}}
=&
-(\partial^A-F^A)(V_A\mathcal R)
+4(\partial^a-F^a)(D_b-F_b)\left[-\partial_b\u\lambda_{ef}D_e\mathcal R^a{}_f-\frac12\partial^c\u\lambda_{ef}F^a{}_{ef}\mathcal R^c{}_b\right]
\\
&{}
-4\left[-\partial_b\u\lambda_{ef}D_e\mathcal R^a{}_f-\frac12\partial^c\u\lambda_{ef}F^a{}_{ef}\mathcal R^c{}_b\right]\mathcal R^a{}_b
+2(\partial^a-F^a)\left[\mathcal R^d{}_b\partial^d\u\lambda_{fg}\o D_bF^a{}_{fg}\right]
\\
&{}
-2(\partial_a-F_a)\left[\mathcal R^b{}_d\partial_d\u\lambda_{fg}\o D_aF^b{}_{fg}\right]
+4(\partial^a-F^a)\left[\mathcal R^b{}_dF^a{}_{fg}\partial_d\u\lambda_{fh}F^b{}_{hg}\right]
\\
&{}
+\frac12(\partial^a-F^a)(D_d-F_d)\left[\partial_d\u\lambda_{fg}F^a{}_{fg}\mathcal R\right]
=
-\delta(\Delta L)
\end{aligned}
\end{equation}
where
\begin{equation}
\begin{aligned}
\Delta L=&\,
4(\partial^a-F^a)(\partial^b-F^b)\left[F^b{}_{ef}D_e\mathcal R^a{}_f\right]
+4(\partial^a-F^a)\left[\mathcal R^{ad}{}_{ef}D_e\mathcal R^d{}_f\right]
-4D_e\mathcal R^a{}_fD_{[e}\mathcal R^a{}_{f]}
\\
&{}
+\mathcal R^a{}_c\mathcal R^b{}_cM^{ab}
+\frac14(\partial^a-F^a)(\partial^b-F^b)\left[M^{ab}\mathcal R\right]\,.
\end{aligned}
\end{equation}
Therefore, taking the field redefinitions into account, we have shown that the Lagrangian
\begin{equation}
\begin{aligned}
\tilde L
=&\,
L_{\mathrm{alt}}+a^2\Delta L+a^2L_{\mathrm{redef}}
=
L
-4a^2D_e\mathcal R^a{}_fD_{[e}\mathcal R^a{}_{f]}
+4a^2(\partial^a-F^a)(\partial^b-F^b)\left[F^b{}_{ef}D_e\mathcal R^a{}_f\right]
\\
&{}
+4a^2(\partial^a-F^a)\left[\mathcal R^{ad}{}_{ef}D_e\mathcal R^d{}_f\right]
-2a^2(\partial^a-F^a)\left[\mathcal R^b{}_dF^a{}_{fg}\o D_dF^b{}_{fg}\right]
+a^2L_{\mathrm{redef}}\,,
\end{aligned}
\end{equation}
where
\begin{equation}
\begin{aligned}
L_{\mathrm{redef}}
=
4(\partial^a-F^a)(D_b-F_b)\Delta E^a{}_b
-4\Delta E^a{}_b\mathcal R^a{}_b
+8(\partial^a-F^a)\partial^a\Delta d
-2\Delta d\mathcal R\,,
\end{aligned}
\end{equation}
with $\Delta E$ and $\Delta d$ given in (\ref{eq:DeltaE}) and (\ref{eq:Deltad}) and $L$ the Lagrangian in (\ref{eq:La2}), is invariant under the double Lorentz transformations with second order correction
\begin{equation}
\begin{aligned}
\delta''E^a{}_b
=&
\frac{a^2}{2}
\left[
\partial_b\partial_c\u\lambda_{ef}\o D_cF^a{}_{ef}
+\partial_b\u\lambda_{ef}\mathcal R^{ca}{}_{eg}F^c{}_{gf}
+\partial_b\u\lambda_{ef}\partial^cF^a{}_{eg}F^c{}_{gf}
+\partial_b(\partial^c\u\lambda_{ef}F^c{}_{eg})F^a{}_{gf}
\right]
\end{aligned}
\end{equation}
and $\delta d=0$, which close off-shell as shown in \cite{Baron:2018lve}. We see that these transformations, and in particular the Lagrangian, are considerably more complicated than the ones presented in the introduction.

}

\section{Reproducing the \texorpdfstring{$\alpha'^2$}{} correction to the heterotic string}\label{sec:match}
In this section we show that the $\alpha'^2$ correction to DFT found above reproduces the known $\alpha'^2$ correction to the tree-level low-energy effective action for the NSNS fields of the heterotic string.

Up to this order the corrected double Lorentz transformations take the form
\begin{equation}
\delta E^{aM}E^b{}_M=\o\lambda^{ab}\,,\qquad
\delta E_a{}^ME_{bM}=\u\lambda_{ab}\,,\qquad
\delta E^{aM}E_{bM}=-\delta E_{bM}E^{aM}=\Delta^a{}_b\,,
\end{equation}
which can also be written as
\begin{equation}
\delta E^{aM}
=
\o\lambda^{ab}E^{bM}
-\Delta^a{}_bE_b{}^M
\,,\qquad
\delta E_a{}^M
=
-\u\lambda_{ab}E_b{}^M
-\Delta^b{}_aE^{bM}\,.
\label{eq:deltaEaM}
\end{equation}
Here $\Delta=a\Delta'+a^2\Delta''$ is the correction to the transformations and from (\ref{eq:deltaEprime}) we have
\begin{equation}
\Delta'^a{}_b=\frac12\partial_b\u\lambda_{cd}F^a{}_{cd}
\label{eq:Deltap}
\end{equation}
and from (\ref{eq:deltaEalt})
\begin{equation}
\begin{aligned}
\Delta''^a{}_b
=&
\frac12(\o D_d-F_d)\o D_d(\partial_b\u\lambda_{fg}F^a{}_{fg})
-\frac12F_{bcd}D_c(\partial_d\u\lambda_{fg}F^a{}_{fg})
+\frac12(\partial^c-F^c)(F^c{}_{bd}\partial_d\u\lambda_{fg}F^a{}_{fg})
\\
&{}
+\frac12F^c{}_{bd}D_d(\partial^c\u\lambda_{fg}F^a{}_{fg})
+\frac12\mathcal R^{ac}{}_{bd}\partial_d\u\lambda_{fg}F^c{}_{fg}
+\frac14F_{bcd}F^e{}_{cd}\partial^e\u\lambda_{fg}F^a{}_{fg}\,.
\end{aligned}
\label{eq:Deltapp}
\end{equation}

Looking at the first equation in (\ref{eq:deltaEaM}), taking the $M$-index to be upper and going to supergravity by setting $e^{(+)}=e^{(-)}=e$ in (\ref{eq:E}), we find the variation of the inverse vielbein
\begin{equation}
\delta\bar e^{am}\bar e^b{}_m=\o\lambda^{ab}-\bar\Delta^{ab}\,,
\qquad
\o\lambda^{ab}+\u\lambda^{ab}=2\bar\Delta^{[ab]}\,.
\label{eq:deltaebar}
\end{equation}
Taking the $M$-index to be lower and using the above relations we find the variation of the $B$-field\footnote{For the supergravity fields we use the usual Einstein summation convention.}
\begin{equation}
\delta\bar B_{mn}=-2\bar\Delta_{[mn]}\,,\qquad\bar\Delta_{mn}=\bar e_m{}^a\bar e_n{}^b\bar\Delta_{ab}\,.
\label{eq:deltaBbar}
\end{equation}
Note that we have put a bar on the supergravity fields to emphasize that these fields, which come from the DFT description, are related by certain field redefinitions to the usual supergravity fields. In fact they differ by terms proportional to $\alpha'$ so we write
\begin{equation}
\bar e^{am}=e^{am}+ae'^{am}+a^2e''^{am}\,,\qquad
\bar e^a{}_m=e^a{}_m-ae'^a{}_m-a^2(e''^a{}_m-e'^a{}_ne'^{bn}e_{bm})\,,
\end{equation}
where $e^{am}$ is the standard inverse vielbein, transforming as $\delta e^{am}=-\lambda^a{}_ce^{cm}$ under local Lorentz transformations, and similarly
\begin{equation}
\bar B_{mn}=B_{mn}+aB'_{mn}+a^2B''_{mn}\,.
\end{equation}
Similarly we allow the transformation parameters to differ at higher orders in $\alpha'$, 
$$
\u\lambda=\lambda+a\u\lambda'+a^2\u\lambda''\,,\qquad
\o\lambda=-\lambda+a\o\lambda'+a^2\o\lambda''\,.
$$
Note the relative sign at the lowest order. {We will now determine the required field redefinitions up to order $\alpha'^2$ (another approach to determine them, potentially to all orders, was discussed in \cite{Baron:2021yqm}).}

Restricting to first order in $\alpha'$ ($a$) we find from (\ref{eq:deltaebar}) and (\ref{eq:Deltap})
\begin{equation}
e'_a{}^me_{bm}=e'_{ab}=\frac18W_{ab}
\,,\qquad
W_{ab}=\tr(\omega^{(-)}_a\omega^{(-)}_b)
\,,\qquad
\o\lambda'_{ab}=\u\lambda'_{ab}=
\frac14\tr(\partial_{[a}\lambda\omega_{b]}^{(-)})
\label{eq:ep1}
\end{equation}
and from (\ref{eq:deltaBbar})
\begin{equation}
B'_{mn}=0\,,
\end{equation}
since
\begin{equation}
\delta\bar B_{mn}=-\frac{a}{2}\tr(\partial_{[m}\lambda\omega^{(-)}_{n]})
\label{eq:deltaBbar1}
\end{equation}
is already the correct transformation for the $B$-field of the heterotic string to first order in $\alpha'$. The non-covariance of the $B$-field being related to the Green-Schwarz anomaly cancellation mechanism.

At the second order we have contributions from both $\Delta'$ and $\Delta''$. The former gives rise to the following terms at order $\alpha'$ from the expansion of the barred fields
\begin{equation}
\begin{aligned}
\Delta'^a{}_b\rightarrow&
-\frac14\tr(\partial_b\lambda\omega^a)
+\frac{a}{4}e'_b{}^m\partial_m\u\lambda_{cd}\omega^{(-)acd}
+\frac{a}{4}\partial_b\u\lambda'_{cd}\omega^{(-)acd}
+\frac{a}{4}\partial_b\u\lambda_{cd}\omega'^{(-)acd}
\\
=&\,{}
-\frac14\tr(\partial_b\lambda\omega^a)
-\frac{a}{32}W_{bc}\tr(\partial^c\lambda\omega^{(-)a})
-\frac{a}{32}W^{ac}\tr(\partial_b\lambda\omega^{(-)}_c)
\\
&{}
+\frac{a}{16}\partial_b\tr(\partial_c\lambda\omega_d^{(-)})\omega^{(-)acd}
+\frac{a}{16}\partial_b\lambda_{cd}\nabla^{(+)c}W^{da}
-\frac{a}{32}\partial_b\lambda^{cd}H_{cde}W^{ea}\,,
\end{aligned}
\label{eq:Deltap2}
\end{equation}
where we used the fact that the correction to the spin connection coming from the first correction to the vielbein takes the form
\begin{equation}
\omega'^{(-)}_{acd}=2\nabla^{(+)}_{[c}e'_{d]a}+(\omega^{(-)}_{ecd}-H_{cde})e'^e{}_a\,.
\label{eq:omegap}
\end{equation}
For $\Delta''$ we find from (\ref{eq:Deltapp})
\begin{equation}
\begin{aligned}
\Delta''^a{}_b
\rightarrow&\,
\frac18(\nabla^d-2\partial^d\Phi)H_{bcd}\tr(\partial^c\lambda\omega^{(-)a})
-\frac18(\nabla^{(+)}_d-2\partial_d\Phi)\nabla^{(+)}_d\tr(\partial_b\lambda\omega^{(-)a})
\\
&{}
+\frac18(\omega^{(-)}-H)_{bcd}\nabla^{(+)}_c\tr(\partial_d\lambda\omega^{(-)a})
-\frac{1}{16}(\omega^{(-)}-H)_{bcd}H_{cde}\tr(\partial_e\lambda\omega^{(-)a})
\\
&{}
-\frac18R^{(-)ac}{}_{bd}\tr(\partial_d\lambda\omega^{(-)c})
\,.
\label{eq:Deltapp2}
\end{aligned}
\end{equation}
Note that the first term is proportional to the lowest order equation of motion for the $B$-field. We do not want this term in the transformation of the vielbein, so we need to remove the part of this term that is symmetric in the free indices $a,b$ as this is the part relevant to the transformation of the vielbein. Because this term is proportional to the equations of motion it can be moved to the transformation of $\bar B$ instead as follows. First we lift the corresponding transformation of the vielbein to a variation of the action using (\ref{eq:deltaR}) which gives
\begin{equation}
\begin{aligned}
(\delta L)_1=&\,
-\frac14(\nabla^a-2\partial^a\Phi)(\nabla^b-2\partial^b\Phi)\left[(\nabla^d-2\partial^d\Phi)H_{bcd}\tr(\partial^c\lambda\omega^{(-)}_a)\right]
\\
&{}
+\frac14(\nabla^d-2\partial^d\Phi)H_{bcd}\left(R^{ab}+2\nabla^a\partial^b\Phi-\frac14H^{aef}H^b{}_{ef}\right)\tr(\partial^c\lambda\omega^{(-)}_a)\,.
\end{aligned}
\end{equation}
The second term is now canceled by modifying the transformation of the $B$-field by adding a term involving the equation of motion for the metric
\begin{equation}
(\delta\bar B_{mn})_1=
-\frac14\left(R_{mk}+2\nabla_m\partial_k\Phi-\frac14H_{mlp}H_k{}^{lp}\right)\tr(\partial_n\lambda\omega^{(-)k})
-(m\leftrightarrow n)
\label{eq:deltaB1}
\end{equation}
and we are left with the following terms in the variation of the Lagrangian
\begin{equation}
\begin{aligned}
(\delta L)'_1=&\,
\frac14(\nabla_a-2\partial_a\Phi)\left[H^{abc}(R_{bd}+2\nabla_b\partial_d\Phi-\frac14H_{bef}H_d{}^{ef})\tr(\partial_c\lambda\omega^{(-)d})\right]
\\
&{}
-\frac14(\nabla^a-2\partial^a\Phi)(\nabla^b-2\partial^b\Phi)\left[(\nabla^d-2\partial^d\Phi)H_{bcd}\tr(\partial^c\lambda\omega^{(-)}_a)\right]\,.
\end{aligned}
\end{equation}
Together with the transformation of the Lagrangian in (\ref{eq:deltaLalt}) its total transformation now becomes
\begin{equation}
\begin{aligned}
\delta L_{alt}
=&\,{}
-\frac14(\nabla^a-2\partial^a\Phi)\left[\left(2\nabla^{(-)b}\partial_d\Phi+R^{(-)bc}{}_{dc}\right)\partial^d\lambda^{ef}R^{(-)}_{abef}\right]
\\
&{}
+\frac12(\nabla_a-2\partial_a\Phi)\left[H^{ab}{}_c(R_{bd}+2\nabla_b\partial_d\Phi-\frac14H_{bef}H_d{}^{ef})\tr(\partial^{(c}\lambda\omega^{(-)d)})\right]
\\
&{}
-\frac18(\nabla_a-2\partial_a\Phi)(\nabla^{(+)}_b-2\partial_b\Phi)\left[\tr(\partial^b\lambda\omega^{(-)a})\mathcal R\right]\,.
\end{aligned}
\end{equation}
All these terms, except for the $H$-contribution from $\nabla^{(+)}$ in the last term, are canceled by adding the following terms to the Lagrangian
\begin{equation}
\begin{aligned}
\Delta L=&\,
\frac14(\nabla^a-2\partial^a\Phi)\left[\left(2\nabla^{(-)b}\partial_d\Phi+R^{(-)bc}{}_{dc}\right)\omega^{(-)def}R^{(-)}_{abef}\right]
+\frac{1}{16}(\nabla^a-2\partial^a\Phi)(\nabla^b-2\partial^b\Phi)\left[W_{ab}\mathcal R\right]
\\
&{}
-\frac14(\nabla_a-2\partial_a\Phi)\left[H^{ab}{}_c\left(R_{bd}+2\nabla_b\partial_d\Phi-\frac14H_{bef}H_d{}^{ef}\right)W^{cd}\right]
\,,
\end{aligned}
\end{equation}
leaving only the transformation
\begin{equation}
\delta(L_{alt}+\Delta L)
=
\frac{1}{16}(\nabla^a-2\partial^a\Phi)\left[H_{abc}\tr(\partial^b\lambda\omega^{(-)c})\mathcal R\right]\,.
\label{eq:deltaLf}
\end{equation}

Going back to the variation of the vielbein in (\ref{eq:deltaebar}) we find that the LHS becomes, at the second order in $\alpha'$,
\begin{equation}
\begin{aligned}
\delta\bar e_a{}^m\bar e_{bm}
\rightarrow&\,
\delta e''_{ab}
+2\lambda_{(a}{}^ce''_{b)c}
-\frac{3}{32}\nabla_{[a}\tr(\partial_c\lambda\omega^{(-)}_{d]})\omega_b^{(-)cd}
-\frac{3}{32}\nabla_{[b}\tr(\partial_c\lambda\omega^{(-)}_{d]})\omega_a^{(-)cd}
\\
&{}
-\frac{1}{64}W_{bc}\tr(\partial_a\lambda\omega^{(-)c})
-\frac{1}{64}W_{bc}\tr(\partial^c\lambda\omega^{(-)}_a)\,,
\end{aligned}
\end{equation}
where we used the first order results (\ref{eq:ep1}). Note the third and fourth term on the RHS which come from the anomalous Lorentz transformation of the $B$-field, (\ref{eq:deltaBbar1}). Using this together with the terms coming from $\Delta$ at this order, (\ref{eq:Deltap2}) and (\ref{eq:Deltapp2}) minus the first term, we find that, after performing another diffeomorphism 
\begin{equation}
\delta e_{(a}{}^me_{b)m}=-\nabla_{(a}v_{b)}\,,\qquad v_a=-\frac{1}{16}H_{acd}\tr(\partial^c\lambda\omega^{(-)d})\,,
\label{eq:extradiff}
\end{equation}
the second order correction to the (inverse) vielbein becomes\footnote{For completeness the correction to the Lorentz parameter is
$$
\begin{aligned}
\o\lambda''_{ab}
=&\,{}
\frac{1}{32}R_{abcd}\tr(\partial^c\lambda\omega^{(-)d})
-\frac{1}{32}\nabla_{(c}H_{d)ab}\tr(\partial^c\lambda\omega^{(-)d})
+\frac{3}{32}\omega_a^{(-)cd}\nabla_{[b}\tr(\partial_c\lambda\omega^{(-)}_{d]})
-\frac18\omega^{(-)}_{acd}\nabla^c\tr(\partial^d\lambda\omega^{(-)}_b)
\\
&{}
-\frac{1}{16}\partial_a\lambda^{cd}\nabla_cW_{bd}
+\frac{3}{32}H_{acd}\nabla^c\tr(\partial^d\lambda\omega^{(-)}_b)
+\frac{1}{32}H_{acd}\nabla^c\tr(\partial_b\lambda\omega^{(-)d})
-\frac{1}{64}H_{ace}\partial_b\lambda^{cd}W_d{}^e
\\
&{}
-\frac{1}{16}\omega_{acd}\omega_b{}^{ce}\tr(\partial_e\lambda\omega^{(-)d})
-\frac{1}{32}\omega_b{}^{cd}H_{ac}{}^e\tr(\partial_{(d}\lambda\omega^{(-)}_{e)})
-\frac{1}{32}H_{acd}H_b{}^{ce}\tr(\partial^d\lambda\omega^{(-)}_e)
\\
&{}
-\frac{1}{64}H_{abe}H^{ecd}\tr(\partial_c\lambda\omega^{(-)}_d)
+\frac{3}{128}W_{ac}\tr(\partial^c\lambda\omega^{(-)}_b)
-\frac{1}{128}W_{ac}\tr(\partial_b\lambda\omega^{(-)c})
-(a\leftrightarrow b)\,.
\end{aligned}
$$
}
\begin{equation}
\begin{aligned}
e''_{ab}
=&\,{}
\frac{1}{32}(\nabla^{(+)}_c-2\partial_c\Phi)\nabla^{(+)c}W_{ab}
+\frac{1}{32}R_{acbd}W^{cd}
-\frac{1}{32}H_a{}^{cd}\Omega_{bcd}
-\frac{1}{32}(\omega^{(-)}-H)_a{}^{cd}\nabla_cW_{db}
\\
&{}
+\frac{1}{64}\omega^{(-)}_{acd}H_b{}^{ce}W_e{}^d
-\frac{3}{128}H_{acd}H_b{}^{ce}W^d{}_e
+\frac{3}{256}W^2_{ab}
+(a\leftrightarrow b)\,.
\end{aligned}
\label{eq:epp}
\end{equation}
Here we have introduced the Lorentz-Chern-Simons form
\begin{equation}
\Omega_{klm}=3\tr(\omega^{(-)}_{[k}\partial_l^{\phantom{(-)}}\omega^{(-)}_{m]})+2\tr(\omega^{(-)}_{[k}\omega^{(-)}_l\omega^{(-)}_{m]})\,.
\label{eq:LCS}
\end{equation}
The extra diffeomorphism (\ref{eq:extradiff}) leads to a total derivative term in the variation of the Lagrangian, which together with (\ref{eq:deltaLf}) gives the total Lorentz transformation of the Lagrangian
\begin{equation}
\begin{aligned}
\delta(L_{alt}+\Delta L)
=&\,{}
(\nabla_a-2\partial_a\Phi)(v^a\mathcal R)
+\frac{1}{16}(\nabla^a-2\partial^a\Phi)\left[H_{acd}\tr(\partial^c\lambda\omega^{(-)d})\mathcal R\right]
=0\,,
\end{aligned}
\end{equation}
so the Lagrangian is now Lorentz invariant, as it should be in supergravity. The diffeomorphism also leads to extra terms in the transformations of $\Phi$ and $B$
\begin{equation}
(\delta\bar B_{mn})_2=v^kH_{mnk}
\,,\qquad
(\delta\Phi)_2=v^m\partial_m\Phi
\,,\qquad v_a=-\frac{1}{16}H_{acd}\tr(\partial^c\lambda\omega^{(-)d})\,.
\label{eq:delta2}
\end{equation}
For the dilaton this implies, together with the transformation (\ref{eq:deltad}), that
\begin{equation}
\begin{aligned}
\delta\bar d
\rightarrow&\,
\frac{1}{16}(\nabla^{(+)a}-2\partial^a\Phi)(\nabla^{(+)b}-2\partial^b\Phi)\tr(\partial^b\lambda\omega^{(-)a})
+\partial_a\Phi v^a
-\frac12\nabla_av^a
\\
=&\,
\frac{1}{32}(\nabla^a-2\partial^a\Phi)(\nabla^b-2\partial^b\Phi)\delta W^{ab}\,.
\end{aligned}
\end{equation}
Expressing this in terms of the dilaton, rather than the generalized dilaton, we find that the DFT dilaton is related to the standard one as
\begin{equation}
\bar\Phi=\Phi+\frac14\ln\left(\frac{\bar G}{G}\right)+\frac{1}{32}(\nabla^a-2\partial^a\Phi)(\nabla^b-2\partial^b\Phi)W^{ab}\,.
\label{eq:phibar}
\end{equation}

Finally we turn to the $B$-field. From (\ref{eq:deltaBbar}) we get, using (\ref{eq:Deltap2}) and (\ref{eq:Deltapp2}) and including the diffeomorphism (\ref{eq:delta2}) and extra modification (\ref{eq:deltaB1}), the second order transformation
\begin{equation}
\begin{aligned}
(\delta\bar B_{mn})^{(2)}
%
=&
\frac18\nabla_m[\omega_{ncd}\tr(\partial^c\lambda\omega^{(-)d})]%
-\frac18(\nabla^d-2\partial^d\Phi)\nabla_d\tr(\partial_m\lambda\omega^{(-)}_n)%
-\frac18R_{mncd}\tr(\partial^c\lambda\omega^{(-)d})%
\\
&{}
-\frac14\left(R^k{}_m+2\nabla^k\partial_m\Phi\right)\tr(\partial_{[n}\lambda\omega^{(-)}_{k]})
+\frac{1}{16}\partial_m\lambda^{cd}\Omega_{cdn}
\\
&{}
\hspace{-20pt}
+\delta\Big[
\frac{1}{16}(\omega^{(-)}-H)_m{}^{cd}\nabla_cW_{dn}%
-\frac{1}{32}\omega_m^{(-)cd}H_{nce}W^e{}_d%
+\frac{1}{32}\nabla_cH_{dmn}W^{cd}%
-\frac{1}{16}\omega_m^{(+)cd}\Omega_{ncd}
\\
&{}
+\frac{1}{16}(\nabla^d-2\partial^d\Phi)H_m{}^{cd}W_{cn}%
-\frac18\left(R^k{}_m+2\nabla^k\partial_m\Phi-\frac14H^{klp}H_{mlp}\right)W_{kn}%
\Big]
-(m\leftrightarrow n)\,.
\end{aligned}
\end{equation}
The first term is canceled by a $B$-field gauge transformation. The second term is trickier but it can be rewritten as follows
\begin{equation}
\begin{aligned}
(\nabla^d-2\partial^d\Phi)\nabla_d\tr(\partial_m\lambda\omega^{(-)}_n)
-(m\leftrightarrow n)
=&\,
3(\nabla^l-2\partial^l\Phi)\nabla_{[l}\tr(\partial_m\lambda\omega^{(-)}_{n]})
\\
&{}
\hspace{-20pt}
-2\nabla_m(\nabla^l-2\partial^l\Phi)\tr(\partial_{[n}\lambda\omega^{(-)}_{l]})
-R_{mn}{}^{kl}\tr(\partial_k\lambda\omega^{(-)}_l)
\\
&{}
\hspace{-20pt}
-2(R_m{}^k+2\nabla_m\partial^k\Phi)\tr(\partial_{[n}\lambda\omega^{(-)}_{k]})
-(m\leftrightarrow n)\,.
\end{aligned}
\end{equation}
Using this fact we find, after removing the $B$-field gauge transformations, that $\bar B=B+a^2B''$ with
\begin{equation}
\begin{aligned}
B''_{mn}
=&\,
\frac{1}{16}(\omega^{(-)}-H)_m{}^{cd}\nabla_cW_{dn}
-\frac{1}{32}\omega_m^{(-)cd}H_{nce}W^e{}_d
+\frac{1}{32}\nabla_cH_{dmn}W^{cd}
\\
&{}
+\frac{1}{16}(\nabla_d-2\partial_d\Phi)H_m{}^{cd}W_{cn}
-\frac18\left(R^k{}_m+2\nabla^k\partial_m\Phi-\frac14H^{klp}H_{mlp}\right)W_{kn}
\\
&{}
-\frac{1}{16}\omega_m^{(+)cd}\Omega_{cdn}
+\frac18(\nabla^l-2\partial^l\Phi)\Omega_{lmn}
-(m\leftrightarrow n)
\end{aligned}
\label{eq:Bpp}
\end{equation}
and that $B$ transforms as
\begin{equation}
\delta B_{mn}
=
-\frac{a}{2}\tr(\partial_{[m}\lambda\omega^{(-)}_{n]})
+\frac{a^2}{8}\tr(\partial_{[m}\lambda\Omega_{n]})
\,,
\end{equation}
which is the correct transformation for the heterotic string (see below).

Putting the contributions to the Lagrangian together the final Lagrangian becomes
\begin{equation}
\tilde L=L-a^2\mathcal R^a{}_c\mathcal R^b{}_cM^{ab}+L_\partial\,,
\end{equation}
where $L$, given in (\ref{eq:La2}), and the second term should be expressed in terms of the usual supergravity fields using (\ref{eq:Fsg}) and (\ref{eq:Rsg}) together with the field redefinitions needed to go from the barred DFT fields to the standard supergravity fields (\ref{eq:ep1}), (\ref{eq:epp}), (\ref{eq:phibar}) and (\ref{eq:Bpp}). Finally, the extra total derivative terms found above take the form
\begin{equation}
\begin{aligned}
L_\partial=&
\frac{a^2}{4}(\nabla_a-2\partial_a\Phi)
\Big[
\left(2\nabla^{(-)b}\partial_d\Phi+R^{(-)bc}{}_{dc}\right)\tr(\omega^{(-)a}\partial^d\omega^{(-)}_b)
\\
&{}
+\left(2\nabla^{(-)b}\partial_d\Phi+R^{(-)bc}{}_{dc}\right)\omega^{(+)d}{}_{be}W^{ae}
+\left(2\nabla^{(-)b}\partial_d\Phi+R^{(-)bc}{}_{dc}\right)\omega^{(-)def}R^{(-)a}{}_{bef}
\\
&{}
+\frac14(\nabla_b-2\partial_b\Phi)[W^{ab}\mathcal R]
-H^{abc}(R_b{}^d+2\nabla_b\partial^d\Phi-\frac14H_{bef}H^{def})W_{cd}
\Big]\,.
\end{aligned}
\end{equation}
With a bit of work\footnote{{The easiest way to do this is to use the fact that the Lagrangian is Lorentz invariant. Therefore one can just drop all terms where the spin connection $\omega^{(-)}$ appears without derivatives (note that dropping $\omega$ or $\omega^{(-)}$ must give the same result) and covariantize the result (replacing $H\rightarrow\hat H$) at the end, which simplifies the calculations a lot. For completeness we have also checked, with the help of Mathematica (in particular the package FieldsX \cite{Frob:2020gdh}), that the same result is obtained without using this trick.}} one finds that $\tilde L$ reduces to the tree-level effective action for the heterotic string up to order $\alpha'^2$ (with the vectors and fermions set to zero), which takes the same form as the action up to order $\alpha'$
\begin{equation}
L_{\mathrm{het}}=
e^{-2\Phi}
\left(
R+4\nabla^a\partial_a\Phi-4\partial^a\Phi\partial_a\Phi-\frac{1}{12}\hat H_{klm}\hat H^{klm}-\frac{a}{8}\hat R^{(-)}_{abcd}\hat R^{(-)abcd}
+\mathcal O(\alpha'^3)
\right)\,,
%
\end{equation}
but where the field strength of the $B$-field $H$ is replaced by the Lorentz invariant $\hat H$, defined recursively through \cite{Bergshoeff:1988nn}
\begin{equation}
\hat H_{klm}=H_{klm}-\frac{a}{2}\hat\Omega_{klm}\,,
\end{equation}
which up to second order gives
\begin{equation}
\hat H_{klm}=H_{klm}
-\frac{a}{2}\Omega_{klm}
+\frac{3a^2}{8}\left(
\partial_{[k}(\Omega_l{}^{np}\omega^{(-)}_{m]np})
+\Omega_{[k}{}^{np}R^{(-)}_{lm]np}
\right)
+\mathcal O(\alpha'^3)
\,.
\end{equation}
This completes the proof that the DFT Lagrangian (\ref{eq:La2}) exactly reproduces the tree-level correction to the heterotic string \cite{Bergshoeff:1988nn} (see also \cite{Metsaev:1986yb}) up to order $\alpha'^2$.

\section{Generalized Bergshoeff-de Roo identification}\label{sec:gBdR}
A different approach to finding the $\alpha'^2$ correction to DFT was introduced in \cite{Baron:2018lve} and dubbed the `generalized Bergshoeff-de Roo identification'. Here we will describe the idea of this approach and relate it to the result of section \ref{sec:R02}. We will define the fields somewhat differently to \cite{Baron:2018lve}, as this leads to some simplifications, but the basic approach will be the same. Then we will find the field redefinitions needed to reproduce the $\alpha'^2$ correction to the DFT action found in section \ref{sec:R02}. For simplicity we consider only the heterotic case.

The idea of \cite{Baron:2018lve} was to start from DFT in a bigger space and perform dimensional reduction in order to get DFT coupled to vectors. We will parametrize the `extended' generalized vielbein of the bigger space as
\begin{equation}
\mathcal E_{\mathcal A}{}^{\mathcal M}
=
\left(
\begin{array}{cc}
m_A{}^B & 0\\
0 & m_{A'}{}^{B'}
\end{array}
\right)
\left(
\begin{array}{cc}
\delta_B{}^C & -A_B^{C'}\\
A_{B'}^C & \delta_{B'}{}^{C'}
\end{array}
\right)
\left(
\begin{array}{cc}
E_C{}^M & 0\\
0 & E_{C'}{}^{M'}
\end{array}
\right)\,.
\label{eq:Ecal}
\end{equation}
This differs from the parametrization used in \cite{Baron:2018lve}, but turns out to lead to some simplifications. Here the index $\mathcal A=(A,\,A')$ runs over the standard DFT directions $A$ and `internal' directions $A'$ and $E_C{}^M$ is the usual generalized vielbein while $E_{C'}{}^{M'}$ is a generalized vielbein for the internal directions. We take the extended generalized metric to have the form
\begin{equation}
\eta_{\mathcal{AB}}=
\left(
\begin{array}{cc}
\eta_{AB} & 0\\
0 & \kappa_{A'B'}
\end{array}
\right)
\,,\qquad
\eta_{\mathcal{MN}}=
\left(
\begin{array}{cc}
\eta_{MN} & 0\\
0 & \kappa_{M'N'}
\end{array}
\right)
\end{equation}
and the condition that these should be related as 
\begin{equation}
\eta_{\mathcal{AB}}=\mathcal E_{\mathcal A}{}^{\mathcal M}\mathcal E_{\mathcal B}{}^{\mathcal N}\eta_{\mathcal{MN}}
\end{equation}
fixes $m_A{}^B$ and $m_{A'}{}^{B'}$ in terms of the vectors $A$ as follows
\begin{equation}
(m^{-2})_{AB}=\eta_{AB}+A_A^{C'}A_{BC'}\,,\qquad
(m^{-2})_{A'B'}=\kappa_{A'B'}+A_{A'}^CA_{B'C}\,.
\label{eq:ms}
\end{equation}
The section condition is
\begin{equation}
\partial^M\otimes\partial_M=0\,,\qquad\partial^M\partial_M=0\,,\qquad\partial_{M'}=0\,,
\end{equation}
corresponding to dimensional reduction of the primed directions.

Because we need the vector fields to be non-abelian we actually have to start from gauged DFT \cite{Grana:2012rr,Geissbuhler:2013uka}, rather than ordinary DFT, in the extended space. The only difference is the appearance of structure constants in some formulas. In particular, under a generalized diffeomorphism and double Lorentz transformation the extended generalized vielbein transforms as
\begin{equation}
\delta\mathcal E_{\mathcal A}{}^{\mathcal M}=
\mathcal V^{\mathcal N}\partial_{\mathcal N}\mathcal E_{\mathcal A}{}^{\mathcal M}
+\mathcal E_{\mathcal A}{}^{\mathcal N}\left(\partial^{\mathcal M}\mathcal V_{\mathcal N}-\partial_{\mathcal N}\mathcal V^{\mathcal M}+gf^{\mathcal M\mathcal K}{}_{\mathcal N}\mathcal V_{\mathcal K}\right)
+\lambda_{\mathcal A}{}^{\mathcal B}\mathcal E_{\mathcal B}{}^{\mathcal M}\,.
\end{equation}
Here $\mathcal V^{\mathcal M}=(V^M,\,V^{M'})$ is an extended generalized vector field. Note the extra structure constant term, proportional to the gauge coupling $g$ of dimension [length]$^{-1}$, which is non-zero only for all indices primed. Restricting to diffeomorphisms and Lorentz transformations involving the `internal' directions, we find that the gauge field transforms as
\begin{equation}
\delta\left[(mA)_A^{B'}E_{B'}{}^{M'}\right]
=
\mathcal E_A{}^{\mathcal N}\left(\partial_{\mathcal N}V^{M'}-gf^{M'K'}{}_{\mathcal N}V_{K'}\right)
-\lambda_A{}^{B'}m_{B'}{}^{C'}E_{C'}{}^{M'}\,.
\end{equation}
Since we are interested in the heterotic case, where we only need to modify one of the double Lorentz factors, we will let the internal indices $A',B',\ldots$ take only lower values $a',b',c',\ldots$, i.e. these are effectively not doubled (however $M',N'$ are still doubled). It then follows that the last term is non-zero only when the index $A$ is lower $a$. We now impose the gauge fixing condition
\begin{equation}
A_{aB'}=0\,,
\end{equation}
which lets us solve for the mixed components of the extended double Lorentz transformations, $\u\lambda_{ab'}$ (note that this gauge implies $m_a{}^B=\eta_a{}^B$). Similarly we may fix the purely internal components $\u\lambda_{a'b'}$ of the extended double Lorentz transformation by taking the internal generalized vielbein to be constant
\begin{equation}
E_{A'}{}^{M'}=\mbox{constant}\,.
\end{equation}
We find\footnote{The components of $m$, $\lambda$ and $V$ are taken with respect to $m_{\mathcal{AB}}=m_{\mathcal A}{}^{\mathcal C}\eta_{\mathcal{CB}}$, $\lambda_{\mathcal{AB}}=\lambda_{\mathcal A}{}^{\mathcal B}\eta_{\mathcal {AB}}$ and $V_{\mathcal C}=E_{\mathcal{CM}}V^{\mathcal M}$.}
\begin{equation}
\begin{aligned}
\u\lambda_{ab'}=\,&{}-\u\lambda_{b'a}=(m^{-1})_{b'c'}\partial_aV_{c'}\,,\\
\u\lambda_{a'b'}=\,&{}(\delta mm^{-1})_{a'b'}
-gm_{a'c'}(m^{-1})_{b'd'}f_{c'd'e'}V_{e'}
+(mA)_{a'}^c\partial^cV_{c'}(m^{-1})_{c'b'}\,.
%
\end{aligned}
\end{equation}
Note that the symmetric part of the RHS of the last equation vanishes automatically, as it should. The fields now transform as follows under double Lorentz transformations and internal generalized diffeomorphisms
\begin{equation}
\begin{aligned}
\delta(mA)^a_{b'}=\,&m^{ac}\partial^cV_{b'}+gf_{b'c'd'}V_{c'}(mA)^a_{d'}+\o\lambda^{ac}(mA)^c_{b'}\,,\\
\delta E_a{}^ME_{bM}=\,&\u\lambda_{ab}\,,\\
\delta E^{aM}E^b{}_M=\,&\o\lambda'^{ab}=-(m^{-1})^{ac}\delta m^{cb}-A^a_{c'}\partial^bV_{c'}+(m^{-1})^{ac}\o\lambda^{cd}m^{db}\,,\\
\delta E_a{}^ME^b{}_M=\,&{}-\delta E^{bM}E_{aM}=\partial_aV_{a'}A_{a'}^b\,.
\end{aligned}
\label{eq:gauge-Lorentz}
\end{equation}
Again the symmetric part of the RHS in the definition of $\o\lambda'^{ab}$ in the third line vanishes, as it must. Note the non-standard induced transformation for the generalized vielbein in the last line.

Let us now compare the transformation of the gauge field found above to the transformation of the generalized flux $\mathcal F^a{}_{\hat b\hat c}$ (the generalized analog of the spin connection), where the hatted lower index runs over both `external' and `internal' directions, $\hat b=(b,b')$. The latter transforms as (cf. (\ref{eq:deltaFs}))\footnote{Contracted hatted indices include an extra sign for the `external' piece, e.g. in the second term
$$
\u\lambda_{[\hat b|\hat d|}\mathcal F^a{}_{|\hat d|\hat c]}
=
-\u\lambda_{[\hat b|d|}\mathcal F^a{}_{|d|\hat c]}
+\u\lambda_{[\hat b|d'|}\mathcal F^a{}_{|d'|\hat c]}\,,
$$
where in the first the contraction is with $\eta$ and in the second with $\kappa$.
}
\begin{equation}
\delta\mathcal F^a{}_{\hat b\hat c}=m^{ad}\partial^d\u\lambda_{\hat b\hat c}
+2\u\lambda_{[\hat b|\hat d|}\mathcal F^a{}_{|\hat d|\hat c]}
+\o\lambda^{ad}\mathcal F^d{}_{\hat b\hat c}\,.
\end{equation}
Assuming the existence of constants
\begin{equation}
t_{a'\hat b\hat c}\,,
\end{equation}
which translate between a primed index and a pair of anti-symmetric indices $\hat b\hat c$, one sees that the identification
\begin{equation}
(mA)^a_{d'}t_{d'\hat b\hat c}=g^{-1}\mathcal F^a{}_{\hat b\hat c}\,,\qquad
V_{c'}t_{c'\hat a\hat b}=g^{-1}\u\lambda_{\hat a\hat b}\,,
\label{eq:GBdR}
\end{equation}
is consistent with the transformations of the fields provided that 
\begin{equation}
t_{a'\hat b\hat c}t_{b'\hat b\hat c}=C\kappa_{a'b'}\,,\qquad [t_{a'},t_{b'}]=f_{a'b'c'}t_{c'}\,,
\end{equation}
for some constant $C$. We see that the $t$'s are generators of $SO(D+K)$ in the fundamental representation, where $K$ is the range of the internal index $a'$. Taking $\kappa$ to be the Killing form
\begin{equation}
\kappa_{a'b'}=-f_{a'c'd'}f_{b'c'd'}\,,
\label{eq:kappa}
\end{equation}
then fixes the constant $C$ to $C=\frac{1}{D+K-2}$. Note that we are identifying $a'$ with $\hat a=(a,\,a')$, which is clearly not possible for finite $K$, so this construction is somewhat formal. However, this does not lead to any issues as long as one truncates at some finite order in $g^{-1}$. We just keep $C$ finite in the calculations and take it to zero only at the end, in this way it acts as a regulator. {The identification (\ref{eq:GBdR}) is implicit since the RHS involves the object on the LHS, but it can be solved recursively order by order in $g^{-1}$.} For further details of this identification we refer the reader to \cite{Baron:2018lve}.

Before we make this identification we will first determine the action. It is given by the usual expression (we take $\hat d=d$)
\begin{equation}
S=\int dX\,e^{-2d}\hat{\mathcal R}\,,
\end{equation}
where the `extended' generalized Ricci scalar is given by
\begin{equation}
\hat{\mathcal R}
=
4\hat\partial^a\mathcal F^a
-2\mathcal F^a\mathcal F^a
-\mathcal F_a{}^{bc}\mathcal F_a{}^{bc}
+\mathcal F_{a'}{}^{bc}\mathcal F_{a'}{}^{bc}
+\frac13\mathcal F^{abc}\mathcal F^{abc}\,.
\label{eq:genR}
\end{equation}
To find the explicit form of the action we need to compute the generalized fluxes. Using the ansatz (\ref{eq:Ecal}) in the definition 
\begin{equation}
\mathcal F_{\mathcal{ABC}}=3\hat\partial_{[\mathcal A}\mathcal E_{\mathcal B}{}^{\mathcal M}\mathcal E_{\mathcal C]\mathcal M}
+g\mathcal E_{\mathcal A}{}^{\mathcal K}\mathcal E_{\mathcal B}{}^{\mathcal L}\mathcal E_{\mathcal C}{}^{\mathcal M}f_{\mathcal{KLM}}
\end{equation}
we find
\begin{equation}
\begin{aligned}
\mathcal F_{ABC}
=
m_{[A}{}^Dm_B{}^Em_{C]}{}^F
\Big(
&{}
F_{DEF}
-gA_{Dd'}A_{Ee'}A_{Ff'}f_{d'e'f'}
+3\partial_DA_{Ee'}A_{Fe'}
\\
&{}
-3(\partial_Dm^{-1}m^{-1})_{EF}
\Big)\,,
\\
\mathcal F_{a'BC}
=
m_{a'd'}m_{[B}{}^Em_{C]}{}^F
\Big(
&{}
A_{d'}^DF_{DEF}
-2\partial_EA_{Fd'}
+gA_{Ee'}A_{Ff'}f_{d'e'f'}
+A_{d'}^G\partial_GA_{Ec'}A_{Fc'}
\\
&{}
-A_{d'}^D(\partial_Dm^{-1}m^{-1})_{EF}
\Big)\,,
\\
\mathcal F_{a'b'C}=
m_{[a'|d'|}m_{b']e'}m_C{}^F
\Big(
&{}
-gA_{Ff'}f_{d'e'f'}
+A_{d'}^DA_{e'}^EF_{DEF}
+2A_{d'}^D\partial_DA_{e'F}
+A_{e'D}\partial_FA_{d'}^D
\\
&{}
-(\partial_Fm^{-1}m^{-1})_{d'e'}
\Big)\,,
\end{aligned}
\label{eq:Fcal}
\end{equation}
{while the expression for $\mathcal F_{a'b'c'}$ will not be needed}. The components entering in the action become
\begin{equation}
\begin{aligned}
\mathcal F_a{}^{bc}
=&\,{}
m^{bd}m^{ce}\left(F_a{}^{de}+\partial_aA^{[d}_{c'}A^{e]}_{c'}-(\partial_am^{-1}m^{-1})^{[de]}\right)\,,
\\
\mathcal F_{a'}{}^{bc}
=&\,{}
-m_{a'd'}m^{bd}m^{ce}
\left(
F^{de}_{d'}
-A_{d'}^f\partial^fA^{[d}_{c'}A^{e]}_{c'}
+A_{d'}^f(\partial^fm^{-1}m^{-1})^{[de]}
\right)\,,
\\
\mathcal F^{abc}
=&\,{}
m^{ad}m^{be}m^{cf}
\left(
\hat F^{def}
+3\partial^{[d}A^e_{c'}A^{f]}_{c'}
-3(\partial^{[d}m^{-1}m^{-1})^{ef]}
\right)\,,
\end{aligned}
\label{eq:Fduu}
\end{equation}
where we have introduced the `field strength' of the gauge field ($D^a$ was defined in (\ref{eq:Ds}), in particular we take it not to act on the primed indices)
\begin{equation}
F^{ab}_{c'}=2D^{[a}A^{b]}_{c'}-gf_{c'd'e'}A^a_{d'}A^b_{e'}
\end{equation}
and defined
\begin{equation}
\hat F^{abc}=F^{abc}-gf_{d'e'f'}A^a_{d'}A^b_{e'}A^c_{f'}\,.
\end{equation}
Finally we need the generalized flux with one index which takes the form
\begin{equation}
\mathcal F^a=m^{ab}F^b-\partial^bm^{ab}
\end{equation}
and a short calculation gives
\begin{equation}
\hat\partial^a\mathcal F^a
=
-(\partial^b-F^b)(m\partial^am)^{ab}
+\partial^aF^b(m^2)^{ab}
+F^{abD}(\partial_Dmm)^{ab}
+\partial^bm^{ac}\partial^cm^{ab}\,.
\end{equation}
Putting these results together we find that many terms cancel and the generalized Ricci scalar (\ref{eq:genR}) simplifies to
\begin{equation}
\begin{aligned}
\hat{\mathcal R}
=&\,{}
-2(\partial^a-F^a)(\partial^b-F^b)(m^2)^{ab}
+2\partial^aF^b(m^2)^{ab}
-F_a{}^{bc}F_a{}^{de}(m^2)^{bd}(m^2)^{ce}
\\
&{}
+\frac13\hat F^{abc}\hat F^{def}(m^2)^{ad}(m^2)^{be}(m^2)^{cf}
+F^{bc}_{a'}F^{de}_{d'}m^2_{a'd'}(m^2)^{bd}(m^2)^{ce}
\\
&{}
+2F^{Abc}\partial_AA^d_{c'}A^e_{c'}(m^2)^{bd}(m^2)^{ce}
+2\partial^cA^b_{c'}A^a_{c'}\partial^dA^e_{d'}A^f_{d'}(m^2)^{ad}(m^2)^{be}(m^2)^{cf}\,.
\end{aligned}
\label{eq:L-gBdR}
\end{equation}
This Lagrangian, together with the vector gauge transformations and double Lorentz transformations in (\ref{eq:gauge-Lorentz}), gives DFT coupled to (non-abelian) vectors.

We now wish to use this action to construct $\alpha'$ corrections to the heterotic DFT action by imposing the generalized Bergshoeff-de Roo identification (\ref{eq:GBdR}). Even before doing this one notices a striking similarity to the DFT action up to second order in $\alpha'$ in (\ref{eq:La2}). The identification tells us to set
\begin{equation}
m^{ad}A^d_{a'}t_{a'\hat b\hat c}=g^{-1}\mathcal F^a{}_{\hat b\hat c}\,.
\end{equation}
The components of the generalized flux appearing on the RHS become, using (\ref{eq:Fcal}),
\begin{equation}
\begin{aligned}
\mathcal F^a{}_{bc}=&\,{}m^{ad}F^d{}_{bc}\,,\qquad
\mathcal F^a{}_{bc'}=m^{ad}m_{c'd'}D_bA^d_{d'}\,,\qquad
\\
\mathcal F^a{}_{b'c'}=&\,{}
m^{ad}m_{[b'|d'|}m_{c']e'}
\left(
-gA^d_{f'}f_{d'e'f'}
+2A_{d'}^eD^eA_{e'}^d
-A_{d'}^e\partial^dA_{e'}^e
-(\partial^dm^{-1}m^{-1})_{d'e'}
\right)\,.
\end{aligned}
\label{eq:Fudd}
\end{equation}
Setting $A^a_{a'}t_{a'\hat b\hat c}\equiv A^a_{\hat b\hat c}$ the identification takes the form
\begin{equation}
\begin{aligned}
A^a{}_{bc}=&\,{}g^{-1}F^a{}_{bc}\,,\qquad
A^a{}_{bc'}=g^{-1}m_{c'd'}D_bA^a_{d'}\,,\qquad
\\
A^a{}_{b'c'}=&\,{}-m_{b'd'}m_{c'e'}A^a_{f'}f_{d'e'f'}
+g^{-1}m_{[b'|d'|}m_{c']e'}
\left(
2A_{d'}^eD^eA_{e'}^a
-A_{d'}^e\partial^aA_{e'}^e
-(\partial^am^{-1}m^{-1})_{d'e'}
\right)\,.
\end{aligned}
\label{eq:Aident}
\end{equation}
The advantage of defining $A$ in the way we did is that the RHS of $A^a{}_{bc}$ is independent of $A$. For the other components $A$ appears on the RHS and the identification must be solved recursively. To see how this works let's compute $A^2$ to lowest order in $g^{-1}$. We have
\begin{equation}
\begin{aligned}
(A^2)^{ab}=&\,{}A^a_{c'}A^b_{c'}=C^{-1}A^a_{\hat c\hat d}A^b_{\hat c\hat d}=
-C^{-1}g^{-2}M^{ab}
-2C^{-1}g^{-2}(mD_cA^a)_{d'}(mD_cA^b)_{d'}
\\
&{}
+C^{-1}A^a_{a'}f_{a'b'c'}(m^2)_{b'd'}(m^2)_{c'e'}f_{d'e'f'}A^b_{f'}
\\
&{}
-2C^{-1}g^{-1}A^a_{a'}f_{a'b'c'}(m^2)_{b'd'}(m^2)_{c'e'}
\left(
2A_{d'}^eD^eA_{e'}^a
-A_{d'}^e\partial^aA_{e'}^e
-(\partial^am^{-1}m^{-1})_{d'e'}
\right)
\\
&{}
+C^{-2}g^{-2}(m^2)_{b'd'}(m^2)_{c'e'}
\left(
2A_{[b'}^eD^eA_{c']}^a
-A_{[b'}^e\partial^aA_{c']}^e
-(\partial^am^{-1}m^{-1})_{[b'c']}
\right)
\\
&{}
\qquad\cdot\left(
2A_{[d'}^eD^eA_{e']}^b
-A_{[d'}^e\partial^bA_{e']}^e
-(\partial^bm^{-1}m^{-1})_{[d'e']}
\right)\,.
\end{aligned}
\label{eq:A2-1}
\end{equation}
Noting that $A$ is of order $g^{-1}$ we have, using (\ref{eq:ms}) and (\ref{eq:kappa}),
\begin{equation}
(A^2)^{ab}=
-C^{-1}g^{-2}M^{ab}
-C^{-1}(A^2)^{ab}
+\mathcal O(g^{-4})
\end{equation}
and rearranging this we find
\begin{equation}
(A^2)^{ab}=-\frac{g^{-2}}{1+C}M^{ab}+\mathcal O(g^{-4})\rightarrow-g^{-2}M^{ab}+\mathcal O(g^{-4})\,,
\end{equation}
where we noted that at the end we should take $K\rightarrow\infty$ which gives $C\rightarrow0$. From the definition of $m$ in (\ref{eq:ms}) we get
\begin{equation}
(m^2)^{ab}=\eta^{ab}-(A^2)^{ab}+\mathcal O(g^{-4})=\eta^{ab}+g^{-2}M^{ab}+\mathcal O(g^{-4})\,,
\end{equation}
which coincides with $\mathcal M^{ab}$ (\ref{eq:Mcal}) provided we set 
\begin{equation}
g^{-2}=-\frac{a}{2}=\frac{\alpha'}{2}\,.
\end{equation}
It is easy to extend this calculation to show that one recovers precisely the action (\ref{eq:La2}) and transformations (\ref{eq:DLT}) at lowest order in $\alpha'$.

It is remarkable that the simple action (\ref{eq:L-gBdR}) can capture, after the generalized Bergshoeff-de Roo identification, an infinite series of $\alpha'$ corrections. However, while it is in principle straightforward to compute the action and transformations to some desired order in $\alpha'$ the expressions quickly become extremely long due to the complicated form of the identification (\ref{eq:Aident}) and the need to apply it recursively. Below we carry out the identification to order $\alpha'^2$ and show that, after suitable field redefinitions, we again recover the action and transformations presented in the introduction. It turns out that the calculations can be simplified by modifying the original Bergshoeff-de Roo identification to
\begin{equation}
\left((mA)^a_{b'}+cg^{-2}\mathcal R^a{}_{b'}\right)t_{b'\hat c\hat d}=g^{-1}\mathcal F^a{}_{\hat c\hat d}\,,
\label{eq:gBdR-mod}
\end{equation}
where $c$ is a constant to be fixed. Here $\mathcal R^a{}_{b'}$ is the mixed components of the extended generalized Ricci tensor which takes the form
\begin{equation}
\mathcal R^a{}_{b'}=m^{ad}\partial^d\mathcal F_{b'}
-(\partial_c-\mathcal F_c)\mathcal F^a{}_{b'c}
+((mA)_{c'}^d\partial^d-\mathcal F_{c'})\mathcal F^a{}_{b'c'}
+\mathcal F_c{}^{da}\mathcal F^d{}_{cb'}
-\mathcal F_{c'}{}^{da}\mathcal F^d{}_{c'b'}
\,,
\label{eq:R-mixed}
\end{equation}
where
\begin{equation}
\mathcal F_a=F_a\,,\qquad
\mathcal F_{a'}=-(\partial^b-F^b)(mA)_{a'}^b
\end{equation}
and the other generalized fluxes are given in (\ref{eq:Fduu}) and (\ref{eq:Fudd}). The fact that $\mathcal R$ is a (extended) generalized diffeomorphism scalar and transforms covariantly under double Lorentz transformations means that $mA+g^{-2}\mathcal R$ transforms in the same way as $mA$, so this form of the identification is also admissible. {Note however that since $A$ appears in the transformation of the generalized vielbein in (\ref{eq:gauge-Lorentz}) changing the identification in this way could result in the algebra only closing on-shell (the modification vanishes on-shell). But this is precisely what we want in order to reproduce the transformations in (\ref{eq:DLT}), which also have this property. For this reason it is natural to expect that a certain non-zero choice of $c$ would reproduce the action and transformations presented in the introduction, whereas the standard identification with $c=0$ should reproduce the more complicated action and transformations found in section \ref{sec:off-shell}. We will now show that this is indeed what happens.}

\subsection{Reproducing the previous \texorpdfstring{$\alpha'^2$}{alpha'**2} result}\label{sec:gBdR2}
Here we will drop terms of order $\alpha'^3$ ($g^5$) or higher. We will use the modified generalized Bergshoeff-de Roo identification (\ref{eq:gBdR-mod}). Using (\ref{eq:R-mixed}) we find
\begin{equation}
\begin{aligned}
\mathcal R^a{}_{b'}=&\,{}
(D_c-F_c)D_cA^a_{b'}
-\partial^a(\partial^b-F^b)A_{b'}^b
+gf_{b'c'd'}A^a_{c'}(\partial^b-F^b)A_{d'}^b
\\
&{}
-gf_{b'c'd'}A_{c'}^b\partial^bA^a_{d'}
+gf_{b'c'd'}A^b_{c'}F^{ab}_{d'}
+\mathcal O(\alpha'^2)\,.
\end{aligned}
\end{equation}
Looking at (\ref{eq:A2-1}) we find that taking $c=C^{-1}$ in (\ref{eq:gBdR-mod}) leads to some cancellations and we find a relatively simple expression for $A^2$
\begin{equation}
(A^2)^{ab}
=
g^{-2}
\left(
-M^{ab}
-(D_c-F_c)D_c(A^aA^b)
+2A^{(a}\partial^{b)}(\partial^c-F^c)A^c
\right)
+\mathcal O(\alpha'^3)
\label{eq:A2a2}
\end{equation}
In the $A^4$ terms in the expansion of the Lagrangian (\ref{eq:L-gBdR}) we effectively replace $A$ by $F^a{}_{cd}$ and it is not hard to see that this reproduces all the corresponding terms at order $\alpha'^2$ in (\ref{eq:La2}). But the identification produces extra terms of the form $\partial^2F^2$ from $A^2$ as above. We will now show that the corresponding terms in the action can be canceled by suitable field redefinitions. These terms come from considering the additional terms produced by the identification in the `order $\alpha'$' terms in the Lagrangian (\ref{eq:L-gBdR}), namely
\begin{equation}
\begin{aligned}
&{}
-2\partial^aF^b(A^2)^{ab}
+2F_a{}^{bc}F_a{}^{bd}(A^2)^{cd}
-F^{abc}F^{abd}(A^2)^{cd}
\\
&{}
-\frac23gF^{abc}A^a[A^b,A^c]
+F^{ab}F^{ab}
+2F^{Abc}\partial_AA^bA^c\,,
\end{aligned}
\label{eq:LA1}
\end{equation}
where we suppressed the primed index on $A^a_{b'}$ for readability. The terms in the first line, which don't involve a derivative of $A$ or a commutator, give rise, using (\ref{eq:A2a2}), to the extra terms (up to total derivatives)
\begin{equation}
\begin{aligned}
&{}
g^{-2}
\Big(
-2D_c\partial^aF^bD_c(A^aA^b)
+4F_a{}^{bc}D_eF_a{}^{bd}D_e(A^cA^d)
-2F^{abc}D_eF^{abd}D_e(A^cA^d)
\\
&{}
-4\partial^{(a}F^{b)}A^a\partial^b(\partial^c-F^c)A^c
+4F_a{}^{bc}F_a{}^{bd}A^c\partial^d(\partial^e-F^e)A^e
-2F^{abc}F^{abd}A^c\partial^d(\partial^e-F^e)A^e
\Big)\,.
\end{aligned}
\end{equation}
For the remaining terms in (\ref{eq:LA1}) the identification (\ref{eq:gBdR-mod}) gives
\begin{equation}
\begin{aligned}
gF^{abc}A^a[A^b,A^c]
=&{}
g^{-2}
\Big(
-2F^{abc}F^a{}_{de}F^b{}_{df}F^c{}_{fe}
-6F^{abc}F^a{}_{de}D_dA^bD_eA^c
-6gF^{abc}A^a[D_dA^b,D_dA^c]
\\
&{}
-3gF^{abc}(D_d-F_d)D_dA^a[A^b,A^c]
+3gF^{abc}\partial^a(\partial^d-F^d)A^d[A^b,A^c]
\Big)
+\mathcal O(\alpha'^3)
\end{aligned}
\end{equation}
and
\begin{equation}
\begin{aligned}
F^{Abc}\partial_AA^bA^c
=&{}
g^{-2}
\Big(
F^{Abc}\partial_AF^b{}_{de}F^c{}_{de}
-2F^{Abc}[\partial_A,D_d]A^bD_dA^c
+2D_dF^{Abc}\partial_AA^bD_dA^c
\\
&{}
+2\partial^{[a}F^{b]}A^a\left((D_c-F_c)D_cA^b-\partial^b(\partial^c-F^c)A^c+g[A^b,(\partial^c-F^c)A^c]\right)
\\
&{}
+2F^{Abc}\partial_AA^b\left(\partial^c(\partial^d-F^d)A^d-g[A^c,(\partial^d-F^d)A^d]\right)
\Big)
\\
&{}
+\mbox{total derivatives}
+\mathcal O(\alpha'^3)\,,
\end{aligned}
\end{equation}
while the term involving the field-strength of $A$ is more involved but eventually one finds (useful intermediate results are given in appendix \ref{app:A})
\begin{equation}
\begin{aligned}
F^{ab}F^{ab}
=&\,{}
g^{-2}
\Big(
4R^{ab}{}_{cd}R^{ab}{}_{cd}
+4R^{ab}{}_{cd}D_cA^aD_dA^b
-8D^{[a}D_cA^{b]}[D^a,D_c]A^b
\\
&{}
+8D^{[a}A^{b]}[D_c,\o D^a]D_cA^b
+4g[D^a,D_c]A^bD_c[A^a,A^b]
-4g[D_c,\o D^a]D_cA^b[A^a,A^b]
\\
&{}
+4D_cF^a{}_{cd}D_dA^bF^{ab}
+4D^aF_cD_cA^bF^{ab}
+2F^a{}_{cd}F_{cde}D_eA^bF^{ab}
\\
&{}
-4D^aA^bF_d{}^{ab}\partial_d(\partial^c-F^c)A^c
-4gD^a\partial^b(\partial^c-F^c)A^c[A^a,A^b]
\Big)
\\
&{}
+\mbox{total derivatives}
+\mathcal O(\alpha'^3)\,.
\end{aligned}
\end{equation}
Putting these results together one finds after some work, evaluating the commutators of derivatives using (\ref{eq:comm-luu}) and in particular the Bianchi identity (\ref{eq:bianchi-luuu}), that the extra terms from (\ref{eq:LA1}) reduce to the following terms proportional to the generalized Ricci tensor
\begin{equation}
4g^{-2}
\left(
\mathcal R^a{}_bD_bA^cF^{ac}
-\u D^a\mathcal R^b{}_cA^aD_cA^b
-\mathcal R^a{}_bA^a\partial_b(\partial^c-F^c)A^c
\right)
+\mbox{total derivatives}
+\mathcal O(\alpha'^3)\,.
\end{equation}
Using (\ref{eq:deltaR}) we see that these terms are canceled by the field redefinition
\begin{equation}
E_A{}^M\rightarrow E_A{}^M+\rho_A{}^BE_B{}^M\,,
\end{equation}
with
\begin{equation}
\rho^a{}_b
=
g^{-2}
\left(
D_bA^cF^{ac}
+(\u D^c-F^c)(A^cD_bA^a)
-A^a\partial_b(\partial^c-F^c)A^c
\right)\,.
\label{eq:gBdR-redef}
\end{equation}
This completes the matching of the Lagrangian (\ref{eq:L-gBdR}) with that in (\ref{eq:La2}) up to the second order in $\alpha'$.

Finally we want to match the form of the corrected double Lorentz transformations in (\ref{eq:DLT}). Before the field redefinition (\ref{eq:gBdR-redef}) the relevant transformation of the generalized vielbein, given by the last line in (\ref{eq:gauge-Lorentz}), is (suppressing the primed index on $V$ and $A^b$ for readability)
\begin{equation}
\begin{aligned}
\delta E_a{}^ME^b{}_M
=\,&{}
\partial_aVA^b
\\
=\,&{}
g^{-2}\partial_a\u\lambda_{cd}F^b{}_{cd}
-2g^{-2}\partial_a\partial_cVD_cA^b
-g^{-1}\partial_a\partial^cV[A^b,A^c]
\\
&{}
-g^{-1}\partial^cV[A^b,\partial_aA^c]
-2g^{-2}\partial_aV\left((D_c-F_c)D_cA^b-\partial^b(\partial^c-F^c)A^c\right)
\\
&{}
-g^{-1}\partial_aV
\left(
2[A^b,(\partial^c-F^c)A^c]
-[A^c,\partial^cA^b]
+[A^c,F^{bc}]
\right)
+\mathcal O(g^{-5})\,.
%
\end{aligned}
\end{equation}
After taking the redefinition (\ref{eq:gBdR-redef}) into account the transformation becomes
\begin{equation}
\delta E_a{}^ME^b{}_M
=
D_aV^b
-D^bV_a
+g^{-2}\partial_a\u\lambda_{cd}F^b{}_{cd}
+4g^{-2}(D_c-F_c)\left(\partial_{[c}V_{|d'|}D_{a]}A^b_{d'}\right)
+\mathcal O(g^{-5})\,.
\end{equation}
The first two terms can be canceled by a generalized diffeomorphism.\footnote{The explicit expressions are
$$
\begin{aligned}
V_a
=\,&
-g^{-2}
\left(
(D^c-F^c)[\partial_aV_{d'}A^c_{d'}]
+\partial_aV_{d'}(\partial^c-F^c)A^c_{d'}
\right)
\,,
\\
V^b=\,&
-g^{-2}
\left(
(D_c-F_c)[\partial_cV_{d'}A^b_{d'}]
+\partial^bV_{d'}(\partial^c-F^c)A^c_{d'}
\right)\,.
\end{aligned}
$$
}
 and we are left with
\begin{equation}
\delta E_a{}^ME^b{}_M
=
g^{-2}\partial_a\u\lambda_{cd}F^b{}_{cd}
+4g^{-4}(D_c-F_c)\left(\partial_{[c}\u\lambda_{|de|}\o D_{a]}F^b{}_{de}\right)
+\mathcal O(\alpha'^3)\,,
\end{equation}
in agreement with (\ref{eq:DLT}) upon setting $g^{-2}=-a/2$ (note that the overall sign compared to (\ref{eq:DLT}) is accounted for by the different index placement).

\section{Second order correction: \texorpdfstring{$\mathcal R^{(1,1)}$}{R(1,1)}}\label{sec:R11}
For completeness we work out also the DFT correction for the case of the bosonic string. This turns out to be considerably more complicated than the heterotic case, due to the the presence of a Riemann cubed correction in this case. This section can be skipped by readers interested only in the heterotic string.

While in the heterotic case the parameter $b$ in (\ref{eq:action}) is zero and the full second order correction is given by $\mathcal R^{(0,2)}$, in the bosonic case we have instead $a=b=-\alpha'$. We have already determined $\mathcal R^{(0,2)}$ and $\mathcal R^{(2,0)}$ is easily obtained from this result since it differs only by raising and lowering indices (and appropriate sign changes). This leaves $\mathcal R^{(1,1)}$ to be determined, which is what we will do in this section. Since the calculations are much longer than for $\mathcal R^{(0,2)}$ we will work only up to total derivatives and relegate some of the longest expressions to the appendix.

For the bosonic string we need to correct both factors in the double Lorentz transformations at order $\alpha'$, both corrections taking the form (\ref{eq:deltaEprime})
\begin{equation}
\u\delta' E^{aM}E_{bM}=a\hat{\u\lambda}^a{}_b=\frac{a}{2}\partial_b\u\lambda_{de}F^a{}_{de}\,,\qquad
\o\delta' E^{aM}E_{bM}=b\hat{\o\lambda}^a{}_b=-\frac{b}{2}\partial^a\o\lambda^{de}F_b{}^{de}\,.
\end{equation}
To derive $\mathcal R^{(0,2)}$ we used the first correction in $\mathcal R^{(0,1)}$. To get $\mathcal R^{(1,1)}$ we must instead use the second correction in $\mathcal R^{(0,1)}$. Using the expression for $\mathcal R^{(0,1)}$ in (\ref{eq:deltaL1}) and dropping total derivatives we find
\begin{equation}
\begin{aligned}
b^{-1}\o\delta'\mathcal R^{(0,1)}
=&\,{}
-2\hat{\o\lambda}_c{}^a\mathcal R^{ab}{}_{de}\partial_cF^b{}_{de}%
-2\hat{\o\lambda}_c{}^a\partial^bF^b{}_{de}\partial_cF^a{}_{de}%
-2\hat{\o\lambda}_c{}^aF^b{}_{de}\partial^b\partial_cF^a{}_{de}%
\\
&{}
-2\hat{\o\lambda}_f{}^aF^b{}_{cf}F^{[a}{}_{de}\partial_cF^{b]}{}_{de}%
+4\hat{\o\lambda}_f{}^cF_f{}^{ab}F^a{}_{de}\partial^{[b}F^{c]}{}_{de}%
\\
&{}
-2\hat{\o\lambda}_c{}^d\partial^bF_c{}^{da}M^{ab}%
-2\hat{\o\lambda}_c{}^dF_c{}^{da}\partial^bM^{ab}%
+\hat{\o\lambda}_c{}^bF^a{}_{cd}\partial_dM^{ab}%
\\
&{}
+2\hat{\o\lambda}_c{}^fF^{fdb}F_c{}^{da}M^{ab}%
+2\hat{\o\lambda}_f{}^dF^b{}_{cf}F_c{}^{da}M^{ab}%
-4\hat{\o\lambda}_f{}^aF_f{}^{bc}M^{abc}%
\\
&{}
+2\hat{\o\lambda}_c{}^aF^b{}_{de}\partial_cF^a{}_{de}F^b%
+2\hat{\o\lambda}_c{}^dF_c{}^{da}M^{ab}F^b%
-2\hat{\o\lambda}_c{}^aM^{ab}\mathcal R^b{}_c%
\\
&{}
+8\partial_{[c}\hat{\o\lambda}_{d]}{}^bD_c\mathcal R^b{}_d%
+4\hat{\o\lambda}_e{}^bF_{ecd}D_c\mathcal R^b{}_d%
+8\hat{\o\lambda}_c{}^eF_d{}^{be}D_c\mathcal R^b{}_d%
\,.
\end{aligned}
\label{eq:deltap-R2}
\end{equation}
One can show, following \cite{Hronek:2021nqk}, that this can not be canceled unless one includes a cubic Riemann term. There is only one cubic Riemann term we can write, namely
\begin{equation}
\mathcal R^3=\frac13\mathcal R^{ab}{}_{cd}\mathcal R^{ae}{}_{cf}\mathcal R^{be}{}_{df}\,.
\end{equation}
We therefore need to compute the leading order $\o\lambda$-variation of this term. We find
\begin{equation}
\begin{aligned}
\o\delta\mathcal R^3
=&
-\partial^g\o\lambda^{ab}F^g{}_{cd}\mathcal R^{ae}{}_{cf}\mathcal R^{be}{}_{df}
=
-3\partial^{[g}\o\lambda^{ab]}F^g{}_{cd}\mathcal R^{ae}{}_{cf}\mathcal R^{be}{}_{df}
+2\partial^a\o\lambda^{bg}F^g{}_{cd}\mathcal R^{ae}{}_{cf}\mathcal R^{be}{}_{df}
\\
=&
-3\partial^{[g}\o\lambda^{ab]}F^g{}_{cd}\mathcal R^{ae}{}_{cf}\partial^bF^e{}_{df}
+3\partial^{[g}\o\lambda^{ab]}F^g{}_{cd}\mathcal R^{ae}{}_{cf}\partial^eF^b{}_{df}
+2\partial^a\o\lambda^{bg}F^g{}_{cd}\mathcal R^{ae}{}_{cf}\partial^bF^e{}_{df}
\\
&{}
-2\partial^a\o\lambda^{bg}F^g{}_{cd}\mathcal R^{ae}{}_{cf}\partial^eF^b{}_{df}
-3\partial^{[g}\o\lambda^{ab]}F^g{}_{cd}\mathcal R^{ae}{}_{cf}\mathcal R'^{be}{}_{df}
+2\partial^a\o\lambda^{bg}F^g{}_{cd}\mathcal R^{ae}{}_{cf}\mathcal R'^{be}{}_{df}
\\
=&
-2\partial^a\o\lambda^{bg}F^g{}_{cd}\partial^a\mathcal R^{be}{}_{cf}F^e{}_{df}%
+3\partial^b\partial^{[g}\o\lambda^{ab]}F^g{}_{cd}\mathcal R^{ae}{}_{cf}F^e{}_{df}%
-2\partial^b\partial^a\o\lambda^{bg}F^g{}_{cd}\mathcal R^{ae}{}_{cf}F^e{}_{df}%
\\
&{}
-\frac32\partial^{[g}\o\lambda^{ab]}\partial^e\mathcal R^{ab}{}_{cf}F^g{}_{cd}F^e{}_{df}%
+3\partial^{[g}\o\lambda^{ab]}\partial^aF^b{}_{cd}\mathcal R^{ge}{}_{cf}F^e{}_{df}%
-\frac32\partial^e\partial^{[g}\o\lambda^{ab]}\mathcal R^{ab}{}_{cf}F^g{}_{cd}F^e{}_{df}%
\\
&{}
-2\partial^g\o\lambda^{ab}\partial^aF^b{}_{cd}\mathcal R^{ge}{}_{cf}F^e{}_{df}%
+\partial^g\o\lambda^{ab}\partial^e\mathcal R^{ab}{}_{cf}F^g{}_{cd}F^e{}_{df}%
-\frac92\partial^{[g}\o\lambda^{ab]}F^g{}_{cd}\partial^{[a}\mathcal R^{be]}{}_{cf}F^e{}_{df}%
\\
&{}
+6\partial^a\o\lambda^{bg}F^g{}_{cd}\partial^{[a}\mathcal R^{be]}{}_{cf}F^e{}_{df}%
-\frac32\partial^{[g}\o\lambda^{ab]}F^g{}_{cd}(\partial^e-F^e)\mathcal R^{ae}{}_{cf}F^b{}_{df}%
\\
&{}
+\partial^a\o\lambda^{bg}F^g{}_{cd}(\partial^e-F^e)\mathcal R^{ae}{}_{cf}F^b{}_{df}%
-F^{beH}\partial_H\o\lambda^{ga}\mathcal R^{ab}{}_{cf}F^g{}_{cd}F^e{}_{df}%
\\
&{}
+\frac14F^{egH}\partial_H\o\lambda^{ab}F^g{}_{cd}\mathcal R^{ab}{}_{cf}F^e{}_{df}%
-\frac14F^{abH}\partial_H\o\lambda^{eg}\mathcal R^{ab}{}_{cf}F^g{}_{cd}F^e{}_{df}%
\\
&{}
-3\partial^{[g}\o\lambda^{ab]}F^g{}_{cd}\mathcal R^{ae}{}_{cf}F^e{}_{df}F^b%
+2\partial^a\o\lambda^{bg}F^g{}_{cd}\mathcal R^{ae}{}_{cf}F^e{}_{df}F^b%
-3\partial^{[g}\o\lambda^{ab]}F^g{}_{cd}\mathcal R^{ae}{}_{cf}\mathcal R'^{be}{}_{df}%
\\
&{}
+2\partial^a\o\lambda^{bg}F^g{}_{cd}\mathcal R^{ae}{}_{cf}\mathcal R'^{be}{}_{df}%
+\mbox{total derivatives}\,.
\end{aligned}
\label{eq:deltaR3-1}
\end{equation}
Here $\mathcal R'^{ab}{}_{cd}$ denotes the subleading terms in $\mathcal R^{ab}{}_{cd}$, i.e.
\begin{equation}
\mathcal R'^{ab}{}_{cd}=\mathcal R^{ab}{}_{cd}-2\partial^{[a}F^{b]}{}_{cd}\,.
\end{equation}
The first thing to note is that all terms in (\ref{eq:deltap-R2}) (except those proportional to equations of motion) contain two `traces' (pairs of contracted anti-symmetrized indices), one explicit and one inside $\hat{\o\lambda}$. Therefore we need to manipulate the terms above and add other pieces to the would be $\mathcal R^{(1,1)}$ until this variation also contain at least two traces. From this point on the calculations become very long. Therefore we will only write explicitly the leading order terms in the number of fields in the following. The subleading terms can be found in appendix \ref{app:R11}. 

To start with we note that the first three terms in (\ref{eq:deltaR3-1}) are the only ones that don't involve a trace at the leading order in fields. The first step is to rewrite the first term, let's call it $\Upsilon,$ as follows
\begin{equation}
\begin{aligned}
\Upsilon
=&\,
-2\partial^a\o\lambda^{bg}F^g{}_{cd}\partial^a\mathcal R^{be}{}_{cf}F^e{}_{df}
=
2\partial^a\o\lambda^{bg}\mathcal R^{be}{}_{cf}\partial^a(F^g{}_{cd}F^e{}_{df})
\\
&{}
+2\partial^a\partial^a\o\lambda^{bg}\mathcal R^{be}{}_{cf}F^g{}_{cd}F^e{}_{df}
-2\partial^a\o\lambda^{bg}\mathcal R^{be}{}_{cf}F^g{}_{cd}F^e{}_{df}F^a
+\mbox{total derivatives}
\\
=&\,
-4\partial^a\o\lambda^{bg}\partial_cF_f{}^{be}\partial^a(F^{[g}{}_{cd}F^{e]}{}_{df})
-2\partial^a\o\lambda^{bg}\mathcal R'_{cf}{}^{be}\partial^a(F^g{}_{cd}F^e{}_{df})
\\
&{}
+2\partial^a\partial^a\o\lambda^{bg}\mathcal R^{be}{}_{cf}F^g{}_{cd}F^e{}_{df}
-2\partial^a\o\lambda^{bg}\mathcal R^{be}{}_{cf}F^g{}_{cd}F^e{}_{df}F^a
+\mbox{total derivatives}
\\
=&\,
4\partial^a\partial_c\o\lambda^{bg}F_f{}^{be}\partial^a(F^{[g}{}_{cd}F^{e]}{}_{df})%
+4\partial^a\o\lambda^{bg}F_f{}^{be}\partial^a\partial_c(F^{[g}{}_{cd}F^{e]}{}_{df})%
\\
&{}
+2\partial^a\partial^a\o\lambda^{bg}\mathcal R^{be}{}_{cf}F^g{}_{cd}F^e{}_{df}%
+4\partial^a\o\lambda^{bg}F_f{}^{be}F_c{}^{aH}\partial_H(F^{[g}{}_{cd}F^{e]}{}_{df})%
\\
&{}
+4F_c{}^{aH}\partial_H\o\lambda^{bg}F_f{}^{be}\partial^a(F^{[g}{}_{cd}F^{e]}{}_{df})%
-2\partial^a\o\lambda^{bg}\mathcal R'_{cf}{}^{be}\partial^a(F^g{}_{cd}F^e{}_{df})%
\\
&{}
-4\partial^a\o\lambda^{bg}F_f{}^{be}\partial^a(F^{[g}{}_{cd}F^{e]}{}_{df})F_c%
-2\partial^a\o\lambda^{bg}\mathcal R^{be}{}_{cf}F^g{}_{cd}F^e{}_{df}F^a%
+\mbox{total derivatives}\,.
%
%
\end{aligned}
\end{equation}
With some work we can write this as the variation of something plus left over terms. Note that whatever we add to $\mathcal R^3$ should be anti-symmetric under exchanging upper and lower indices and adding an extra minus sign for each pair of contracted indices. This is equivalent to symmetry under exchanging the $P_\pm$ projections (\ref{eq:Ppm}) and ensures that the action contains only even powers of $H_{abc}$. Defining
\begin{equation}
\begin{aligned}
L_1=&\,
\partial^a(F_c{}^{bg}F_f{}^{eb})\partial^a(F^{[g}{}_{cd}F^{e]}{}_{df})
-\u D^a(F_c{}^{bg}F_f{}^{eb})\u D^a(F^{[g}{}_{cd}F^{e]}{}_{df})
\\
&{}
-\o D_a(F_c{}^{bg}F_f{}^{eb})\o D_a(F^{[g}{}_{cd}F^{e]}{}_{df})
-8F_{[a}{}^{bh}F_{c]}{}^{bg}\mathcal R^{he}{}_{af}F^{[g}{}_{cd}F^{e]}{}_{df}\,,
\end{aligned}
\end{equation}
we find
\begin{equation}
\Upsilon=
\o\delta L_1%
+\partial^a\partial^a\o\lambda^{bg}\mathcal R^{be}{}_{cf}F^g{}_{cd}F^e{}_{df}%
+\Psi
+\Upsilon_{4,0}
+\Upsilon_{5,0}
+\Upsilon_{4,0,\Phi}
+\Upsilon_{5,0,\Phi}
+\Upsilon_{\mathrm{e.o.m.}}
+\mbox{total derivatives}\,.
\label{eq:upsilon}
\end{equation}
Here we have defined the combination of terms
\begin{equation}
\begin{aligned}
\Psi
=&\,{}
-4\partial_a\o\lambda^{bg}\partial_aF_f{}^{be}\o D_c(F^{[g}{}_{cd}F^{e]}{}_{df})
-2\partial^a\partial^a\o\lambda^{bg}F_f{}^{be}\o D_c(F^{[g}{}_{cd}F^{e]}{}_{df})
\\
&{}
+2\partial_f\o\lambda^{bg}\partial_aF_a{}^{be}\o D_c(F^{[g}{}_{cd}F^{e]}{}_{df})
+2\partial_f\o\lambda^{bg}F_a{}^{bh}F_a{}^{he}\partial_c(F^{[g}{}_{cd}F^{e]}{}_{df})
\\
&{}
-2\partial_f\o\lambda^{bh}F_a{}^{bg}F_a{}^{he}\partial_c(F^{[g}{}_{cd}F^{e]}{}_{df})
-4\partial_a\o\lambda^{bh}F_a{}^{gh}F_f{}^{eb}\partial_c(F^{[g}{}_{cd}F^{e]}{}_{df})
\\
&{}
-4\partial_a\o\lambda^{bg}F_a{}^{eh}F_f{}^{hb}\partial_c(F^{[g}{}_{cd}F^{e]}{}_{df})
+4\partial^a\o\lambda^{bg}F_h{}^{eb}F^a{}_{fh}\partial_c(F^{[g}{}_{cd}F^{e]}{}_{df})\,,
\end{aligned}
\end{equation}
which, in particular, contains all the terms cubic in the fields. The remaining terms are grouped according to the number of fields and the number of traces, e.g. $\Upsilon_{4,0}$ contains the quartic terms with no traces. The terms containing the dilaton, i.e. $F_a$ ($F^a$) are listed separately, e.g. quartic terms containing the dilaton and without traces are contained in $\Upsilon_{4,0,\Phi}$. Finally, terms proportional to the equations of motion are contained in $\Upsilon_{\mathrm{e.o.m.}}$. The explicit expressions can be found in appendix \ref{app:R11}.

The next step is to manipulate the $\Psi$-terms. After some work one finds that they can be organized as follows
\begin{equation}
\begin{aligned}
\Psi
=&\,{}
\Psi_{3,1}
+\Psi_{3,2}
+\Psi_{4,0}
+\Psi_{4,1}
+\Psi_{4,2}
+\Psi_{5,0}
+\Psi_{5,1}
+\Psi_{3,0,\Phi}
+\Psi_{3,1,\Phi}
+\Psi_{4,0,\Phi}
+\Psi_{4,1,\Phi}
\\
&{}
+\Psi_{\mathrm{e.o.m.}}
+\mbox{total derivatives}\,.
\end{aligned}
\label{eq:Psi}
\end{equation}
We will give only the cubic terms here. The rest can be found in appendix \ref{app:R11}. The terms with one trace are
\begin{equation}
\begin{aligned}
\Psi_{3,1}=&\,{}
\partial_e\o\lambda^{cf}\partial_eF_g{}^{fd}\left(F_{abg}\mathcal R^{cd}{}_{ab}-2F^{[c}{}_{ab}\partial_gF^{d]}{}_{ab}\right)%
+\frac12\partial^e\partial^e\o\lambda^{cf}F_g{}^{fd}\left(F_{abg}\mathcal R^{cd}{}_{ab}-2F^{[c}{}_{ab}\partial_gF^{d]}{}_{ab}\right)%
\\
&{}
-\frac12\partial_g\o\lambda^{cf}\partial_eF_e{}^{fd}\left(F_{abg}\mathcal R^{cd}{}_{ab}-2F^{[c}{}_{ab}\partial_gF^{d]}{}_{ab}\right)\,,%
\end{aligned}
\end{equation}
while those with two traces take the form
\begin{equation}
\begin{aligned}
\Psi_{3,2}=&\,
2\partial_a\partial^{[e|}\o\lambda^{bg}\partial_aF_f{}^{be}F^{|g]}{}_{cd}F_{cdf}%
-\partial_f\partial^{[e|}\o\lambda^{bg}\partial_aF_a{}^{be}F^{|g]}{}_{cd}F_{cdf}%
+\partial^a\partial^a\partial^{[e|}\o\lambda^{bg}F_f{}^{be}F^{|g]}{}_{cd}F_{cdf}%
\\
&{}
+\frac12\partial_g\o\lambda^{ab}\partial_g\partial_fF^{abe}F^e{}_{cd}F_{cdf}%
-\frac14\partial_f\o\lambda^{ab}\partial_g\partial_gF^{abe}F^e{}_{cd}F_{cdf}%
+\frac14\partial^g\partial^g\o\lambda^{ab}\partial_fF^{abe}F^e{}_{cd}F_{cdf}%
\\
&{}
-\frac12\partial_g\o\lambda^{ab}\partial_g\u D^eF_f{}^{ab}F^e{}_{cd}F_{cdf}%
+\frac14\partial_f\o\lambda^{ab}\partial_g\u D^eF_g{}^{ab}F^e{}_{cd}F_{cdf}%
-\frac14\partial^g\partial^g\o\lambda^{ab}\u D^eF_f{}^{ab}F^e{}_{cd}F_{cdf}\,.%
\end{aligned}
\end{equation}
The cubic terms containing the dilaton ($F^a$ or $F_a$) and no trace are
\begin{equation}
\Psi_{3,0,\Phi}=
4\partial_a\o\lambda^{bg}\partial_aF_f{}^{be}F^{[e}{}_{df}D^{g]}F_d
+2\partial^a\partial^a\o\lambda^{bg}F_f{}^{be}F^{[e}{}_{df}D^{g]}F_d
-2\partial_f\o\lambda^{bg}\partial_aF_a{}^{be}F^{[e}{}_{df}D^{g]}F_d
\end{equation}
and those with one trace
\begin{equation}
\Psi_{3,1,\Phi}=
-\partial_a\o\lambda^{bg}F^g{}_{cd}F_{cdf}\partial_aD_fF^b%
+\frac12\partial_f\o\lambda^{bg}F^g{}_{cd}F_{cdf}\partial_aD_aF^b%
-\frac12\partial^a\partial^a\o\lambda^{bg}F^g{}_{cd}F_{cdf}D_fF^b\,.%
\end{equation}

Finally one finds, after a lot of work, that
\begin{equation}
ab\o\delta\mathcal R^{(1,1)}+a\o\delta'\mathcal R^{(0,1)}
=
-4\o\delta''E^a{}_b\mathcal R^a{}_b
-2\o\delta''d\mathcal R\,,
\end{equation}
where
\begin{equation}
\mathcal R^{(1,1)}=\o{\mathcal R}^{(1,1)}+\u{\mathcal R}^{(1,1)}
\end{equation}
with
\begin{equation}
\begin{aligned}
\o{\mathcal R}^{(1,1)}
=&\,
\frac16\mathcal R^{ab}{}_{cd}\mathcal R^{ae}{}_{cf}\mathcal R^{be}{}_{df}
-\mathcal R^{ab}{}_{cd}[F,F]^{be}{}_{cd}\partial^{(a}F^{e)}
-\frac12M^{abc}F^{abD}\partial_DF^c
\\
&{}
+\left(\mathcal R^{ab}{}_{[c|g}F^b{}_{g|d]}-(D_{[c}+\o D_{[c}+F_{[c})\mathcal R^a{}_{d]}-\frac12(\u D^b-F^b)[F,F]^{ab}{}_{cd}\right)
\\
&{}
\qquad\times\left(\frac12F^{aef}\mathcal R^{ef}{}_{cd}+2F_c{}^{ae}\mathcal R^e{}_d+F_c{}^{ef}\u D^aF_d{}^{ef}\right)
-\frac18\partial_e[F,F]_{ab}{}^{cd}\partial_e[F,F]^{cd}{}_{ab}
\\
&{}
+\frac14\u D^e[F,F]_{ab}{}^{cd}\u D^e[F,F]^{cd}{}_{ab}
-[F,F]_{ab}{}^{cd}(\mathcal R^{ce}{}_{af}-\frac{1}{32}\eta^{ce}\eta_{af}\mathcal R)[F,F]^{de}{}_{bf}
\\
&{}
-\frac12\partial_a(FF)_b{}^c\partial_a(FF)^c{}_b
+\u D^a(FF)_b{}^c\u D^a(FF)^c{}_b
\\
&{}
-(FF)_a{}^c(\mathcal R^{cd}{}_{ab}-\frac18\eta_{ab}\eta^{cd}\mathcal R)(FF)^d{}_b
+\frac{1}{32}\partial_gM_{aa}\partial_gM^{bb}
\\
&{}
-\frac{1}{128}M_{aa}M^{bb}\mathcal R
+\frac18\partial_g\big(F^aF^a+\frac16F^{abe}F^{abe}\big)\partial_g\big(F_cF_c+\frac16F_{cdf}F_{cdf}\big)
\\
&{}
+\frac{1}{32}(F^aF^a+\frac16F^{abc}F^{abc})(F_dF_d+\frac16F_{def}F_{def})\mathcal R
-\frac18\partial^aM_{bb}F^a{}_{cd}(\partial^e-F^e)F^e{}_{cd}
\\
&{}
+\frac12(FF)^a{}_b\u D^a\big(F_b{}^{cd}(\partial_e-F_e)F_e{}^{cd}\big)
+\frac14\u D^gF_e{}^{ab}[F,F]_{ef}{}^{ab}F^g{}_{cd}F_{cdf}
+\o L_{\mathrm{e.o.m.}}\,,
\end{aligned}
\label{eq:oR11}
%
\end{equation}
where we have introduced the following shorthand notation
\begin{equation}
\begin{aligned}
&[F,F]^{ab}{}_{cd}=2F^{[a}{}_{ce}F^{b]}{}_{ed}\,,\qquad
[F,F]_{ab}{}^{cd}=2F_{[a}{}^{ce}F_{b]}{}^{ed}\,,
\\
&(FF)_a{}^b\equiv F_a{}^{bc}F^c+\frac12F_a{}^{cd}F^{cdb}\,,\qquad
(FF)^a{}_b\equiv F^a{}_{bc}F_c-\frac12F^a{}_{cd}F_{cdb}\,.
\end{aligned}
\end{equation}
The extra terms proportional to the equations of motion take the form
\begin{equation}
\begin{aligned}
\o L_{\mathrm{e.o.m.}}
=&{}
-2\partial^aF^b\mathcal R^a{}_c\mathcal R^b{}_c
-\frac32\mathcal R^{ab}{}_{cd}\left(\mathcal R^a{}_c\mathcal R^b{}_d+\frac{1}{16}\mathcal R^{ab}{}_{cd}\mathcal R\right)
+2F^{cdf}\mathcal R^f{}_bD^c\mathcal R^d{}_b
\\
&{}
+F^{cdf}\mathcal R^{cd}{}_{ab}\o D_a\mathcal R^f{}_b
+2\mathcal R^{gh}{}_{cd}\u D^bF_c{}^{gh}\mathcal R^b{}_d
+[F,F]^{ab}{}_{cd}\left(\mathcal R^a{}_c\mathcal R^b{}_d-\frac{1}{16}\mathcal R^{ab}{}_{cd}\mathcal R\right)
\\
&{}
-\frac14M^{aa}\left(\mathcal R^c{}_b\mathcal R^c{}_b-\frac{1}{16}\mathcal R^2\right)
+\o D_aF^d{}_{bc}M_{ab}\mathcal R^d{}_c
+\frac14M^{aa}F_d{}^{bc}D^b\mathcal R^c{}_d
\\
&{}
-\frac12F_d{}^{ab}M^{abc}\mathcal R^c{}_d
+\frac12F_e{}^{df}F^d{}_{ab}(\partial^c-F^c)F^c{}_{ab}\mathcal R^f{}_e
+2F_{[c}{}^{bg}\mathcal R^{[b}{}_{d]}F^{e]}{}_{cf}\mathcal R^{eg}{}_{fd}
\\
&{}
+\frac18F^{cdf}\mathcal R^{cd}{}_{ab}F_{abh}\mathcal R^f{}_h
-\frac18F_e{}^{ab}\mathcal R^{ab}{}_{cd}F^f{}_{cd}\mathcal R^f{}_e
+\frac12F^aF_b\mathcal R^{ac}{}_{bd}\mathcal R^c{}_d
\\
&{}
-\frac18F_bF^a\mathcal R^a{}_b\mathcal R
-(FF)^c{}_bD^c(\mathcal R^d{}_bF^d)
+\frac12(F^cF^c+\frac16F^{cdf}F^{cdf})(FF)^a{}_h\mathcal R^a{}_h
\\
&{}
-(FF)_c{}^a[F,F]^{ab}{}_{cd}\mathcal R^b{}_d
+F^a[F,F]^{ab}{}_{cd}D_c\mathcal R^b{}_d
+\frac18F_c{}^{ab}(\partial_d-F_d)F_d{}^{ab}\mathcal RF_c
\\
&{}
-\frac12F^a{}_{cd}\partial_d\mathcal R\mathcal R^a{}_c
-\frac18F^{abg}\mathcal R^{ab}{}_{cd}F^g{}_{cd}\mathcal R
+\frac18F_c{}^{ab}\u D^eF_d{}^{ab}F^e{}_{cd}\mathcal R
\\
&{}
-\frac18F_c{}^{ef}F^e{}_{cg}\mathcal R^f{}_g\mathcal R\,.
\end{aligned}
\end{equation}
These terms, which could be canceled by field redefinitions, are included in order to simplify the form of the corresponding $\alpha'^2$ correction to the double Lorentz transformations which, with this choice, takes the form
\begin{equation}
\o\delta''d=
\frac{ab}{16}
\Big(
4\partial_c\o\lambda^{ab}F_d{}^{ab}\partial_{(c}F_{d)}
+2\partial^c\o\lambda^{de}F^c{}_{ab}F^d{}_{ab}F^e
+\partial_f\o\lambda^{cd}F_f{}^{ce}F^e{}_{ab}F^d{}_{ab}
-\partial^f\o\lambda^{cd}F^{cef}F^e{}_{ab}F^d{}_{ab}
\Big)
\end{equation}
and
\begin{equation}
\begin{aligned}
\o\delta''E^a{}_b
=&
-\frac{ab}{8}
\Big(
2\partial^a\o\lambda^{ef}F^e{}_{cd}\o D_bF^f{}_{cd}
+8(\o D_c-F_c)(\partial^d\o\lambda^{ef}F_{(c}{}^{ef}F_{b)}{}^{ad})
\\
&{}
-4(D_c-\o D_c)(\partial_{[b}\o\lambda^{ef}F_{d]}{}^{ef}F^a{}_{cd})
-2\partial_b(\partial_c\o\lambda^{ef}F_d{}^{ef})F^a{}_{cd}
-2\partial_c\o\lambda^{ef}\mathcal R'_{bd}{}^{ef}F^a{}_{cd}
\\
&{}
+\partial_c\o\lambda^{ef}\mathcal R'^{ef}{}_{bd}F^a{}_{cd}
-\partial_c\o\lambda^{ef}\mathcal R'^{ef}{}_{cd}F^a{}_{bd}
+\partial^g\o\lambda^{ef}F_c{}^{ef}F^h{}_{bc}F^{agh}
\\
&{}
-\partial_h\o\lambda^{ge}F_h{}^{ea}F^g{}_{cd}F_{bcd}
+\partial_h\o\lambda^{ae}F_h{}^{eg}F^g{}_{cd}F_{bcd}
+4\partial^a\o\lambda^{ef}\mathcal R^{ef}{}_{bc}F_c
\Big)\,.
\end{aligned}
\end{equation}
The expressions for $\u{\mathcal R}^{(1,1)}$ and $\u L_{\mathrm{e.o.m.}}$ are related to the ones above by exchanging upper and lower indices and including a sign for each pair of contracted indices plus one, for completeness the explicit expressions are given in appendix \ref{app:R11}. Of course, the form of the action and transformations presented here is probably not the most economical, though it seems to have an interesting structure.

\section{Conclusions}
We have constructed the two-parameter DFT action and transformations (\ref{eq:action}) to order $\alpha'^2$. We did this by direct calculations which led to vastly simpler expressions than those found in \cite{Baron:2020xel} using the generalized Bergshoeff-de Roo identification. We also showed that, at least in the heterotic case, the two approaches give the same result, the two being related by rather complicated field redefinitions, {generalized diffeomorphisms and modifications of the transformations and Lagrangian}. We have also demonstrated that in the heterotic case the DFT action reproduces the tree-level string effective action to order $\alpha'^2$, extending the results of \cite{Hronek:2021nqk} beyond the leading order in the fields.

The simple form for the DFT action to second order in $\alpha'$ presented here should be useful for various applications. One example would be to find the $\alpha'^2$ correction to (generalized) T-duality transformations and related integrable deformations of strings following \cite{Borsato:2020bqo} and \cite{Borsato:2020wwk,Hassler:2020tvz,Codina:2020yma} (for early works on corrections to standard abelian T-duality see \cite{Bergshoeff:1995cg,Kaloper:1997ux}).

While the generalized Bergshoeff-de Roo identification seems more complicated in this case, it is still very interesting since it captures an infinite series of $\alpha'$ corrections. It would be very interesting to try to use it beyond order $\alpha'^2$, for example to see that one recovers the quartic Riemann terms found in \cite{Bergshoeff:1989de} at order $\alpha'^3$. Another interesting question is including the gauge vectors, and fermions and supersymmetry, \cite{Baron:2018lve} in the DFT description of the heterotic string \cite{Hohm:2011ex}, see \cite{Lescano:2021guc} for a detailed account at order $\alpha'$.

\vspace{2cm}

\section*{Acknowledgments}
This work is supported by the grant ``Integrable Deformations'' (GA20-04800S) from the Czech Science Foundation (GA\v CR). The work of SH was also supported by Operational Program Research, Development and Education -- ``Project Internal Grant Agency of Masaryk University'' (No. CZ.02.2.69/0.0/0.0/19\_073/0016943).

\vspace{4cm}

\appendix

\section{Details of comparison to generalized Bergshoeff-de Roo identification}\label{app:A}
Here we give some useful intermediate results for section \ref{sec:gBdR2}. The identification (\ref{eq:gBdR-mod}) gives
\begin{equation}
\begin{aligned}
D^{[a}A^{b]}D^aA^b
=&{}
g^{-2}
\Big(
\o D^{[a}F^{b]}{}_{cd}\o D^aF^b{}_{cd}
-2\o D^{[a}D_cA^{b]}\o D^aD_cA^b
\\
&{}
\hspace{-40pt}
-2D^{[a}A^{b]}D^a\left((D_c-F_c)D_cA^b-\partial^b(\partial^c-F^c)A^c+g[A^b,(\partial^c-F^c)A^c]\right)
\Big)
+\mathcal O(\alpha'^3)\,.
\end{aligned}
\end{equation}
We also find
\begin{equation}
\begin{aligned}
gD^{[a}A^{b]}[A^a,A^b]
=&\,{}
g^{-2}
\Big(
-2\o D^{[a}F^{b]}{}_{cd}F^a{}_{ce}F^b{}_{ed}
-2\o D^{[a}F^{b]}{}_{cd}D_cA^aD_dA^b
\\
&{}
\hspace{-40pt}
+4\o D^{[a}D_cA^{b]}F^a{}_{ce}D_eA^b
-2g\o D^{[a}D_cA^{b]}D_c[A^a,A^b]
-2gD^aA^b[D_eA^a,D_eA^b]
\\
&{}
\hspace{-40pt}
-gD^a\left((D_c-F_c)D_cA^b-\partial^b(\partial^c-F^c)A^c+g[A^b,(\partial^c-F^c)A^c]\right)[A^a,A^b]
\\
&{}
\hspace{-40pt}
-2gD^{[a}A^{b]}[A^a,(D_c-F_c)D_cA^b-\partial^b(\partial^c-F^c)A^c+g[A^b,(\partial^c-F^c)A^c]]
\Big)
+\mathcal O(\alpha'^3)
\end{aligned}
%
\end{equation}
and
\begin{equation}
\begin{aligned}
g^2[A^a,A^b][A^a,A^b]
=\,&{}
g^{-2}
\Big(
2F^{[a}{}_{cf}F^{b]}{}_{fd}F^a{}_{ce}F^b{}_{ed}
+4F^{[a}{}_{ce}F^{b]}{}_{ed}D_cA^aD_dA^b
-8F^{[a}_{cf}D_fA^{b]}F^a{}_{ce}D_eA^b
\\
&{}
\hspace{-40pt}
-2g^2D_c[A^a,A^b]D_c[A^a,A^b]
+8gF^a{}_{ce}D_eA^bD_c[A^a,A^b]
-4g^2[A^a,A^b][D_eA^a,D_eA^b]
\\
&{}
\hspace{-40pt}
-4g^2[A^a,A^b][A^a,(D_c-F_c)D_cA^b-\partial^b(\partial^c-F^c)A^c]
\Big)
+\mathcal O(\alpha'^3)\,.
\end{aligned}
\end{equation}

\section{Details of \texorpdfstring{$\mathcal R^{(1,1)}$}{R(1,1)} calculation}\label{app:R11}
The terms of higher order in fields in (\ref{eq:upsilon}) take the form
\begin{equation}
\begin{aligned}
\Upsilon_{4,0}=&\,{}
-4\partial^a\o\lambda^{bg}\mathcal R^{be}{}_{fh}F^a{}_{ch}F^{[g}{}_{cd}F^{e]}{}_{df}
+2\partial^a\o\lambda^{bg}F_h{}^{be}\partial_hF^a{}_{cf}F^{[g}{}_{cd}F^{e]}{}_{df}
\\
&{}
+2\partial_a\o\lambda^{bh}F_a{}^{gh}\mathcal R^{eb}{}_{cf}F^g{}_{cd}F^e{}_{df}
+2\partial_a\o\lambda^{bg}F_a{}^{eh}\mathcal R^{hb}{}_{cf}F^g{}_{cd}F^e{}_{df}
+\partial^a\o\lambda^{bg}\partial_hF_h{}^{ge}M^{abe}
\\
&{}
+2\partial_a\o\lambda^{bg}\partial_aF_h{}^{be}F^g{}_{cd}F^e{}_{df}F_{cfh}
-\partial_h\o\lambda^{bg}\partial_aF_a{}^{be}F^g{}_{cd}F^e{}_{df}F_{cfh}
\\
&{}
+\partial_a\partial_a\o\lambda^{bg}F_h{}^{be}F^g{}_{cd}F^e{}_{df}F_{cfh}\,,
\end{aligned}
\end{equation}

\begin{equation}
\begin{aligned}
\Upsilon_{5,0}=&\,{}
-4\partial^a\o\lambda^{bg}F^a{}_{ch}F_h{}^{bk}F_f{}^{ke}F^{[g}{}_{cd}F^{e]}{}_{df}
+4\partial^a\o\lambda^{bh}F^a{}_{ck}F_k{}^{be}F_f{}^{gh}F^{[g}{}_{cd}F^{e]}{}_{df}
\\
&{}
+2\partial^a\o\lambda^{bg}F^a{}_{kh}F_h{}^{be}F^g{}_{cd}F^e{}_{df}F_{cfk}
+2\partial_f\o\lambda^{bg}F_a{}^{bh}F_a{}^{he}\o D'_c(F^{[g}{}_{cd}F^{e]}{}_{df})
\\
&{}
-2\partial_f\o\lambda^{bh}F_a{}^{bg}F_a{}^{he}\o D'_c(F^{[g}{}_{cd}F^{e]}{}_{df})
-4\partial_a\o\lambda^{bh}F_a{}^{gh}F_f{}^{eb}\o D'_c(F^{[g}{}_{cd}F^{e]}{}_{df})
\\
&{}
-4\partial_a\o\lambda^{bg}F_a{}^{eh}F_f{}^{hb}\o D'_c(F^{[g}{}_{cd}F^{e]}{}_{df})
+\partial^g\o\lambda^{ab}F_h{}^{ae}F_h{}^{ek}M^{bgk}
+\partial^g\o\lambda^{ab}F_h{}^{ae}F_h{}^{bk}M^{gek}
\\
&{}
+2\partial^a\o\lambda^{bg}F_h{}^{ak}F_h{}^{be}M^{kge}
+2\partial_a\o\lambda^{bh}F_a{}^{gh}F_k{}^{eb}F^{[g}{}_{cd}F^{e]}{}_{df}F_{cfk}
\\
&{}
+2\partial_a\o\lambda^{bg}F_a{}^{eh}F_k{}^{hb}F^{[g}{}_{cd}F^{e]}{}_{df}F_{cfk}
+\partial_k\o\lambda^{bg}F_a{}^{bh}F_a{}^{he}F^{[g}{}_{cd}F^{e]}{}_{df}F_{fck}
\\
&{}
-\partial_k\o\lambda^{bh}F_a{}^{bg}F_a{}^{he}F^{[g}{}_{cd}F^{e]}{}_{df}F_{fck}\,,
\end{aligned}
\end{equation}
where $D'$ denotes the connection terms only, i.e. $D'_c=D_c-\partial_c$,
\begin{equation}
\begin{aligned}
\Upsilon_{4,0,\Phi}=&\,{}
4\partial_a\o\lambda^{bg}\partial_aF_f{}^{be}F^{[g}{}_{cd}F^{e]}{}_{df}F_c
+2\partial_a\partial_a\o\lambda^{bg}F_f{}^{be}F^{[g}{}_{cd}F^{e]}{}_{df}F_c
\\
&{}
-2\partial_f\o\lambda^{bg}\partial_aF_a{}^{be}F^{[g}{}_{cd}F^{e]}{}_{df}F_c
-2\partial_a\o\lambda^{bg}F_f{}^{be}\o D'_c(F^{[g}{}_{cd}F^{e]}{}_{df})F_a
\\
&{}
-2\partial_f\o\lambda^{bg}F_h{}^{be}\partial_c(F^{[g}{}_{cd}F^{e]}{}_{df})F_h
+\partial_a\o\lambda^{bg}\mathcal R^{be}{}_{cf}F^g{}_{cd}F^e{}_{df}F_a
\\
&{}
-\partial^a\o\lambda^{bg}\mathcal R^{be}{}_{cf}F^g{}_{cd}F^e{}_{df}F^a
+2\partial^a\o\lambda^{bg}F_f{}^{be}\o D_c(F^{[g}{}_{cd}F^{e]}{}_{df})F^a\,,
\end{aligned}
\end{equation}

\begin{equation}
\begin{aligned}
\Upsilon_{5,0,\Phi}
=&\,{}
4\partial_a\o\lambda^{bh}F_a{}^{gh}F_f{}^{eb}F^{[g}{}_{cd}F^{e]}{}_{df}F_c
+4\partial_a\o\lambda^{bg}F_a{}^{eh}F_f{}^{hb}F^{[g}{}_{cd}F^{e]}{}_{df}F_c
\\
&{}
+2\partial_c\o\lambda^{bg}F_a{}^{bh}F_a{}^{he}F^{[g}{}_{cd}F^{e]}{}_{df}F_f
-2\partial_c\o\lambda^{bh}F_a{}^{bg}F_a{}^{he}F^{[g}{}_{cd}F^{e]}{}_{df}F_f
\\
&{}
-2\partial_c\o\lambda^{bh}F_f{}^{eb}F_a{}^{gh}F^{[g}{}_{cd}F^{e]}{}_{df}F_a
-2\partial_c\o\lambda^{bg}F_f{}^{eh}F_a{}^{hb}F^{[g}{}_{cd}F^{e]}{}_{df}F_a
\\
&{}
-4\partial^a\o\lambda^{bg}F_h{}^{be}F^a{}_{ch}F^{[g}{}_{cd}F^{e]}{}_{df}F_f
-\partial^h\o\lambda^{bg}F_a{}^{be}F^h{}_{cf}F^{[g}{}_{cd}F^{e]}{}_{df}F_a
\\
&{}
-\partial_h\o\lambda^{bg}F_a{}^{be}F^g{}_{cd}F^e{}_{df}F_{cfh}F_a
-\partial^a\o\lambda^{bg}F_h{}^{be}F^g{}_{cd}F^e{}_{df}F_{cfh}F^a
\\
&{}
+\partial_a\o\lambda^{bg}F_h{}^{be}F^g{}_{cd}F^e{}_{df}F_{cfh}F_a
-2\partial_a\o\lambda^{bg}F_f{}^{be}F^{[g}{}_{cd}F^{e]}{}_{df}F_aF_c
\\
&{}
+2\partial_f\o\lambda^{bg}F_a{}^{be}F^{[g}{}_{cd}F^{e]}{}_{df}F_aF_c\,,
\end{aligned}
\end{equation}
while the equations of motion terms are
\begin{equation}
\Upsilon_{\mathrm{e.o.m.}}
=
4\partial_c\o\lambda^{bg}F^{[g}{}_{cd}F^{e]}{}_{df}D^{[e}\mathcal R^{b]}{}_f
+4\partial^a\o\lambda^{bg}F_f{}^{be}\mathcal R^a{}_cF^{[g}{}_{cd}F^{e]}{}_{df}
+2\partial^h\o\lambda^{bg}F_f{}^{eb}\mathcal R^h{}_cF^{[g}{}_{cd}F^{e]}{}_{df}\,.
\end{equation}

The terms of higher order in the fields in (\ref{eq:Psi}) take the form
\begin{equation}
\Psi_{4,0}=
-2\partial_a\o\lambda^{bg}\partial_aF_f{}^{be}F^g{}_{ch}F^e{}_{hd}F_{cdf}%
-\partial^a\partial^a\o\lambda^{bg}F_f{}^{be}F^g{}_{ch}F^e{}_{hd}F_{cdf}%
+\partial_f\o\lambda^{bg}\partial_aF_a{}^{be}F^g{}_{ch}F^e{}_{hd}F_{cdf}\,,%
\end{equation}

\begin{equation}
\begin{aligned}
\Psi_{4,1}
=&\,{}
\partial^a\o\lambda^{bg}F_h{}^{eb}F^a{}_{fh}\big(F_{cdf}\mathcal R^{ge}{}_{cd}-2F^{[g}{}_{cd}\partial_fF^{e]}{}_{cd}\big)%
-\partial_a\o\lambda^{bg}F_a{}^{eh}F_f{}^{hb}\big(F_{cdf}\mathcal R^{ge}{}_{cd}-2F^{[g}{}_{cd}\partial_fF^{e]}{}_{cd}\big)%
\\
&{}
-\partial_a\o\lambda^{bh}F_a{}^{gh}F_f{}^{eb}\big(F_{cdf}\mathcal R^{ge}{}_{cd}-2F^{[g}{}_{cd}\partial_fF^{e]}{}_{cd}\big)%
-\frac12\partial_f\o\lambda^{bh}F_a{}^{bg}F_a{}^{he}\big(F_{cdf}\mathcal R^{ge}{}_{cd}-2F^{[g}{}_{cd}\partial_fF^{e]}{}_{cd}\big)%
\\
&{}
+\frac12\partial_f\o\lambda^{bg}F_a{}^{bh}F_a{}^{he}\big(\mathcal R^{ge}{}_{cd}F_{cdf}-2F^{[g}{}_{cd}\partial_fF^{e]}{}_{cd}\big)%
-2\partial_a\o\lambda^{bg}\partial_aF_f{}^{be}F_f{}^{h[g}F^{e]}{}_{cd}F^h{}_{cd}%
\\
&{}
-\partial^a\partial^a\o\lambda^{bg}F_f{}^{be}F_f{}^{h[g}F^{e]}{}_{cd}F^h{}_{cd}%
+\partial_f\o\lambda^{bg}\partial_aF_a{}^{be}F_f{}^{h[g}F^{e]}{}_{cd}F^h{}_{cd}%
+\partial^g\o\lambda^{ab}F_h{}^{be}\mathcal R^{ae}{}_{fh}F^g{}_{cd}F_{cdf}%
\\
&{}
+\partial^b\o\lambda^{ag}F_h{}^{be}\partial_fF_h{}^{ae}F^g{}_{cd}F_{cdf}%
-\frac12\partial^e\partial_f\o\lambda^{bg}F_a{}^{bh}F_a{}^{he}F^g{}_{cd}F_{cdf}%
+\frac12\partial^h\o\lambda^{bg}\partial^aF^{ahe}F_f{}^{be}F^g{}_{cd}F_{cdf}%
\\
&{}
+\frac12\partial^h\o\lambda^{bg}\partial_aF_a{}^{be}F_f{}^{eh}F^g{}_{cd}F_{cdf}%
-\partial_a\o\lambda^{bh}F_f{}^{eb}\partial^gF_a{}^{he}F^g{}_{cd}F_{cdf}%
-\frac12\partial_f\o\lambda^{bh}F_a{}^{he}\partial^gF_a{}^{eb}F^g{}_{cd}F_{cdf}%
\\
&{}
-\partial_a\o\lambda^{bg}\mathcal R^{be}{}_{af}F^{egh}F^h{}_{cd}F_{cdf}%
-3\partial^{[a}\o\lambda^{bg]}F_h{}^{be}\mathcal R^{ae}{}_{fh}F^g{}_{cd}F_{cdf}
-\partial_h\o\lambda^{be}F^{abg}\mathcal R^{ae}{}_{fh}F^g{}_{cd}F_{cdf}\,,
\end{aligned}
\end{equation}

\begin{equation}
\begin{aligned}
\Psi_{4,2}=&\,{}
-\frac12\partial_g\o\lambda^{ab}\partial_g(F_f{}^{eh}F^{abh})F^e{}_{cd}F_{cdf}%
+\frac14\partial_f\o\lambda^{ab}\partial_g(F_g{}^{eh}F^{abh})F^e{}_{cd}F_{cdf}%
\\
&{}
-\frac14\partial^g\partial^g\o\lambda^{ab}F_f{}^{eh}F^{abh}F^e{}_{cd}F_{cdf}%
-2\partial^{[e|}\partial^a\o\lambda^{bg}F_h{}^{eb}F^a{}_{fh}F^{|g]}{}_{cd}F_{cdf}%
\\
&{}
-2\partial^a\o\lambda^{bg}\partial^{[e|}F_h{}^{eb}F^a{}_{fh}F^{|g]}{}_{cd}F_{cdf}%
+\partial^a\partial_h\o\lambda^{bg}F_f{}^{be}F_h{}^{ga}F^e{}_{cd}F_{cdf}%
\\
&{}
-\frac12\partial^h\o\lambda^{bg}\partial^aF^{abh}F_f{}^{ge}F^e{}_{cd}F_{cdf}%
-\frac12\partial^h\o\lambda^{bg}\partial_aF_a{}^{be}F_f{}^{gh}F^e{}_{cd}F_{cdf}%
\\
&{}
+\partial^h\o\lambda^{bg}\partial_aF_f{}^{be}F_a{}^{gh}F^e{}_{cd}F_{cdf}%
+2\partial^{[e|}\partial_a\o\lambda^{bh}F_a{}^{gh}F_f{}^{eb}F^{|g]}{}_{cd}F_{cdf}%
\\
&{}
+\partial^{[e|}\partial_f\o\lambda^{bh}F_a{}^{bg}F_a{}^{he}F^{|g]}{}_{cd}F_{cdf}%
-\partial_a\o\lambda^{bh}F_a{}^{ge}\partial^hF_f{}^{eb}F^g{}_{cd}F_{cdf}%
\\
&{}
+\frac12\partial_f\o\lambda^{bg}F_a{}^{he}\partial^gF_a{}^{bh}F^e{}_{cd}F_{cdf}%
+\partial^a\o\lambda^{bg}\mathcal R^{ab}{}_{fh}F_h{}^{eg}F^e{}_{cd}F_{cdf}\,,%
\end{aligned}
\end{equation}

\begin{equation}
\begin{aligned}
\Psi_{5,0}=&\,{}
4\partial^a\o\lambda^{bg}F^a{}_{ch}F_h{}^{bh}F_f{}^{he}F^{[g}{}_{cd}F^{e]}{}_{df}%
-4\partial^a\o\lambda^{bh}F^a{}_{ch}F_h{}^{be}F_f{}^{gh}F^{[g}{}_{cd}F^{e]}{}_{df}%
\\
&{}
-2\partial^a\o\lambda^{bg}F^a{}_{kh}F_h{}^{be}F^g{}_{cd}F^e{}_{df}F_{cfk}%
-2\partial_f\o\lambda^{bg}F_a{}^{bh}F_a{}^{he}\o D'_c(F^{[g}{}_{cd}F^{e]}{}_{df})%
\\
&{}
+2\partial_f\o\lambda^{bh}F_a{}^{bg}F_a{}^{he}\o D'_c(F^{[g}{}_{cd}F^{e]}{}_{df})%
+4\partial_a\o\lambda^{bh}F_a{}^{gh}F_f{}^{eb}\o D'_c(F^{[g}{}_{cd}F^{e]}{}_{df})%
\\
&{}
+4\partial_a\o\lambda^{bg}F_a{}^{eh}F_f{}^{hb}\o D'_c(F^{[g}{}_{cd}F^{e]}{}_{df})%
-2\partial_a\o\lambda^{bh}F_a{}^{gh}F_f{}^{eb}F^g{}_{ck}F^e{}_{kd}F_{cdf}%
\\
&{}
-2\partial_a\o\lambda^{bg}F_a{}^{eh}F_f{}^{hb}F^g{}_{ck}F^e{}_{kd}F_{cdf}%
+\partial_f\o\lambda^{bg}F_a{}^{bh}F_a{}^{he}F^g{}_{ck}F^e{}_{kd}F_{cdf}%
\\
&{}
-\partial_f\o\lambda^{bh}F_a{}^{bg}F_a{}^{he}F^g{}_{ck}F^e{}_{kd}F_{cdf}\,,%
\end{aligned}
\end{equation}

\begin{equation}
\begin{aligned}
\Psi_{5,1}=&\,{}
\partial^a\o\lambda^{bg}F_h{}^{eb}F_f{}^{ek}F^a{}_{fh}F^g{}_{cd}F^k{}_{cd}%
-\partial^a\o\lambda^{bg}F_h{}^{eb}F_f{}^{gk}F^a{}_{fh}F^e{}_{cd}F^k{}_{cd}%
\\
&{}
+\partial_f\o\lambda^{bg}F_a{}^{bh}F_a{}^{he}F^{[g}{}_{cd}F_f{}^{e]h}F^h{}_{cd}%
-\partial_f\o\lambda^{bh}F_a{}^{bg}F_a{}^{he}F^{[g}{}_{cd}F_f{}^{e]h}F^h{}_{cd}%
\\
&{}
-2\partial_a\o\lambda^{bh}F_a{}^{gh}F_f{}^{eb}F^{[g}{}_{cd}F_f{}^{e]h}F^h{}_{cd}%
-2\partial_a\o\lambda^{bg}F_a{}^{eh}F_f{}^{hb}F^{[g}{}_{cd}F_f{}^{e]h}F^h{}_{cd}%
\\
&{}
-\partial^a\o\lambda^{bg}F_h{}^{ae}\mathcal R'_{hf}{}^{be}F^g{}_{cd}F_{cdf}%
+\partial^a\o\lambda^{bg}F_h{}^{be}\mathcal R'^{ae}{}_{fh}F^g{}_{cd}F_{cdf}%
\\
&{}
-\partial^a\o\lambda^{bg}F_h{}^{be}\mathcal R'^{ag}{}_{fh}F^e{}_{cd}F_{cdf}%
+\partial^a\o\lambda^{bg}F_h{}^{eb}F^a{}_{fh}F^{gek}F^k{}_{cd}F_{cdf}%
\\
&{}
+2\partial^a\o\lambda^{bg}F_h{}^{eb}F^a{}_{fh}F^{[g}{}_{cd}F^{e]}{}_{fk}F_{cdk}%
+\frac12\partial^a\o\lambda^{bg}F_f{}^{ba}F_k{}^{gh}F_k{}^{he}F^e{}_{cd}F_{cdf}%
\\
&{}
-\frac12\partial^a\o\lambda^{bg}F_f{}^{be}F_h{}^{ak}F_h{}^{ke}F^g{}_{cd}F_{cdf}%
+\frac12\partial^a\o\lambda^{bg}F_f{}^{be}F_h{}^{ak}F_h{}^{kg}F^e{}_{cd}F_{cdf}%
\\
&{}
+\partial^a\o\lambda^{bg}F_f{}^{hb}F_k{}^{ga}F_k{}^{eh}F^e{}_{cd}F_{cdf}%
-\frac12\partial_f\o\lambda^{bh}F_a{}^{he}F_a{}^{gk}F^{ebk}F^g{}_{cd}F_{cdf}%
\\
&{}
-\partial_a\o\lambda^{bh}F_f{}^{eb}F_a{}^{gk}F^{ekh}F^g{}_{cd}F_{cdf}%
+\frac12\partial_f\o\lambda^{bg}F_a{}^{bh}\u D'^eF_a{}^{he}F^g{}_{cd}F_{cdf}%
\\
&{}
-\partial_f\o\lambda^{bh}F_a{}^{bg}\u D'^eF_a{}^{he}F^g{}_{cd}F_{cdf}%
-\partial_f\o\lambda^{bg}F_h{}^{be}\o D'_aF^{[e}{}_{ah}F^{g]}{}_{cd}F_{cdf}%
\\
&{}
-\frac32\partial_f\o\lambda^{bh}F_a{}^{he}\u D'^{[e}F_a{}^{bg]}F^g{}_{cd}F_{cdf}%
+\partial_a\o\lambda^{bg}F_a{}^{eh}F_k{}^{hb}F^e{}_{fk}F^g{}_{cd}F_{cdf}%
\\
&{}
-\partial_a\o\lambda^{bh}F_a{}^{ge}F_k{}^{eb}F^h{}_{fk}F^g{}_{cd}F_{cdf}%
-\partial_a\o\lambda^{bh}F_a{}^{he}F_k{}^{bg}F^e{}_{fk}F^g{}_{cd}F_{cdf}%
\\
&{}
-\frac12\partial_k\o\lambda^{bg}F_a{}^{bh}F_a{}^{he}F^e{}_{fk}F^g{}_{cd}F_{cdf}%
+\partial_k\o\lambda^{bh}F_a{}^{bg}F_a{}^{he}F^e{}_{fk}F^g{}_{cd}F_{cdf}%
\\
&{}
+\partial_k\o\lambda^{bh}F_f{}^{eb}F_a{}^{he}F^g{}_{ak}F^g{}_{cd}F_{cdf}%
+\partial_k\o\lambda^{bh}F_f{}^{eb}F_a{}^{gh}F^e{}_{ak}F^g{}_{cd}F_{cdf}\,,%
\end{aligned}
\end{equation}

\begin{equation}
\begin{aligned}
\Psi_{4,0,\Phi}=&\,{}
-2\partial_f\o\lambda^{bg}F_a{}^{bh}F_a{}^{he}F^{[e}{}_{df}D^{g]}F_d
+2\partial_f\o\lambda^{bh}F_a{}^{bg}F_a{}^{he}F^{[e}{}_{df}D^{g]}F_d
\\
&{}
+4\partial_a\o\lambda^{bh}F_a{}^{gh}F_f{}^{eb}F^{[e}{}_{df}D^{g]}F_d
+4\partial_a\o\lambda^{bg}F_a{}^{eh}F_f{}^{hb}F^{[e}{}_{df}D^{g]}F_d
\\
&{}
-4\partial^a\o\lambda^{bg}F_h{}^{eb}F^a{}_{fh}F^{[e}{}_{df}D^{g]}F_d\,,
\end{aligned}
\end{equation}

\begin{equation}
\begin{aligned}
\Psi_{4,1,\Phi}=&\,{}
-\partial_a\partial^g\o\lambda^{bg}F_a{}^{eh}F_f{}^{hb}F^e{}_{cd}F_{cdf}%
+\frac12\partial_f\partial^g\o\lambda^{bg}F_a{}^{bh}F_a{}^{he}F^e{}_{cd}F_{cdf}%
\\
&{}
-\frac12\partial_f\o\lambda^{bg}F_a{}^{bh}\u D^eF_a{}^{he}F^g{}_{cd}F_{cdf}%
+\partial_f\o\lambda^{bh}F_a{}^{bg}\u D^eF_a{}^{he}F^g{}_{cd}F_{cdf}%
\\
&{}
+\partial_a\o\lambda^{bg}F_f{}^{hb}\u D^eF_a{}^{eh}F^g{}_{cd}F_{cdf}%
-\partial_a\o\lambda^{bh}F_f{}^{eb}\u D^gF_a{}^{gh}F^e{}_{cd}F_{cdf}%
\\
&{}
+\partial_a\o\lambda^{bh}F_a{}^{gh}\u D^eF_f{}^{eb}F^g{}_{cd}F_{cdf}%
-\partial_h\o\lambda^{bg}F_f{}^{be}\u D^aF_h{}^{[e|a|}F^{g]}{}_{cd}F_{cdf}%
\\
&{}
+\partial_f\o\lambda^{bg}F_h{}^{be}\o D_aF^{[e}{}_{ah}F^{g]}{}_{cd}F_{cdf}%
+\partial_f\o\lambda^{bg}F_a{}^{bh}F_a{}^{he}F^{[g}{}_{cd}F^{e]}F_{cdf}%
\\
&{}
-\partial_f\o\lambda^{bh}F_a{}^{bg}F_a{}^{he}F^{[g}{}_{cd}F^{e]}F_{cdf}%
-2\partial_a\o\lambda^{bh}F_a{}^{gh}F_f{}^{eb}F^{[g}{}_{cd}F^{e]}F_{cdf}%
\\
&{}
-2\partial_a\o\lambda^{bg}F_a{}^{eh}F_f{}^{hb}F^{[g}{}_{cd}F^{e]}F_{cdf}%
+2\partial^a\o\lambda^{bg}F_h{}^{eb}F^a{}_{fh}F^{[g}{}_{cd}F^{e]}F_{cdf}%
\\
&{}
-2\partial_a\o\lambda^{bg}\partial_aF_f{}^{be}F^{[g}{}_{cd}F^{e]}F_{cdf}%
-\partial^a\partial^a\o\lambda^{bg}F_f{}^{be}F^{[g}{}_{cd}F^{e]}F_{cdf}%
\\
&{}
+\partial_f\o\lambda^{bg}\partial_aF_a{}^{be}F^{[g}{}_{cd}F^{e]}F_{cdf}\,,%
\end{aligned}
\end{equation}
while the equations of motion terms are
\begin{equation}
\begin{aligned}
\Psi_{\mathrm{e.o.m.}}=&\,{}
-4\partial_a\o\lambda^{bg}\partial_aF_f{}^{be}\mathcal R^{[g}{}_dF^{e]}{}_{df}%
-2\partial^a\partial^a\o\lambda^{bg}F_f{}^{be}\mathcal R^{[g}{}_dF^{e]}{}_{df}%
\\
&{}
+2\partial_f\o\lambda^{bg}\partial_aF_a{}^{be}\mathcal R^{[g}{}_dF^{e]}{}_{df}%
+\partial_a\o\lambda^{bg}\partial_a\mathcal R^b{}_fF^g{}_{cd}F_{cdf}%
\\
&{}
-\frac12\partial_f\o\lambda^{bg}\partial_a\mathcal R^b{}_aF^g{}_{cd}F_{cdf}%
+\frac12\partial^a\partial^a\o\lambda^{bg}\mathcal R^b{}_fF^g{}_{cd}F_{cdf}%
\\
&{}
+2\partial_f\o\lambda^{bg}F_a{}^{bh}F_a{}^{he}\mathcal R^{[g}{}_dF^{e]}{}_{df}%
-2\partial_f\o\lambda^{bh}F_a{}^{bg}F_a{}^{he}\mathcal R^{[g}{}_dF^{e]}{}_{df}%
\\
&{}
-4\partial_a\o\lambda^{bh}F_a{}^{gh}F_f{}^{eb}\mathcal R^{[g}{}_dF^{e]}{}_{df}%
-4\partial_a\o\lambda^{bg}F_a{}^{eh}F_f{}^{hb}\mathcal R^{[g}{}_dF^{e]}{}_{df}%
\\
&{}
+4\partial^a\o\lambda^{bg}F_h{}^{eb}F^a{}_{fh}\mathcal R^{[g}{}_dF^{e]}{}_{df}\,.%
\end{aligned}
\end{equation}

The underlined version of (\ref{eq:oR11}) is
\begin{equation}
\begin{aligned}
\u{\mathcal R}^{(1,1)}
=&\,
\frac16\mathcal R^{ab}{}_{cd}\mathcal R^{ae}{}_{cf}\mathcal R^{be}{}_{df}
-\mathcal R^{cd}{}_{ab}[F,F]_{be}{}^{cd}\partial_{(a}F_{e)}
+\frac12M_{abc}F_{abD}\partial^DF_c
\\
&{}
+\left(\mathcal R^{[c|g}{}_{ab}F_b{}^{g|d]}+(D^{[c}+\u D^{[c}+F^{[c})\mathcal R^{d]}{}_a+\frac12(\o D_b-F_b)[F,F]_{ab}{}^{cd}\right)
\\
&{}
\qquad\times\left(\frac12F_{aef}\mathcal R^{cd}{}_{ef}+2F^c{}_{ae}\mathcal R^d{}_e-F^c{}_{ef}\o D_aF^d{}_{ef}\right)
-\frac18\partial_e[F,F]_{ab}{}^{cd}\partial_e[F,F]^{cd}{}_{ab}
\\
&{}
+\frac14\o D_e[F,F]_{ab}{}^{cd}\o D_e[F,F]^{cd}{}_{ab}
-[F,F]_{ab}{}^{cd}(\mathcal R^{ce}{}_{af}-\frac{1}{32}\eta^{ce}\eta_{af}\mathcal R)[F,F]^{de}{}_{bf}
\\
&{}
-\frac12\partial_a(FF)_b{}^c\partial_a(FF)^c{}_b
+\o D_a(FF)_b{}^c\o D_a(FF)^c{}_b
\\
&{}
-(FF)_a{}^b(\mathcal R^{bd}{}_{ac}-\frac18\eta_{ac}\eta^{bd}\mathcal R)(FF)^d{}_c
+\frac{1}{32}\partial_gM_{aa}\partial_gM^{bb}
\\
&{}
-\frac{1}{128}M_{aa}M^{bb}\mathcal R
+\frac18\partial_g\big(F^aF^a+\frac16F^{abe}F^{abe}\big)\partial_g\big(F_cF_c+\frac16F_{cdf}F_{cdf}\big)
\\
&{}
+\frac{1}{32}(F^aF^a+\frac16F^{abc}F^{abc})(F_dF_d+\frac16F_{def}F_{def})\mathcal R
-\frac18\partial_aM^{bb}F_a{}^{cd}(\partial_e-F_e)F_e{}^{cd}
\\
&{}
-\frac12(FF)_a{}^b\o D_a(F^b{}_{cd}(\partial^e-F^e)F^e{}_{cd})
-\frac14\o D_gF^e{}_{ab}[F,F]^{ef}{}_{ab}F_g{}^{cd}F^{cdf}
+\u L_{\mathrm{e.o.m.}}\,,
\end{aligned}
%
%
\end{equation}
where
\begin{equation}
\begin{aligned}
\u L_{\mathrm{e.o.m.}}
=&
{}
-2\partial_aF_b\mathcal R^c{}_a\mathcal R^c{}_b
-\frac32\mathcal R^{ab}{}_{cd}\left(\mathcal R^a{}_c\mathcal R^b{}_d+\frac{1}{16}\mathcal R^{ab}{}_{cd}\mathcal R\right)
-2F_{cdf}\mathcal R^b{}_fD_c\mathcal R^b{}_d
\\
&{}
-F_{cdf}\mathcal R^{ab}{}_{cd}\u D^a\mathcal R^b{}_f
-2\mathcal R^{cd}{}_{gh}\o D_bF^c{}_{gh}\mathcal R^d{}_b
+[F,F]_{cd}{}^{ab}\left(\mathcal R^a{}_c\mathcal R^b{}_d-\frac{1}{16}\mathcal R^{ab}{}_{cd}\mathcal R\right)
\\
&{}
-\frac14M_{aa}\left(\mathcal R^b{}_c\mathcal R^b{}_c-\frac{1}{16}\mathcal R^2\right)
-\u D^aF_d{}^{bc}M^{ab}\mathcal R^c{}_d
-\frac14M_{aa}F^d{}_{bc}D_b\mathcal R^d{}_c
\\
&{}
-\frac12F^d{}_{ab}M_{abc}\mathcal R^d{}_c
-\frac12F^e{}_{df}F_d{}^{ab}(\partial_c-F_c)F_c{}^{ab}\mathcal R^e{}_f
+2F_{[c}{}^{bg}\mathcal R^{[b}{}_{d]}F^{e]}{}_{cf}\mathcal R^{eg}{}_{fd}
\\
&{}
+\frac18F^{cdf}\mathcal R^{cd}{}_{ab}F_{abh}\mathcal R^f{}_h
-\frac18F_e{}^{ab}\mathcal R^{ab}{}_{cd}F^f{}_{cd}\mathcal R^f{}_e
+\frac12F^aF_b\mathcal R^{ac}{}_{bd}\mathcal R^c{}_d
\\
&{}
-\frac18F_bF^a\mathcal R^a{}_b\mathcal R
+(FF)_c{}^bD_c(\mathcal R^b{}_dF_d)
-\frac12(F_cF_c+\frac16F_{cdf}F_{cdf})(FF)_a{}^h\mathcal R^h{}_a
\\
&{}
+(FF)^c{}_a[F,F]_{ab}{}^{cd}\mathcal R^d{}_b
+F_a[F,F]_{ab}{}^{cd}D^c\mathcal R^d{}_b
+\frac18F^c{}_{ab}(\partial^d-F^d)F^d{}_{ab}\mathcal RF^c
\\
&{}
-\frac12F_c{}^{ab}\partial^a\mathcal R\mathcal R^b{}_c
-\frac18F_{abg}\mathcal R^{cd}{}_{ab}F_g{}^{cd}\mathcal R
-\frac18F^c{}_{ab}\o D_eF^d{}_{ab}F_e{}^{cd}\mathcal R
\\
&{}
-\frac18F_c{}^{ef}F^e{}_{cg}\mathcal R^f{}_g\mathcal R\,.
\end{aligned}
\end{equation}

\bibliographystyle{nb}
\bibliography{biblio}{}
\end{document}